\documentclass[floats,floatfix,showpacs,amssymb,prd,twocolumn,superscriptaddress,nofootinbib,nolongbibliography,reprint]{revtex4-2}

\usepackage{amssymb,amsmath,verbatim,mathtools,needspace,enumitem,etoolbox,graphicx,physics,microtype,afterpage,bigints,gensymb,tabularx,mathtools}

\usepackage[dvipsnames, usenames]{xcolor}
\definecolor{linkcolor}{rgb}{0.0,0.3,0.5}
\definecolor{dodgerblue}{HTML}{1E90FF}
\usepackage[unicode, colorlinks=true, linkcolor=linkcolor, citecolor=linkcolor, filecolor=linkcolor,urlcolor=linkcolor, pdfusetitle]{hyperref}
\usepackage[all]{hypcap}
\usepackage[T1]{fontenc}
\usepackage[utf8]{inputenc}
\usepackage{orcidlink}
\usepackage{soul}
\newcommand{\ssim}{\mathchar"5218\relax\,}

\makeatletter
\newcommand*{\balancecolsandclearpage}{\close@column@grid \cleardoublepage \twocolumngrid}
\makeatother

\usepackage{lipsum}

\newcommand{\milan}{\affiliation{Dipartimento di Fisica ``G. Occhialini'', Universit\'a degli Studi di Milano-Bicocca, Piazza della Scienza 3, 20126 Milano, Italy}}
\newcommand{\infn}{\affiliation{INFN, Sezione di Milano-Bicocca, Piazza della Scienza 3, 20126 Milano, Italy}}

\begin{document}

\title{Residual eccentricity as a systematic uncertainty on \\ the formation channels of binary black holes}

\author{Giulia Fumagalli$\,$\orcidlink{}}
\email{g.fumagalli47@campus.unimib.it}
\milan \infn 

\author{Isobel Romero-Shaw$\,$\orcidlink{}}
\affiliation{Department of Applied Mathematics and Theoretical Physics, Cambridge CB3 0WA, United Kingdom}
\affiliation{Kavli Institute for Cosmology Cambridge, Madingley Road Cambridge CB3 0HA, United Kingdom}

\author{Davide Gerosa$\,$\orcidlink{0000-0002-0933-3579}}
\milan \infn

\author{Viola De Renzis$\,$\orcidlink{0000-0001-7038-735X}}
\milan \infn 

\author{Konstantinos Kritos$\,$\orcidlink{0000-0002-0212-3472}}
\affiliation{William H. Miller III Department of Physics and Astronomy, Johns Hopkins University, Baltimore, Maryland 21218, USA}

\author{Aleksandra Olejak$\,$\orcidlink{0000-0002-6105-6492}}
\affiliation{Max Planck Institut f\"ur Astrophysik, Karl-Schwarzschild-Stra{\ss}e 1, 85748 Garching bei München, Germany}


\date{\today}

\begin{abstract}

Resolving the formation channel(s) of merging binary black holes is a key goal in gravitational-wave astronomy.
The orbital eccentricity is believed to be a precious tracer of the underlying formation pathway, but is largely dissipated  %
during the usually long inspiral between black hole formation and merger. Most gravitational-wave sources are thus expected to enter the sensitivity windows of current detectors on configurations that are compatible with quasi-circular orbits. 
In this paper, we investigate the impact of ``negligible'' residual eccentricity ---lower than %
currently detectable by %
infer the formation history of binary black holes, focusing in particular on their spin orientations.
We trace the evolution of both %
observed and synthetic gravitational-wave events backward in time, while resampling their residual eccentricities to values that are below the detectability threshold. Eccentricities in-band as low as $\ssim 10^{-4}$ can lead to significant biases when reconstructing the spin directions, especially in the case of loud, highly precessing systems.
Residual eccentricity thus act like a systematic uncertainty for our astrophysical inference. As a mitigation strategy, one can marginalize the %
posterior distribution over the residual eccentricity using astrophysical predictions.

\end{abstract}

\maketitle

\section{Introduction}

Inferring the formation and evolutionary processes of merging stellar-mass black holes (BHs) is one of the most pressing questions %
 in modern high-energy astrophysics.  LIGO, Virgo, and KAGRA (LVK) %
  are delivering 
  hundreds of gravitational-wave (GW) detections \cite{2019PhRvX...9c1040A,2021PhRvX..11b1053A,2024PhRvD.109b2001A,2023PhRvX..13d1039A} and, despite some initial optimism, the ``formation-channel'' problem is still far from solved \cite{2021hgwa.bookE..16M,2021NatAs...5..749G,2022PhR...955....1M,2022LRR....25....1M,2017ApJ...846...82Z, 2023ApJ...955..127C}. %
BHs are simple objects in general relativity, which implies we have a limited number of  observables at our disposal, notably masses, spins, merger rates, and potentially the orbital eccentricity. Among these, spin directions~\cite{2013PhRvD..87j4028G,2016ApJ...832L...2R, 2017CQGra..34cLT01V,2017MNRAS.471.2801S,2018PhRvD..98h4036G,2020A&A...635A..97B,2021ApJ...921L...2O,2023ApJ...953...80B} 
 and eccentricity \cite{2014ApJ...784...71S,2018PhRvD..97j3014S, 2021ApJ...921L..43Z} %
 are believed to be clean indicators of the BH progenitors and their formation mechanisms, with large spin misalignments and large eccentricities pointing to dynamical assembly in\ new{densely populated environments}. %

Spin precession and orbital eccentricity are deeply intertwined. From a signal perspective, they both introduce %
 on timescales that are longer than that of the orbit \cite{1994PhRvD..49.6274A,1995PhRvD..52..821K, 2019MNRAS.490.5210R}. Such waveform features can be comparable, and isolating the two effects in current GW data presents challenges, at least for sufficiently short signals \cite{2023MNRAS.519.5352R}.
As for the binary evolution, couplings between the spin and eccentricity sectors of the binary dynamics introduces a non-trivial phenomenology. Sources formed with identical parameters but different eccentricities might enter the %
 LVK band with spin evolution patterns that are considerably different~\cite{2023PhRvD.108l4055F}.

This interplay between spins and eccentricity requires careful consideration when reconstructing the history of BH binaries from their  parameters, which are inferred from their GW signals. %
Propagating spin directions from formation to detection~\cite{2015PhRvD..92f4016G, 2023PhRvD.108b4042G} is crucial to transfer astrophysical predictions to the regime where sources are observable \cite{2013PhRvD..87j4028G, 2016ApJ...832L...2R,2018PhRvD..98h4036G}, thus allowing for a meaningful comparison with the data. The reverse operation, %
 ``back-propagating'' binaries from detection to formation, has key applications in single-event inference~\cite{2022PhRvD.105b4076M,2022PhRvD.106b3001J,2024PhRvD.109d3002K},  population studies \cite{2022PhRvD.105b4076M, 2023PhRvX..13a1048A}, and archival searches for, say, the LISA mission~\cite{2018PhRvL.121y1102W,2019PhRvD..99j3004G, 2021PhRvD.103b3025E,2022PhRvD.106j4034T}. 
In this paper, we show that eccentricity is a source of systematics that might affect, perhaps significantly, these downstream applications.

The BH binaries observed by %
LVK have been consistently reported to be quasi-circular \cite{2019ApJ...883..149A, 2023arXiv230803822T}, perhaps with a few exceptions \cite{2022ApJ...940..171R,2024arXiv240414286G}. This is not surprising, as leading-order post-Newtonian (PN) effects tend to remove eccentricity from binary systems on a timescale that is shorter than that of the inspiral itself \cite{1964PhRv..136.1224P}, leading to quasi-circular mergers even for systems that initially formed on highly eccentric orbits. Current ground-based interferometers are capable of detecting eccentricities as low as $e_{\rm thr}\simeq 0.05$ for GW150914-like binaries \cite{2018PhRvD..98h3028L}, %
meaning that binaries entering the sensitivity band of our instruments with eccentricities lower than this threshold will be reported as quasi-circular. In reality, GW sources have some residual eccentricity $e_{\rm res} < e_{\rm thr}$ which cannot be captured. 
For typical astrophysical environments, current models predict values of $e_{\rm res} \lesssim 10^{-8}$ for field binaries and  $\sim 10^{-7}\lesssim e_{\rm res} \lesssim 10^{-4}$ for the majority of binaries formed in  clusters (cf. Sec.~\ref{Marginalization over eres with astrophysical models} below). %
For binaries that are assembled dynamically, models also predict a small fraction of sources merging with  $e_{\rm res}>e_{\rm thr}$ which, if detected, are poised to be  highly informative \cite{2018PhRvD..97j3014S}. 
Quasi-circular orbits act as dynamical attractors for the forward evolution of BH binaries \cite{1964PhRv..136.1224P}, which unfortunately implies they act as repulsor where sources are back-propagated: for binaries on eccentric orbits, small variations of their parameters at GW detection implies large variations at BH formation. %
As first identified in previous work by some of us~\cite{2023PhRvD.108l4055F}, this issue is particularly concerning when considering the spin directions.
 Do residual, below-threshold eccentricities significantly affect our inference on the spin of BH binaries and thus their formation mechanism? How can this source of systematic be mitigated?  What are the consequences when inferring the formation channels of GW sources? 
We tackle these questions by applying the PN formalism of Refs.~\cite{2023PhRvD.108b4042G,2023PhRvD.108l4055F} to real %
LVK observations, synthetic GW events, and predictions from astrophysical population-synthesis simulations.
For each event, instead of making the common assumption that %
$e_{\rm res}=0$, %
we resample the residual eccentricity  $e_{\rm res}< e_{\rm thr}$ and back-propagate the resulting spin evolution to a common large separation. %
We compare the resulting spin distributions %
against those obtained  assuming quasi-circularity for the entire inspiral.
For current events up to GWTC-3 \cite{2019PhRvX...9c1040A,2021PhRvX..11b1053A,2024PhRvD.109b2001A,2023PhRvX..13d1039A}, we find that residual eccentricity introduces a rather mild systematic effect. This is due to  the large uncertainties on the spin directions as well as the weak or absent evidence of spin precession in most of the signals. We are safe, for now. Using a set of synthetic injections~\cite{2022PhRvD.106h4040D} with %
higher %
 signal-to-noise ratios (SNR) and prominent two-spin effects, 
we %
observe severe divergences between the eccentric and quasi-circular spin predictions. Unaccounted (and, at present, unaccountable) eccentricity at detection can introduce substantial variation in our  inference of spins at BH binary formation,%
 which implies that residual eccentricity is effectively a systematic one should take into account when inferring the origin of GW events.
We show that astrophysical models of BH binary formation \cite{2022MNRAS.516.2252O, 2020ApJS..247...48K,2022arXiv221010055K} 
can be used to heuristically ``marginalize'' over this systematic, thereby mitigating its impact.
Our paper is organized as follows: in Sec. \ref{Eccentric back propagations} we summarize the back-propagation procedure, %
the adopted statistical tools, and the targeted GW events; in Sec. \ref{Results} we present our results in terms of the bias induced %
and propose a strategy to marginalize over residual eccentricity using astrophysical predictions; %
in Sec. \ref{conclusion} we discuss the implications of our findings.

\section{Back propagation}\label{Eccentric back propagations}

\subsection{Black-hole binary dynamics}\label{Black-hole binary dynamics}

BH binaries are characterized by masses $m_{1,2}$, mass ratio $q=m_2/m_1\leq 1$,  total mass $M=m_1+m_2$, dimensionless spin magnitudes $\chi_{1,2} \in [0,1]$, polar spin angles $\theta_{1,2} \in [0, \pi]$ measured from the orbital angular momentum, and azimuthal spin angles $\Phi_{1,2}$ measured in the orbital plane. As long as the orbital timescale is much shorter than the inspiral timescale, orbits can be described  using Keplerian notions for the semi-major axis $a$ and the orbital eccentricity $e \in [0, 1)$. Hereafter we set $c=G=1$. 
Spin information is often condensed into the parameters $\chi_{\rm eff}$ (which includes information about the aligned components of the spins \cite{2008PhRvD..78d4021R}) and $\chi_{\rm p}$ (which instead captures spin precession \cite{2015PhRvD..91b4043S,2021PhRvD.103f4067G}). In the following, we use the ``average'' definition of  $\chi_{\rm p}$ put forward in Refs.~\cite{2021PhRvD.103f4067G,2022PhRvD.106h4040D}.
In particular, one has $\chi_{\text{p}} \in [0,2]$ where the lower bound implies aligned spins and $\chi_{\rm p}>1$ implies that both spins are precessing. %
Such definition can be trivially extended to eccentric orbits using the mapping detailed in Ref.~\cite{2023PhRvD.108l4055F}. In particular it is sufficient to substitute $r \to a(1-e^2)$ in the calculations of Ref.~\cite{2021PhRvD.103f4067G}.
We evolve binaries along their inspiral using the precession-averaged PN approach of Refs.~\cite{2023PhRvD.108b4042G, 2023PhRvD.108l4055F}; we refer to those previous papers for extensive derivation and validation of the formalism. In brief, we first neglect radiation reaction and solve the spin-precession problem semi-analytically by exploiting constants of motion~\cite{2015PhRvL.114h1103K,2008PhRvD..78d4021R}. GW emission is then introduced quasi-adiabatically, modeling the spin evolution as a continuous series of those semi-analytic solutions. Initially restricted to circular sources~\cite{2015PhRvD..92f4016G,2023PhRvD.108b4042G}, precession-averaged PN evolutions have been recently extended to small-to-moderate eccentricities $e\lesssim 0.6$ \cite{2023PhRvD.108l4055F}, where the limitation is set by the validity of the underlying orbit-averaged PN equations~\cite{1964PhRv..136.1224P}. %
Posterior samples are provided at the reference GW frequency $f_{\rm ref}=20~{\rm Hz}$%
, with the exception of GW190521 where instead $f_{\rm ref}=11~{\rm Hz}$ \cite{2021PhRvX..11b1053A}. From these, we calculate the semi-major axis $a$ using the PN expression reported in Eq. (4.13) of Ref. \cite{1995PhRvD..52..821K}, which results in separations of $\mathcal{O}(10 M)$. This conversion neglects the presence of eccentricity, which is appropriate for $e_{\rm res}<e_{\rm thr}$. %

\subsection{Residual eccentricity}

Residual eccentricity in-band might go unnoticed. We capture this effect by assigning each posterior sample a new value of the eccentricity that is below distinguishability threshold. In the following, we pursue two strategies:
\begin{enumerate}[label=(\roman*)]
\item  For a rather agnostic approach, we consider  thermal distributions $f(e) \propto e$ which naturally arise in statistical physics. We set $e\in[0,e_{\rm max}]$ such that $e_{\rm max}=0$ corresponds to assuming  binaries that evolved on quasi-circular orbits since formation. In the following, we use $e_{\text{max}}=10^{-4},10^{-3},10^{-2}$ such that all samples have an eccentricity $e 
\leq e_{\rm max}<e_{\rm thr}\simeq 0.05$ in band.  Note that eccentricities are resampled at $f_{\rm ref}=20, 11$~Hz while Ref.~\cite{2018PhRvD..98h3028L} quotes the threshold $e_{\rm thr}\sim 0.05$ at 10 Hz. %
We check that our resampled values of $e_{\rm res}$ are %
 lower than $e_{\rm thr}$: using \citeauthor{1964PhRv..136.1224P}' equations \cite{1964PhRv..136.1224P}, a binary %
 $e=10^{-2}$ at $20$ Hz reaches $e\sim0.02$ at $10$ Hz. For a more in-depth study, one should consider an eccentricity threshold that depends on the binary parameters (for instance: it is  easier to infer the eccentricity for binaries with lower masses because they complete more cycles in band \cite{2021ApJ...921L..31R}). We leave this to future work.
\item For a more astrophysical motivated approach, we attempt a direct modeling of the residual eccentricity predicted by state-of-the-art population-synthesis codes. We make use of simulations of both binaries formed in isolation \cite{2020A&A...636A.104B} and binaries assembled dynamically in dense stellar clusters~\cite{2022arXiv221010055K, 2021ApJ...921L..43Z}. We draw eccentricities directly from their distributions as predicted at GW detection; see Sec. \ref{Marginalization over eres with astrophysical models} for details. In particular, distributions are truncated at $e_{\rm res}<e_{\rm thr}$. This mimics a scenario where  highly eccentric sources predicted can be identified as such and are not affected by the systematic uncertainty targeted here.
\end{enumerate}

Given this initial configuration, we propagate binaries backward in time until the separation reaches $a=10^4 M$.  %
 This is rather conservative as BH binary formation typically happens as separations as large as $\sim 10^6 M$~\cite{2020FrASS...7...38M,2020MNRAS.492.2936A, 2022arXiv221010055K}. %
If the residual eccentricity in band is  $\neq 0$,  binaries will reach this large separation with orbits that are considerably more eccentric; this is due to  the repulsive character of  \citeauthor{1964PhRv..136.1224P}'~\cite{1964PhRv..136.1224P} equations. 
When back propagating from eccentricities $e_{\rm res}$ of $\mathcal{O}(10^{-2})$, the resulting eccentricities at $a=10^4 M$ can be larger than $0.6$, which was quoted as a conservative limit for validity of our formalism \cite{2023PhRvD.108l4055F}. Extending the applicability of PN integrations in the high-eccentricity regime is conceptual problem which is outside of the scope of this paper and will be addressed elsewhere \cite{giulianickdavidematteo}. For simplicity, here we nonetheless use the  \citeauthor{1964PhRv..136.1224P}' equations for all our sources, which is %
 in line with common practice in the astrophysical community. %

\subsection{Gravitational-wave signals}\label{Data}
We consider both current %
LVK observations as well as synthetic injections. Together, these sources cover a broad range of SNRs and degrees of spin precession. %

\begin{enumerate}[label=(\roman*)]

\item 
We consider 69 binary BH mergers as reported in the currently available GWTC catalog %
\cite{2019PhRvX...9c1040A,2021PhRvX..11b1053A,2024PhRvD.109b2001A,2023PhRvX..13d1039A} , selecting events with false alarm rate $<1\,\rm{yr}^{-1}$ and astrophysical probability $p_{\rm astro} > 0.5$. We use samples labeled as Mixed-Cosmo which combine results from the
\textsc{IMRPhenomXPHM}~\cite{2021PhRvD.103j4056P}  and \textsc{SEOBNRv4PHM}~\cite{2020PhRvD.102d4055O} waveform approximants. %
For event GW200129$\_$065458 we also consider posterior samples obtained with \textsc{NRSur7dq4}~\cite{2019PhRvR...1c3015V} from Ref.~\cite{2022Natur.610..652H}, which present a much stronger evidence for spin precession  (see also Ref.~\cite{2022PhRvL.128s1102V} for similar conclusions and  Ref.~\cite{2022PhRvD.106j4017P} for caveats related to  data quality). 

\item We also use 100 publicly available \cite{violadata} software injections which specifically target highly precessing systems. These were first presented in Figs. $4$ and $5$ of Ref.~\cite{2022PhRvD.106h4040D}. The parameters of these sources are distributed  by reweighting the uninformative prior used in %
LVK parameter estimation in favor of a uniform distribution in $\chi_{\rm p}\in [0,2]$ and applying a SNR threshold of 20.
In particular, these binaries have  masses $m_{1,2}\in [5,100]M_{\odot}$ constrained to $q \in [1/8,1]$ and $\mathcal{M}_c \in[10, 60]M_{\odot}$ (where $\mathcal{M}_c$ is the detector-frame chirp mass), spins $\chi_{1,2}\in[0,0.99]$, and luminosity distances $D_L \in [100, 5000] $ Mpc. Source are injected and recovered with the \textsc{IMRPhenomXPHM}~\cite{2021PhRvD.103j4056P} approximant, assuming  detector performances representative of the 4th observing run of %
LVK  and neglecting non-stationary noise realizations (for details see Ref.~\cite{2022PhRvD.106h4040D}). %
\end{enumerate}
Crucially, all these sources, both real and synthetic,  have been analyzed assuming BHs on quasi-circular orbits. We do not have posterior distributions for the residual eccentricity in band, which thus acts as a systematic uncertainty.

\subsection{Hellinger distance}\label{helldist}

Quantifying the impact of residual  eccentricities requires a notion of distance between probability distributions. Among the many available options \cite{chung1989measures}, we opt for the Hellinger distance %
 \cite{2021PhRvD.104h3008M}. This is defined as 

\begin{align}
d_{\rm H}(p,q)=\sqrt{1-\int  \sqrt{p(x) q(x)}\,d  x}\,,
\label{hdist}
\end{align}
where $p$ and $q$ are probability density functions of a (possibly multi-dimensional) variable $x$. 
In our case, these are the back-propagated posteriors obtained assuming either quasi-circularity or resampled eccentricities, respectively. We integrate over $x=(\cos\theta_1,\cos\theta_2)$ and have verified that $x=\chi_{\rm p}$ yields results that are largely indistinguishable. We evaluate probability density functions using Kernel Density Estimation and compute the integral in Eq.~(\ref{hdist}) via Monte Carlo.

 The Hellinger distance has the desirable properties of being symmetric, i.e. $d_{\rm H}(p,q)=d_{\rm H}(q,p)$, and defined in $[0,1]$ such that $d_{\rm H}=0$ implies that the two distributions are identical and $d_{\rm H}=1$ implies that  their supports are disjoint.
For some intuition, one can convert Hellinger-distance values into $\sigma$ levels by considering two one-dimensional Gaussians separated by  $n$ standard deviations.  The result is  %
\begin{equation} 
d_{\rm H}\big[\mathcal{N}(\mu, \sigma), \mathcal{N}(\mu+n\sigma, \sigma)\big] = \sqrt{1- \exp\left(- \frac{n^2}{8}\right)}\,.
\end{equation}
Some evaluations are reported in Table~\ref{hdsigma}. 

\begin{table}
\renewcommand{\arraystretch}{1.3}
\centering
\begin{tabular}{c@{\hspace{1em}}|@{\hspace{1em}}c@{\hspace{1em}}c@{\hspace{1em}}c@{\hspace{1em}}c@{\hspace{1em}}ccc}
& $1\sigma$ & $2\sigma$ & $3\sigma$ & $4\sigma$ & $5\sigma$ &\dots  & $10\sigma$
\\
\hline
$d_{\rm H}$ & 0.343 & 0.627 & 0.822 & 0.930 & 0.978 && $\sim1\!-\!10^{-6} $
\end{tabular}
\caption{Values of the Hellinger distance for two identical Gaussian distributions separated by an increasing number $n$ of standard deviations $\sigma$.}
\label{hdsigma}
\renewcommand{\arraystretch}{1}
\end{table}

\section{Systematic uncertainty}
\label{Results}

\subsection{A few examples} 

\label{A few examples}

\begin{figure*}[p]
\includegraphics[scale=0.47]{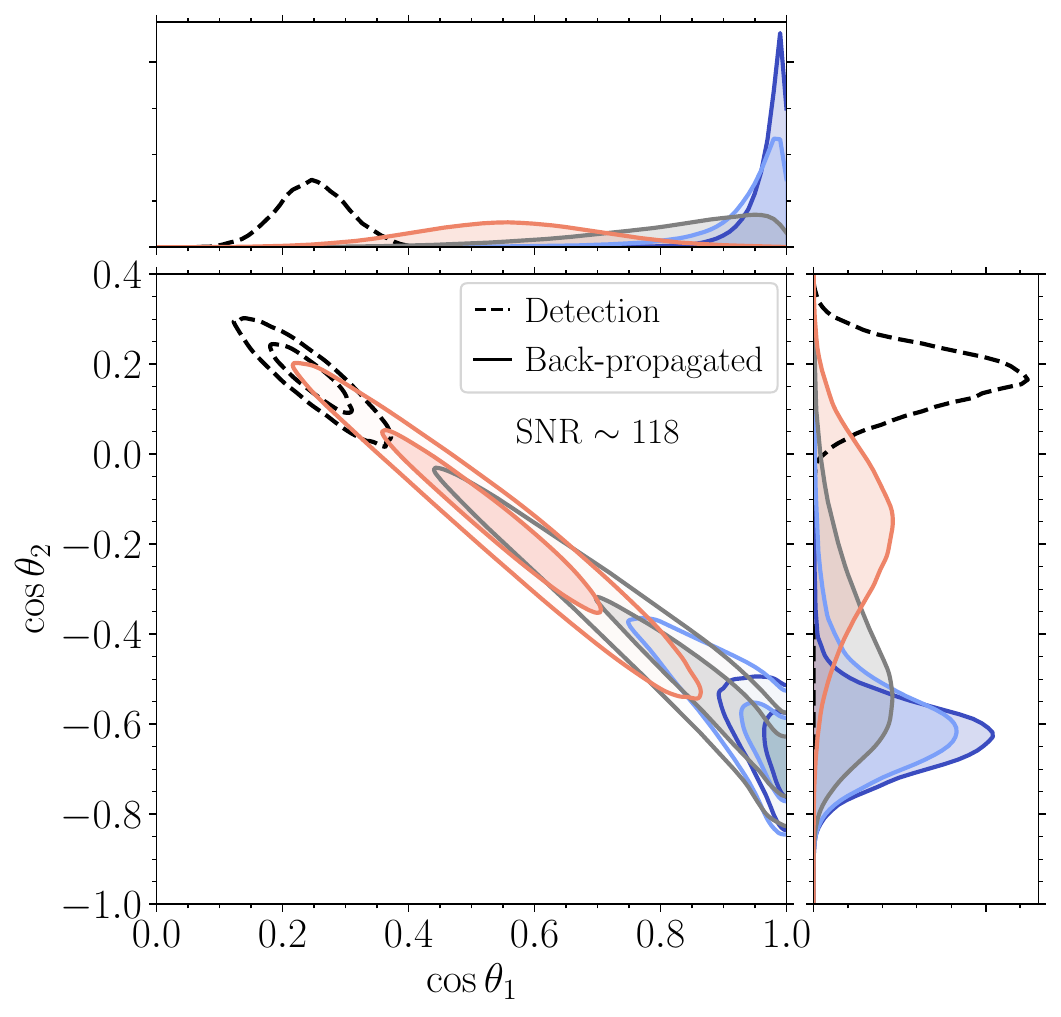}
$\quad$
\includegraphics[scale=0.47]{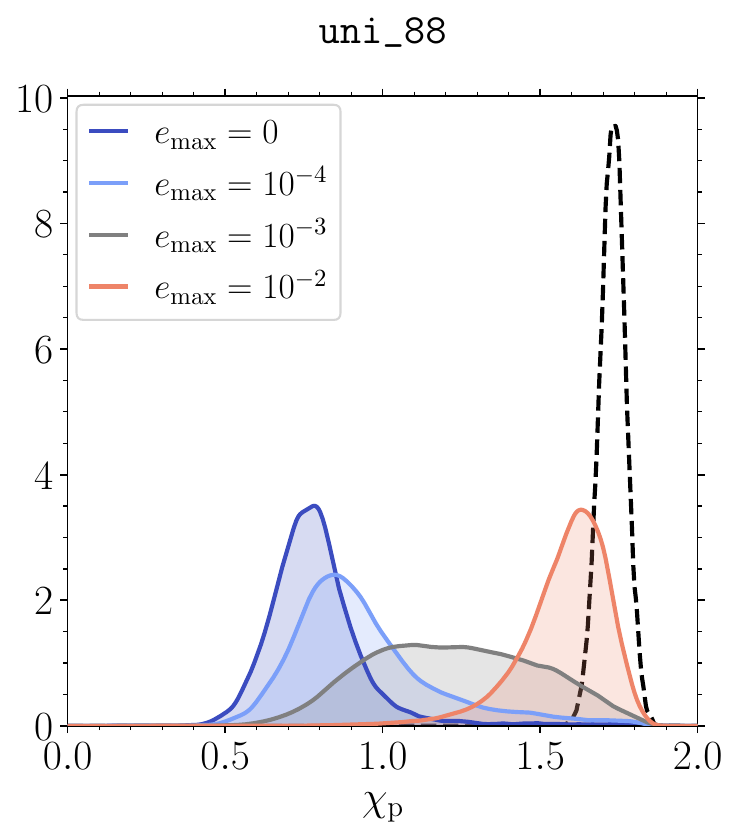}

\vspace{0.5cm}$\quad$

\includegraphics[scale=0.47]{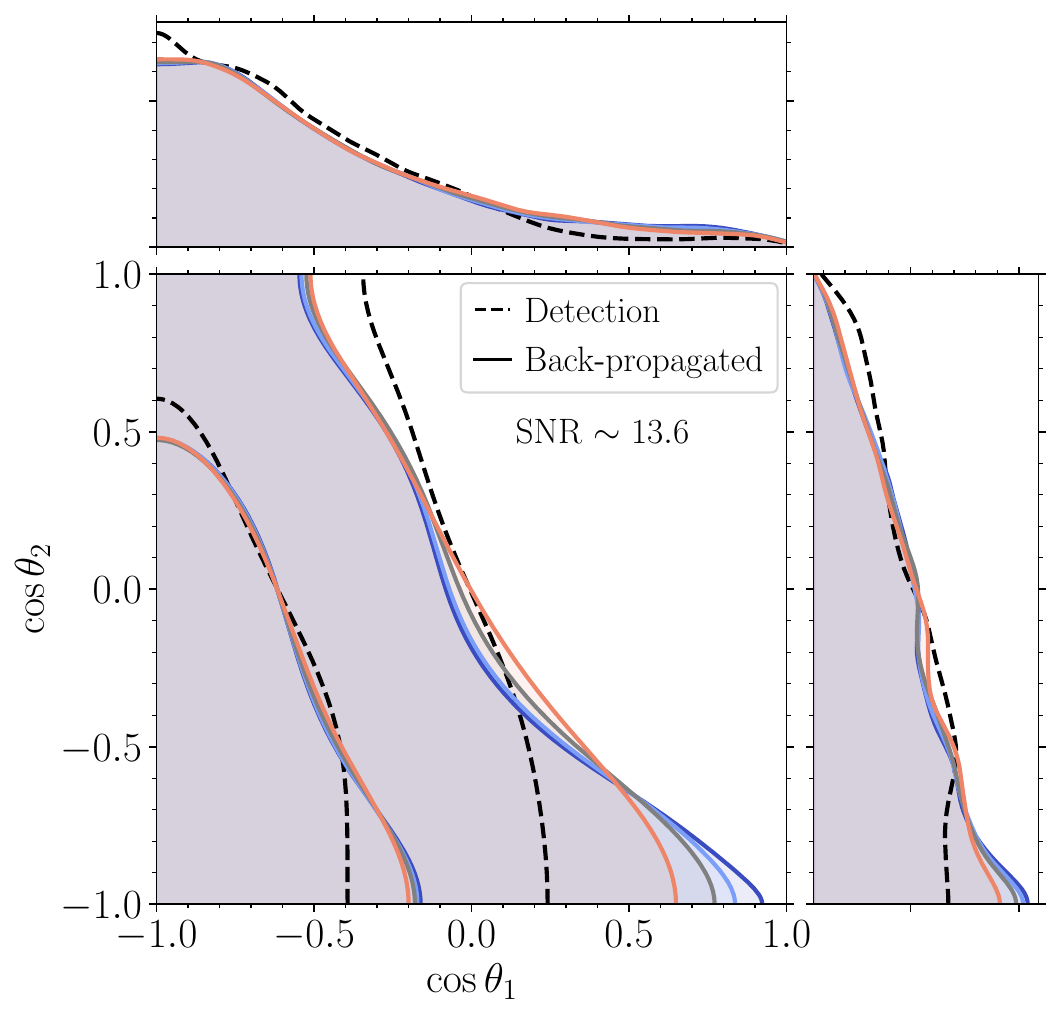}
$\quad$
\includegraphics[scale=0.47]{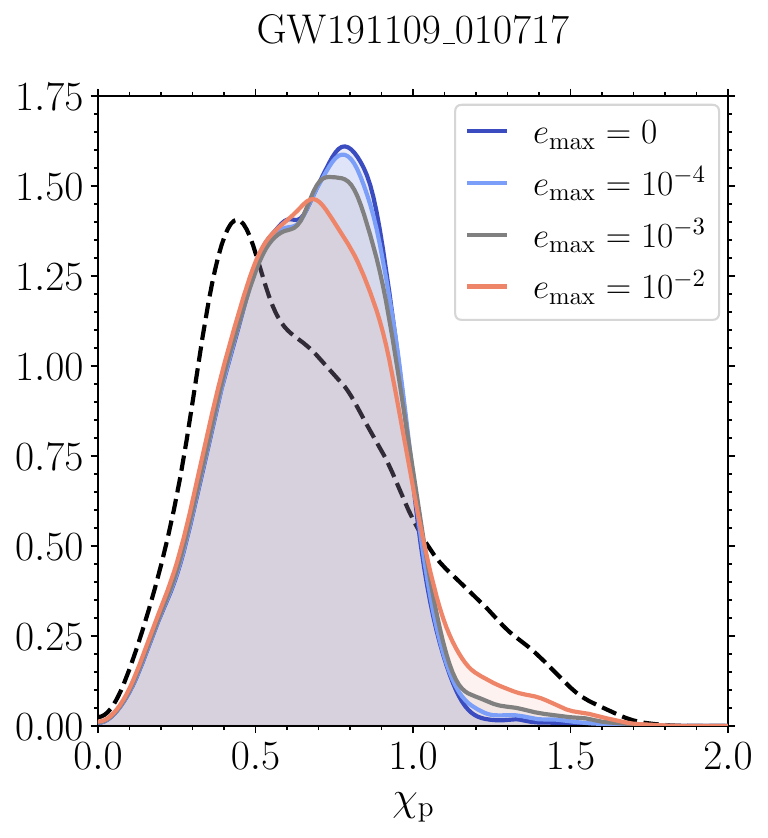}
\caption{
Back-propagated posterior distributions assuming both eccentric and quasi-circular configurations. The top panels show results for the synthetic signal labeled \texttt{uni\_88} in the dataset of Ref.~\cite{violadata}; the bottom panel shows event event GW$191109\_010717$~\cite{2023PhRvX..13d1039A}.
In both cases, left panels shows the joint posterior distribution of the spin orientations $(\theta_{1},\theta_{2})$ and right panels shows the posterior distribution of $\chi_{\rm p}$. Dashed curves show the distributions at detection ($f_{\rm ref}=20$~Hz); solid curves are obtained by back-propagating posterior samples to $a=10^{4} M$. %
The dark blue distributions ($e_{\rm max}=0$) assumes sources evolved on quasi-circular orbit throughout their inspiral. The other curves assume some residual eccentricity at detection. These are drawn from a thermal distribution $f(e)\propto e$ truncated at $e_{\rm max}= 10^{-4}$ (light blue), $10^{-3}$ (gray), and $10^{-2}$ (red). Contours in the 2-dimensional distributions on the left correspond to 50\% and 90\% credible intervals. An animated version of this figure is available at \href{https://davidegerosa.com/spinprecession/}{www.davidegerosa.com/spinprecession}.}

\label{best}
\end{figure*}

The top panels of Fig.~\ref{best} show posterior distributions of the spin angles ($\theta_1$, $\theta_2$) and  the effective precession parameter $\chi_{\rm p}$ for the synthetic event labeled as \texttt{uni\_88} in the dataset of Ref.~\cite{violadata}. This signal was handpicked for illustrative purposes because it has the largest Hellinger distance between the quasi-circular and eccentric reconstructions. This is a loud (${\rm SNR}\simeq 118$) system with BHs  of detector-frame masses $m_1=49.5^{+1.2}_{-1.8} M_\odot$, $m_2=45.3^{+1.1}_{-1.1} M_\odot $ and spin magnitudes $\chi_1=0.95^{+0.03}_{-0.06}$, $\chi_2=0.94^{+0.04}_{-0.07}$ (hereafter we report medians and 90\% credible intervals). %
We show posterior distributions at detection ($f_{\rm ref}=20$~Hz) as well as those resulting from our back-propagation procedure ($a=10^4M$) assuming residual eccentricities extracted from a thermal distribution truncated at some $e_{\rm max}$.
An animated version of Fig.~\ref{best} is available at \href{https://davidegerosa.com/spinprecession/}{www.davidegerosa.com/spinprecession} and shows the  evolution of such posteriors as a function of $a$.
 
As the adopted residual eccentricity increases, the back-propagated posterior distributions show larger deviations from the expected circular predictions, possibly leading to biased estimations of the spin parameters.
For this event, the Hellinger distance between the back-propagated posterior obtained with $e_{\rm max }=10^{-2}$  and its circular counterpart with $e_{\rm max }= 0$ is $\ssim 0.79$ (i.e. $\gtrsim 2 \sigma$ levels). 
 
It is informative to compare the locations of the back-propagated distributions against those at detection as a function of the residual eccentricity. When projected in the $(\cos\theta_1,\cos\theta_2)$ plane, all distributions lie roughly on the same diagonal line; this is because $\chi_{\rm eff}$ is a constant of motion for both the eccentric and the quasi-circular problem  \cite{2008PhRvD..78d4021R} (at least at 2PN in spin precession, which is the order considered here). But crucially, larger residual eccentricities imply spin orientations that are closer to those at detection. 

While this might seem counterintuitive at first, it is a direct consequence of the PN equations of motion \cite{2021arXiv210610291K,2023PhRvD.108l4055F}.
The evolution of the spin orientations depends on semi-major axis $a$ and eccentricity $e$ through the orbital angular momentum $L$. This acts much like a time coordinate, with sources evolving from large $L$ to small $L$. From the Newtonian scaling $L\propto \sqrt{a (1-e)}$, eccentric binaries have a smaller angular momentum than circular binaries for a given value of $a$. The posteriors shown in Fig.~\ref{best} are naturally ordered by angular momentum, with the eccentric back-propagated posteriors seated between the  back-propagated circular posteriors (which have the largest orbital angular momentum) and the posteriors at detection (which have the smallest orbital angular momentum). %

The bottom panels of Fig.~\ref{best} show back-propagated posteriors  for the %
LVK  event GW191109\_010717; once more, this system was chosen for its most pronounced differences between eccentric and circular back-propagated results across the current GW catalog.  GW191109\_010717 has an SNR of $\ssim 13.6$ and is consistent with a binary BH with $m_1=81^{+13}_{-9} M_\odot$, $m_2=60^{+16}_{-17} M_\odot$, $\chi_1=0.65^{+0.32}_{-0.58}$, and $\chi_2=0.82^{+0.15}_{-0.57}$. %
Although subtler, the same features we %
discussed for our simulated system are also present for GW191109\_010717; this is, however, characterized by much larger uncertainties (the SNR is about 10 times lower). In this case, we report a distance $d_{\rm H} \sim 0.01$ between back-propagated posteriors with $e_{\rm max}=10^{-2}$ and $e_{\rm max}=0$. Overall, this shows that, for current events, the systematic bias induced by neglecting residual eccentricity in band is  %
 mild. It might, however, become important for exceptionally loud events and/or with the next leap in sensitivity of our detectors. 
Despite the promising evidence for precession in GW200129\_065458~\cite{2022Natur.610..652H,2022PhRvL.128s1102V}, we find $d_{\rm H} \sim 10^{-5}$ between back-propagated posteriors with $e_{\rm max}=10^{-2}$ and $e_{\rm max}=0$ when using the samples of Ref.~\cite{2022Natur.610..652H}. This is likely attributed to the largely unconstrained posterior for the secondary spin.

\subsection{Parameter-space exploration}\label{Bias on spin parameters across all events}
 \begin{figure}[ht!]\label{fig:HD}
\includegraphics[width=\columnwidth]{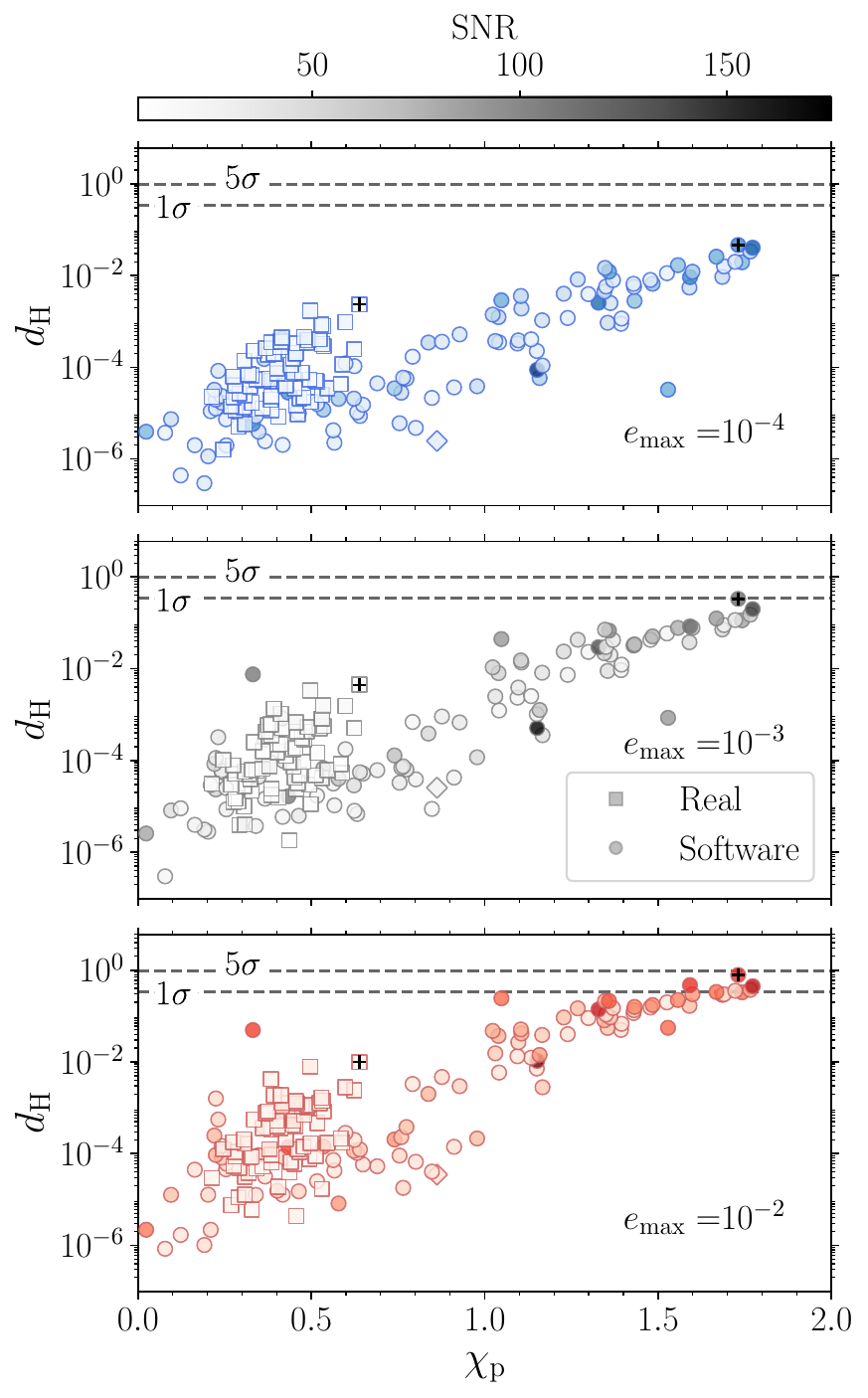}
 \caption{Hellinger distances between circular and eccentric back-propagated posterior distributions of $(\cos\theta_1,\cos\theta_2)$ as a function of the median value of $\chi_{\rm p}$ at detection. We consider 69 real events detected by %
 LVK  (squares) and 100 synthetic signals (circles).
Diamonds mark results obtained using the posteriors samples of Ref.~\cite{2022Natur.610..652H} for event GW200129$\_$065458. Crosses indicate the events used in Figs.~\ref{best} and \ref{fig:rapster}. 
Panels and colors refer to different assumptions for the residual eccentricity in band. This is extracted from a thermal distribution truncated at $e_{\rm max}=10^{-4}$ (blue, top), $e_{\rm max}=10^{-3}$ (grey, middle), and $e_{\rm max}=10^{-2}$ (red, bottom). %
Dashed horizontal lines mark the Hellinger distances values corresponding $1$ and $5$ $\sigma$-levels (cf. Table \ref{hdsigma}).%
}
\end{figure}

We now present a broader exploration of the parameter space using the Hellinger distance  (Sec.~\ref{helldist}) as our summary statistic.  Figure \ref{fig:HD} shows the values of $d_{\rm H}$ across  all the signals of Sec.~\ref{Data}, including both real and synthetic events.
 We compute Hellinger distances between the three eccentric back-propagated distributions $e_{\rm max}=10^{-2}, 10^{-3}, 10^{-4}$ and the circular case $e_{\rm res}= e_{\rm max}=0$. The distance is computed over $x=(\cos \theta_1,\cos \theta_2)$  and plotted against %
 the median value of $\chi_{\rm p}$ at detection. %

The Hellinger distance between our circular and eccentric back-propagated prediction increases with $\chi_{\rm p}$, indicating that the systematic uncertainty associated with residual eccentricity is strongly correlated with the amount of precession in the signal. More trivially and as already illustrated in Fig.~\ref{best}, we find that the distance increases with~$e_{\rm max}$.

On top of these main trends, Fig. \ref{fig:HD} shows substantial variability, even across several orders of magnitude in $d_{\rm H}$. This is not surprising given the high dimensionality of the BH binary parameter space and our sparse coverage. For a given value of $\chi_{\rm p}$ and $e_{\rm max}$, we find that sources at lower (higher) SNR tend to have larger (smaller) values of $d_{\rm H}$, corresponding to the regime where statistical systematic uncertainties dominate the error budget. This can also be seen by comparing the top and bottom panels of Fig.~\ref{best}.

 \subsection{Marginalization with astrophysical models}\label{Marginalization over eres with astrophysical models}

The use of thermal distributions truncated at $e_{\rm max} $ in the previous sections was arbitrary. We now present a more motivated strategy which makes direct use of state-of-the-art population-synthesis predictions. We interpret this as an ``astrophysical marginalization'' over the residual eccentricity: the back-propagated posteriors on the spin directions one infers are broader than one would obtain by naively assuming that binaries evolved on ideal quasi-circular orbits for their entire inspiral.  %

Eccentricity at merger is a distinct signature of binaries that formed in dynamical environments, notably dense stellar clusters. 
We use predictions obtained with \textsc{Rapster} \cite{2022arXiv221010055K} and \textsc{CMC}  \cite{2020ApJS..247...48K}, which are two of the current state-of-the-art cluster codes in the field. 
These are compared against a population of BH binaries formed in isolation as predicted with the \textsc{StarTrack} \cite{2020A&A...636A.104B,2022MNRAS.516.2252O} %
code.
For binaries formed in clusters, we label sources as  (i) ``{ejected}'' for BHs that merge outside of the cluster; (ii) ``{in-cluster}'' for BHs that merge inside the cluster following binary formation and (iii) ``{GW capture}'' which also merge inside the cluster but abruptly. %

The \textsc{Rapster} population is generated based on the assumptions detailed in Sec.~III.~A of Ref.~\cite{2024PhRvD.109l4029Y}. For \textsc{StarTrack}, we use the population referred to as ``default model'' in Ref. \cite{2024arXiv240412426O}.  Results from \textsc{CMC} are extracted from Fig.~1 of Ref. \cite{2021ApJ...921L..43Z}, which refers to detectable populations of sources at $f_{\rm ref}=10$~Hz. We eyeball their figure and consider skewed log-normal distributions with means of $10^{-6.8}$, $10^{-5.5}$, and $10^{-2.5}$, standard deviations of $10^{-1}$, and skewness parameters of $9$, $6$, and $1$  for the ejected, in-cluster and GW capture subchannels, respectively. We weight each of these distributions as reported in Table \ref{f}. %
For the \textsc{Rapster} and \textsc{StarTrack} populations, eccentricity and spin distributions are provided at BH formation, which we forward-propagate to  $f_{\rm ref}=10$~Hz as described in Sec.~\ref{Black-hole binary dynamics}. 
For these two codes, we post-process information for the GW detectability by considering a single LIGO instrument and a SNR threshold of 8.

Signals are computed using the \textsc{IMRPhenomXPHM}~\cite{2021PhRvD.103j4056P} waveform model and a noise power spectral density that is representative of the 4th observing run. This is consistent with the injections from Ref. \cite{2022PhRvD.106h4040D} introduced in Sec.~\ref{Data}. %
We analytically marginalize over the extrinsic parameters \cite{1993PhRvD..47.2198F,2024CQGra..41l5002G} as implemented in  Ref.~\cite{gwdet}. This results in a set of weights $p_{\rm det}$ for each source in our simulated populations, which is then normalized, i.e. we set $\sum_i {\rm p_{\rm det,i}}=1$.  For \textsc{Rapster}, the fractional contributions of each of the three subchannels is reported in Table~\ref{f}.  %

\begin{table}
\renewcommand{\arraystretch}{1.3}
\centering

\begin{tabular}{c@{\hspace{0.5em}}|@{\hspace{0.5em}}cc@{\hspace{0.5em}}|@{\hspace{0.5em}}cc@{\hspace{0.5em}}|@{\hspace{0.5em}}cc}

&\multicolumn{2}{c|@{\hspace{0.5em}}}{Ejected} & \multicolumn{2}{c|@{\hspace{0.5em}}}{In-cluster}  &\multicolumn{2}{c}{GW-capture}  
\\
\hline
\textsc{Rapster} &0.71& (0.70) &0.20 & (0.20) & 0.09 & (0.002)\\
\textsc{CMC}&  0.70& (0.70)&  0.20 & (0.20) &  0.10 & (0.079)\\
\end{tabular}
\caption{Fractional contributions to the BH merger rate for the three subchannels of the dynamical formation channel using predictions from \textsc{Rapster} and \textsc{CMC}. Numbers in parenthesis refer to sources with $e_{\rm res}<e_{\rm thr}=0.05$.%
}
\label{f}
\renewcommand{\arraystretch}{1}
\end{table}

  \begin{figure}\label{fig:eccentricity_distr}
\includegraphics[width=\columnwidth]{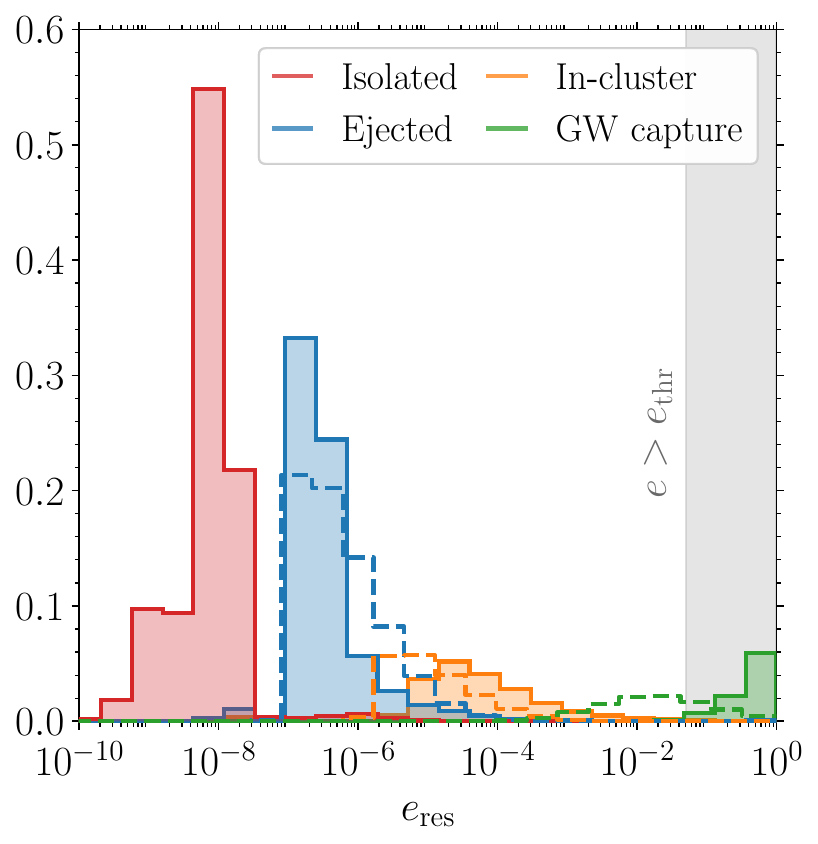}
 \caption{
 Distributions of eccentricities at the reference frequency $f_{\rm ref}=10$~Hz  for some representative astrophysical populations of BH binaries from stellar physics simulations. Distributions related to three subchannels within the two dynamically formed populations are shown in blue for ejected binaries merging outside the cluster, orange for binaries merging inside the cluster, and green for binaries formed via GW captures.
For these, solid and dashed histograms show predictions from the \textsc{Rapster} and \textsc{CMC} codes, respectively. %
The isolated-binary population from \textsc{StarTrack} is shown in red. The grey area to the left mark systems with eccentricity larger than the resolvability threshold, here set to $e_{\rm thr}=0.05$.
Binaries are weighted by their GW detectability and the resulting histograms are normalized to the cumulative detection probability, i.e. the sum of the bin heights for the isolated and dynamical channels is equal to $1$. The contributions provided by each of the subchannels is reported in Table \ref{f}.
}
\end{figure}

Figure~\ref{fig:eccentricity_distr} shows the resulting residual eccentricities at $f_{\rm ref}=10$~Hz. %
The distributions predicted for each formation (sub)channel are quite distinct and differ by orders of magnitudes. Predictions from the two populations of dynamically assembled binaries are in broad agreement. %
With the notable exception of GW captures, residual eccentricities fall well below the current distinguishability threshold $e_{\rm thr}\simeq 0.05$ \cite{2018PhRvD..98h3028L}. Even third-generation detectors, which might reach  $e_{\rm thr}\simeq 10^{-3}$~\cite{2024MNRAS.528..833S},  %
will not be of much help here. In the following, we truncate the distributions of Fig.~\ref{fig:eccentricity_distr} to $e\leq e_{\rm thr}\simeq 0.05$, which mimics a scenario where highly eccentric sources can be identified as such and thus do not pollute our inference of binaries that appear quasi-circular. Note that  Fig.~\ref{fig:eccentricity_distr} is likely to overestimate the importance of highly eccentric sources because of the related difficulties with estimating GW detectability; this is partly motivates the difference between \textsc{CMC} and \textsc{Rapster} for the GW-capture subchannel.  %

\begin{figure*}
\includegraphics[scale=0.47]{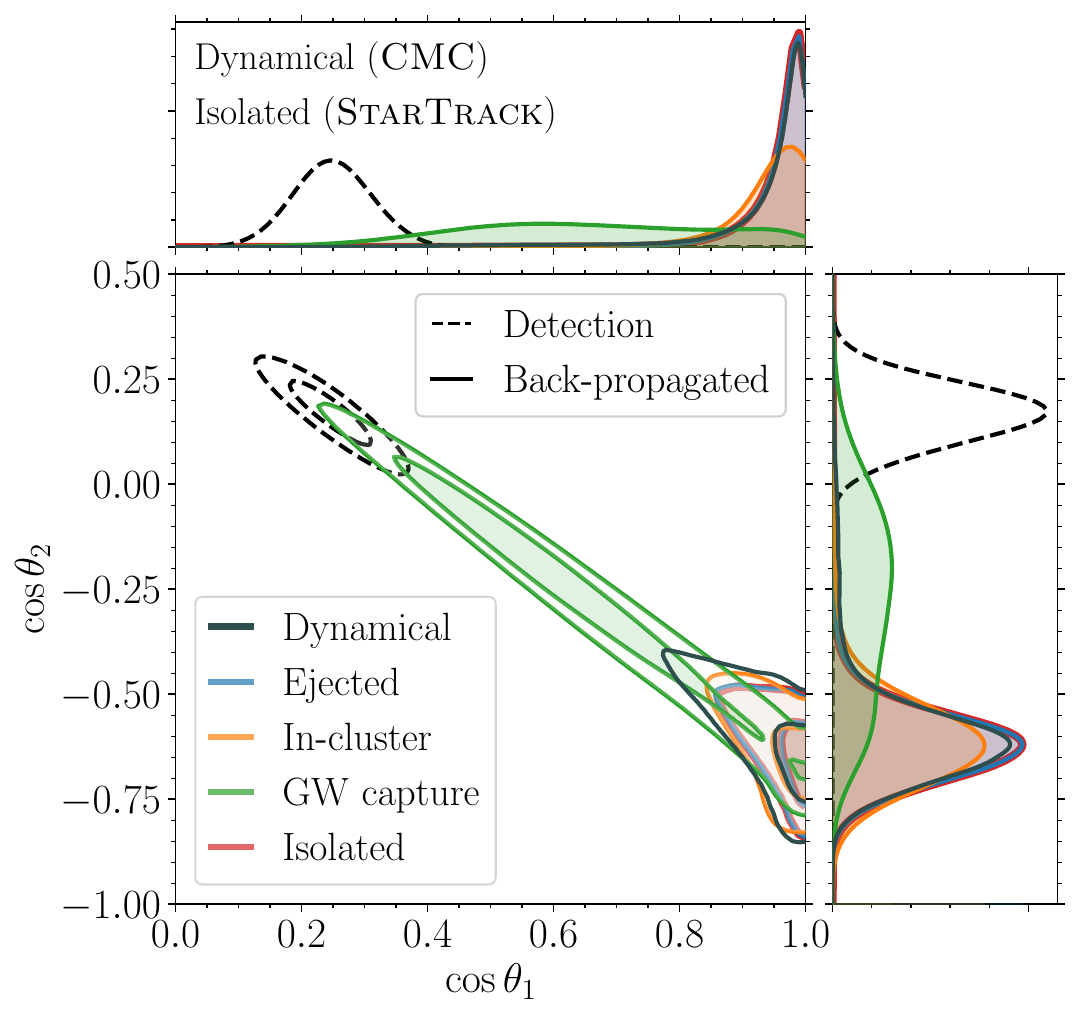}
$\quad$
\includegraphics[scale=0.47]{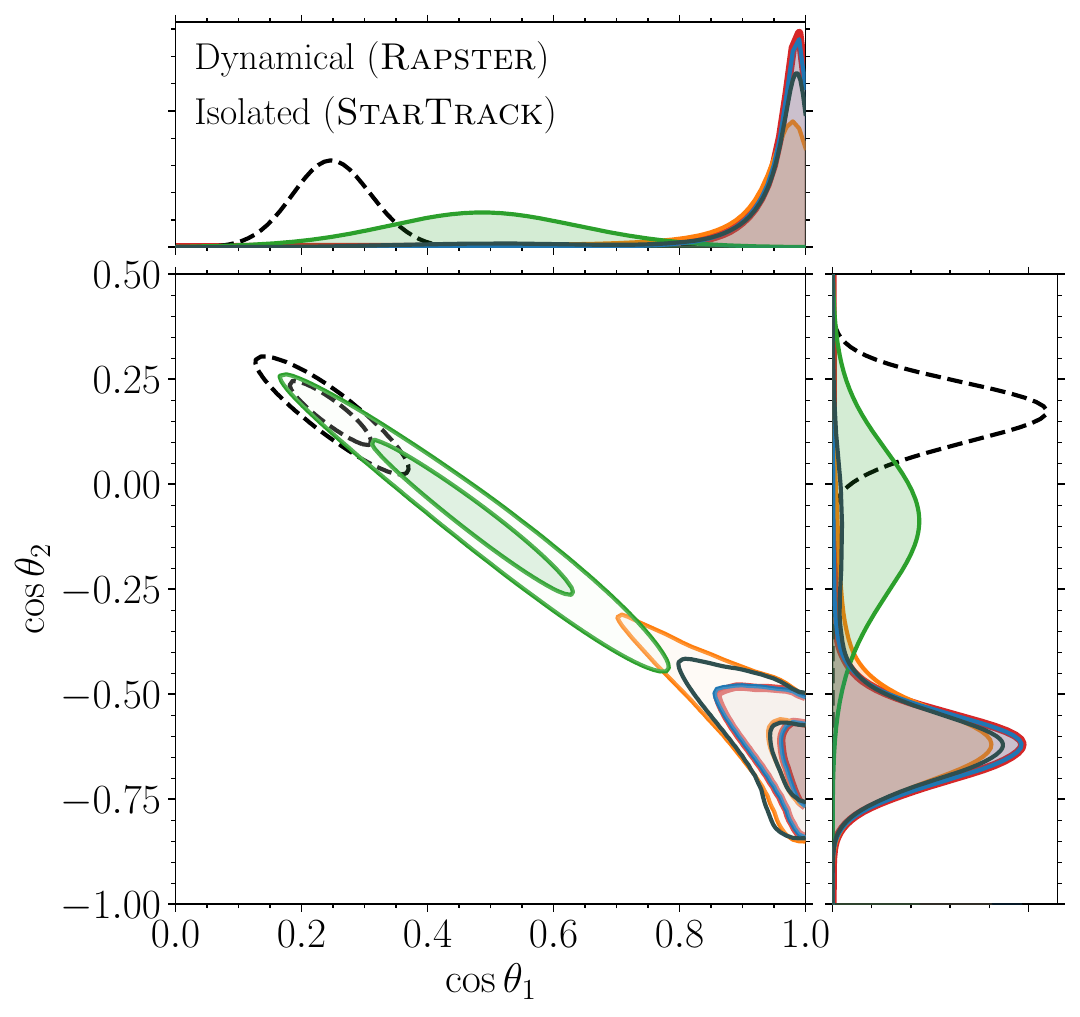}
\caption{Back-propagated posteriors of the spin directions assuming  astrophysically motivated distributions of residual eccentricity. We consider the SNR$\sim 100$ synthetic signal \texttt{uni\_88} from the dataset of Ref.~\cite{violadata} (cf. top panel of Fig.~\ref{best}). Dashed curves show  distributions at detection ($f_{\rm ref}=20$~Hz) while solid curves shows distribution a the common large separation ($a=10^{4} M$). %
Contours in the 2-dimensional distributions correspond to 50\% and 90\% credible intervals.
The left (right) panel uses predictions for dynamically assembled BHs by the \textsc{Rapster} (\textsc{CMC}) code, see text for details. %
We show predictions from individual subchannels (blue, orange, green)  as well as a joint predictions that keeps into account the relative mixing fractions (dark grey).
For reference both panels report results from an isolated-star distribution from \textsc{StarTrack} (red), which are virtually indistinguishable from those obtained with the ejected subchannel (blue). An animated version of this figure is available at \href{https://davidegerosa.com/spinprecession/}{www.davidegerosa.com/spinprecession}.
}
\label{fig:rapster}
\end{figure*}

Given these predictions, we perform the same operation discussed in Sec.~\ref{Results} while drawing residual eccentricities from the distributions of Fig.~\ref{fig:eccentricity_distr}. %
Notably, this does not provide an exhaustive picture of all possible eccentric mergers. Eccentric binaries are expected to form in environments other than globulars, such as active galactic nuclei \cite{2022Natur.603..237S} and different types of star clusters \cite{2022MNRAS.511.1362T}. %
 For clarity, we limit our analysis to the populations described above but stress that the methods presented here can be straightforwardly applied to other predictions. 

Figure \ref{fig:rapster} shows results for the same software injection considered in top panels of Fig.~\ref{best}. 
We first propagate the GW posteriors forward from $f_{\rm ref}=20$~Hz (where the injections was performed) to $10$~Hz (where the residual eccentricity is provided, see above). We then back-propagate spin directions to the joint large separation $a=10^4$M. The detailed evolution of these posteriors as a function of $a$ is provided in an animated version of Fig.~\ref{fig:rapster}, which is available at \href{https://davidegerosa.com/spinprecession/}{www.davidegerosa.com/spinprecession}.

The results of Fig.~\ref{fig:rapster} should be interpreted as our best estimate of the spin orientations  at (or, more accurately, close to) BH formation given the detected GW data and \emph{assuming} a specific formation pathway. Figure~\ref{fig:rapster} shows the same trends highlighted in Sec.~\ref{A few examples}: distributions with the lowest (largest) residual eccentricity such as isolated binaries (GW captures) have back-propagated spin directions that are further (closer) to those at detection, and the main reason is that those binaries have a larger (lower) angular momentum.
When considering contributions from all the dynamical subchannels, our predictions sit close to the distributions for the ejected and in-cluster subchannels because those two classes dominate the dynamical formation merger rate, see Fig.~\ref{fig:eccentricity_distr} and Table~\ref{f} (recall that we are truncating the distribution at $e_{\rm thr}$, which further diminishes the GW-capture contribution).

Crucially, the procedure we present heavily relies on the adopted astrophysical population, which is subject to significant uncertainties. Furthermore, our results do not capture correlations between the residual eccentricity and the other parameters (say the masses or the spins) of the back-propagated signal. While our results are indicative even for fixed populations,  we ultimately envision this approach to be used in synchronicity with full GW population fits, the results of which can then be repurposed to shed light on individual detections, see e.g. Ref.~\cite{2021PhRvD.104h3008M}.

\section{Conclusions}\label{conclusion}

Some of the key observables in GW astronomy are time dependent, i.e. their values change, sometime significantly, as sources inspiral. Among these are the spin directions and the orbital eccentricity. Capturing their coupled evolution between BH formation and GW detection is important for an unbiased  reconstruction of formation pathway of stellar-mass BH binaries. 
Current ground-based detectors are not able to distinguish eccentricities in BHs binaries which are $\lesssim 0.05$ at $10$ Hz. Those sources are thus typically reported as compatibile with binaries on quasi-circular orbits.

Eccentricity and spin inclinations are coupled \cite{2023PhRvD.108l4055F} and, as illustrated in this paper, mismodeling in the former at detection translates into biased distributions of the latter at BH formation. %
Residual eccentricity is a systematic uncertainty for the spin directions, which should be considered in addition to the statistical uncertainty due to detector noise. In particular, our study illustrates that:

\begin{enumerate}[label=(\roman*)]

 \item The systematic bias on the spin directions due to residual eccentricity increases with both the residual eccentricity itself as well as the amount of spin precession in the signals. This illustrate that this effect is indeed due to couplings between the precession and eccentricity sectors of the BH dynamics.
 \item For the events reported by %
LVK  so far, the systematic uncertainty due to residual eccentricity is subdominant compared to the statistical uncertainty of the spin directions ($d_{\rm H}\lesssim 0.01$).
 \item The situation is reversed for putative sources with SNR $\sim100$ and precessing spins, which might present spin posteriors at BH formation that are essentially disjoint from those obtained assuming quasi-circular orbits ($d_{\rm H}\sim 0.8$).
 \end{enumerate}
 
 Residual eccentricity influences not only the spin predictions of individual events but also those of population-level analyses, which we plan to investigate in future work. Features in the inferred population of the BH spins are used to constrain specific mechanisms behind the assembly of BH binaries such as supernova kicks, tidal interactions, mass transfer, and even the symmetry of the environment~\cite{2013PhRvD..87j4028G,2017MNRAS.471.2801S,2021PhRvD.103f3007S,2021PhRvD.103f3032S,2022MNRAS.514.3886M,2022ApJ...938...66T,2022MNRAS.517.2738M,2023PhRvD.108h3033S,2023ApJ...955..127C,2024ApJ...964L...6A,2024arXiv240412426O}. %
 GW population fits that include spin directions are now performed after binaries have been back-propagated to past time infinity (i.e. $f_{\rm ref}=0$~Hz) \emph{assuming quasi-circular orbits} \cite{2022PhRvD.105b4076M,2022PhRvD.106b3001J,2023PhRvX..13a1048A}. 
The evolution of eccentric binaries to past time infinity is still an open problem because the commonly employed averaging techniques~\cite{1964PhRv..136.1224P,2023PhRvD.108l4055F} are expected to break down~\cite{2023PhRvD.108b4042G}. In this paper we sidestepped the issue by halting our evolution at a large but finite separation, $a=10^4$M. Extrapolating from what we presented here, we speculate that importance of the systematic bias due to residual eccentricity might increase if evolutions are extended to larger separations, though we stress this requires the development of an appropriate PN-evolution strategy \cite{giulianickdavidematteo} in conjunction with a population fit in hierarchical Bayesian statistics. %

Residual eccentricity might also impact archival searches of stellar-mass BH binaries in LISA data It has been argued that ground-based detections of merging binaries might be used to ``know where to look for,'' thus digging deeper into the LISA noise and increasing the sensitivity of our searches~\cite{2018PhRvL.121y1102W,2021PhRvD.103b3025E}. This paper shows that residual eccentricity is an important caveat to this statement; we will not really know where to look for.
The uncertainty is further exacerbated by the known degeneracy between eccentricity and time to merger~\cite{2021PhRvD.104d4065B}. %
More broadly, residual eccentricity will have an impact whenever the spin directions are considered; this systematic effect will become more severe for larger SNRs, notably including massive BHs observed by LISA \cite{2024arXiv240207571C} and stellar-mass BHs observed by 3rd-generation ground-based observatories~\cite{2019BAAS...51g..35R}. 
Still operating at the single-event level, we explored a potential strategy to quantify the systematic error due to residual eccentricity when inferring the BH spin directions, which relies on pre-computed astrophysical distributions. While admittedly model-dependent, the procedure highlighted in this paper mitigates the eccentricity systematics by folding astrophysical modeling into the spin-orientation inference. The adopted astrophysical distribution of $e_{\rm res}$ is yet another systematic, but this is arguably better than blindly assuming $e_{\rm res}=0$ for all posterior samples of all the events as has been done so far.

\acknowledgements

We thank Matthew Mould, Philippa Cole, Arianna Renzini, Ssohrab Borhanian, Nick Loutrel, Chris Moore, Ulrich Sperhake, Johan Samsing, and Fabio Antonini for discussions.
G.F., D.G., and V.D.R. are supported by ERC Starting Grant No.~945155--GWmining, 
Cariplo Foundation Grant No.~2021-0555, MUR PRIN Grant No.~2022-Z9X4XS, 
and the ICSC National Research Centre funded by NextGenerationEU.  
G.F. is supported by Sigma Xi Grant No. G20230315-4609 and an Erasmus+ scholarship.
I.R.S. acknowledges support received from the Herchel Smith Postdoctoral Fellowship Fund.
D.G. is supported by MSCA Fellowships No.~101064542--StochRewind and  No.~101149270--ProtoBH.
K.K. is supported by Onassis Foundation Scholarship No.~FZT041-1/2023-2024, NSF Grants No. AST-2006538, PHY-2207502, PHY-090003 and PHY-20043, NASA Grants No. 20-LPS20-0011 and 21-ATP21-0010, John Templeton Foundation Grant No.~62840, Italian Ministry of Foreign Affairs and International Cooperation Grant No.~PGR01167, and the Simons Foundation.
A.O. is supported by the Foundation for Polish Science (FNP).
Computational work was performed at CINECA with allocations 
through INFN and Bicocca.

\bibliography{residual}

\begin{thebibliography}{83}%
\makeatletter
\providecommand \@ifxundefined [1]{%
 \@ifx{#1\undefined}
}%
\providecommand \@ifnum [1]{%
 \ifnum #1\expandafter \@firstoftwo
 \else \expandafter \@secondoftwo
 \fi
}%
\providecommand \@ifx [1]{%
 \ifx #1\expandafter \@firstoftwo
 \else \expandafter \@secondoftwo
 \fi
}%
\providecommand \natexlab [1]{#1}%
\providecommand \enquote  [1]{``#1''}%
\providecommand \bibnamefont  [1]{#1}%
\providecommand \bibfnamefont [1]{#1}%
\providecommand \citenamefont [1]{#1}%
\providecommand \href@noop [0]{\@secondoftwo}%
\providecommand \href [0]{\begingroup \@sanitize@url \@href}%
\providecommand \@href[1]{\@@startlink{#1}\@@href}%
\providecommand \@@href[1]{\endgroup#1\@@endlink}%
\providecommand \@sanitize@url [0]{\catcode `\\12\catcode `\$12\catcode
  `\&12\catcode `\#12\catcode `\^12\catcode `\_12\catcode `\%12\relax}%
\providecommand \@@startlink[1]{}%
\providecommand \@@endlink[0]{}%
\providecommand \url  [0]{\begingroup\@sanitize@url \@url }%
\providecommand \@url [1]{\endgroup\@href {#1}{\urlprefix }}%
\providecommand \urlprefix  [0]{URL }%
\providecommand \Eprint [0]{\href }%
\providecommand \doibase [0]{https://doi.org/}%
\providecommand \selectlanguage [0]{\@gobble}%
\providecommand \bibinfo  [0]{\@secondoftwo}%
\providecommand \bibfield  [0]{\@secondoftwo}%
\providecommand \translation [1]{[#1]}%
\providecommand \BibitemOpen [0]{}%
\providecommand \bibitemStop [0]{}%
\providecommand \bibitemNoStop [0]{.\EOS\space}%
\providecommand \EOS [0]{\spacefactor3000\relax}%
\providecommand \BibitemShut  [1]{\csname bibitem#1\endcsname}%
\let\auto@bib@innerbib\@empty
%</preamble>
\bibitem [{\citenamefont {{Abbott}}\ \emph
  {et~al.}(2019{\natexlab{a}})\citenamefont {{Abbott}}, \citenamefont
  {{Abbott}}, \citenamefont {{Abbott}}, \citenamefont {{Abraham}},
  \citenamefont {{Acernese}}, \citenamefont {{Ackley}}, \citenamefont
  {{Adams}}, \citenamefont {{Adhikari}}, \citenamefont {{Adya}}, \citenamefont
  {{Affeldt}},\ and\ \citenamefont {et~al.}}]{2019PhRvX...9c1040A}%
  \BibitemOpen
  \bibfield  {author} {\bibinfo {author} {\bibfnamefont {B.~P.}\ \bibnamefont
  {{Abbott}}}, \bibinfo {author} {\bibfnamefont {R.}~\bibnamefont {{Abbott}}},
  \bibinfo {author} {\bibfnamefont {T.~D.}\ \bibnamefont {{Abbott}}}, \bibinfo
  {author} {\bibfnamefont {S.}~\bibnamefont {{Abraham}}}, \bibinfo {author}
  {\bibfnamefont {F.}~\bibnamefont {{Acernese}}}, \bibinfo {author}
  {\bibfnamefont {K.}~\bibnamefont {{Ackley}}}, \bibinfo {author}
  {\bibfnamefont {C.}~\bibnamefont {{Adams}}}, \bibinfo {author} {\bibfnamefont
  {R.~X.}\ \bibnamefont {{Adhikari}}}, \bibinfo {author} {\bibfnamefont
  {V.~B.}\ \bibnamefont {{Adya}}}, \bibinfo {author} {\bibfnamefont
  {C.}~\bibnamefont {{Affeldt}}},\ and\ \bibinfo {author} {\bibnamefont
  {et~al.}},\ }\href {https://doi.org/10.1103/PhysRevX.9.031040} {\bibfield
  {journal} {\bibinfo  {journal} {Phys. Rev. X}\ }\textbf {\bibinfo {volume}
  {9}},\ \bibinfo {eid} {031040} (\bibinfo {year} {2019}{\natexlab{a}})},\
  \Eprint {https://arxiv.org/abs/1811.12907} {arXiv:1811.12907 [astro-ph.HE]}
  \BibitemShut {NoStop}%
\bibitem [{\citenamefont {{Abbott}}\ \emph {et~al.}(2021)\citenamefont
  {{Abbott}}, \citenamefont {{Abbott}}, \citenamefont {{Abraham}},
  \citenamefont {{Acernese}}, \citenamefont {{Ackley}}, \citenamefont
  {{Adams}}, \citenamefont {{Adams}}, \citenamefont {{Adhikari}}, \citenamefont
  {{Adya}}, \citenamefont {{Affeldt}},\ and\ \citenamefont
  {et~al.}}]{2021PhRvX..11b1053A}%
  \BibitemOpen
  \bibfield  {author} {\bibinfo {author} {\bibfnamefont {R.}~\bibnamefont
  {{Abbott}}}, \bibinfo {author} {\bibfnamefont {T.~D.}\ \bibnamefont
  {{Abbott}}}, \bibinfo {author} {\bibfnamefont {S.}~\bibnamefont {{Abraham}}},
  \bibinfo {author} {\bibfnamefont {F.}~\bibnamefont {{Acernese}}}, \bibinfo
  {author} {\bibfnamefont {K.}~\bibnamefont {{Ackley}}}, \bibinfo {author}
  {\bibfnamefont {A.}~\bibnamefont {{Adams}}}, \bibinfo {author} {\bibfnamefont
  {C.}~\bibnamefont {{Adams}}}, \bibinfo {author} {\bibfnamefont {R.~X.}\
  \bibnamefont {{Adhikari}}}, \bibinfo {author} {\bibfnamefont {V.~B.}\
  \bibnamefont {{Adya}}}, \bibinfo {author} {\bibfnamefont {C.}~\bibnamefont
  {{Affeldt}}},\ and\ \bibinfo {author} {\bibnamefont {et~al.}},\ }\href
  {https://doi.org/10.1103/PhysRevX.11.021053} {\bibfield  {journal} {\bibinfo
  {journal} {Phys. Rev. X}\ }\textbf {\bibinfo {volume} {11}},\ \bibinfo {eid}
  {021053} (\bibinfo {year} {2021})},\ \Eprint
  {https://arxiv.org/abs/2010.14527} {arXiv:2010.14527 [gr-qc]} \BibitemShut
  {NoStop}%
\bibitem [{\citenamefont {{Abbott}}\ \emph {et~al.}(2024)\citenamefont
  {{Abbott}}, \citenamefont {{Abbott}}, \citenamefont {{Acernese}},
  \citenamefont {{Ackley}}, \citenamefont {{Adams}}, \citenamefont
  {{Adhikari}}, \citenamefont {{Adhikari}}, \citenamefont {{Adya}},
  \citenamefont {{Affeldt}}, \citenamefont {{Agarwal}},\ and\ \citenamefont
  {et~al.}}]{2024PhRvD.109b2001A}%
  \BibitemOpen
  \bibfield  {author} {\bibinfo {author} {\bibfnamefont {R.}~\bibnamefont
  {{Abbott}}}, \bibinfo {author} {\bibfnamefont {T.~D.}\ \bibnamefont
  {{Abbott}}}, \bibinfo {author} {\bibfnamefont {F.}~\bibnamefont
  {{Acernese}}}, \bibinfo {author} {\bibfnamefont {K.}~\bibnamefont
  {{Ackley}}}, \bibinfo {author} {\bibfnamefont {C.}~\bibnamefont {{Adams}}},
  \bibinfo {author} {\bibfnamefont {N.}~\bibnamefont {{Adhikari}}}, \bibinfo
  {author} {\bibfnamefont {R.~X.}\ \bibnamefont {{Adhikari}}}, \bibinfo
  {author} {\bibfnamefont {V.~B.}\ \bibnamefont {{Adya}}}, \bibinfo {author}
  {\bibfnamefont {C.}~\bibnamefont {{Affeldt}}}, \bibinfo {author}
  {\bibfnamefont {D.}~\bibnamefont {{Agarwal}}},\ and\ \bibinfo {author}
  {\bibnamefont {et~al.}},\ }\href
  {https://doi.org/10.1103/PhysRevD.109.022001} {\bibfield  {journal} {\bibinfo
   {journal} {Phys. Rev. D}\ }\textbf {\bibinfo {volume} {109}},\ \bibinfo
  {eid} {022001} (\bibinfo {year} {2024})},\ \Eprint
  {https://arxiv.org/abs/2108.01045} {arXiv:2108.01045 [gr-qc]} \BibitemShut
  {NoStop}%
\bibitem [{\citenamefont {{Abbott}}\ \emph
  {et~al.}(2023{\natexlab{a}})\citenamefont {{Abbott}}, \citenamefont
  {{Abbott}}, \citenamefont {{Acernese}}, \citenamefont {{Ackley}},
  \citenamefont {{Adams}}, \citenamefont {{Adhikari}}, \citenamefont
  {{Adhikari}}, \citenamefont {{Adya}}, \citenamefont {{Affeldt}},
  \citenamefont {{Agarwal}},\ and\ \citenamefont
  {et~al.}}]{2023PhRvX..13d1039A}%
  \BibitemOpen
  \bibfield  {author} {\bibinfo {author} {\bibfnamefont {R.}~\bibnamefont
  {{Abbott}}}, \bibinfo {author} {\bibfnamefont {T.~D.}\ \bibnamefont
  {{Abbott}}}, \bibinfo {author} {\bibfnamefont {F.}~\bibnamefont
  {{Acernese}}}, \bibinfo {author} {\bibfnamefont {K.}~\bibnamefont
  {{Ackley}}}, \bibinfo {author} {\bibfnamefont {C.}~\bibnamefont {{Adams}}},
  \bibinfo {author} {\bibfnamefont {N.}~\bibnamefont {{Adhikari}}}, \bibinfo
  {author} {\bibfnamefont {R.~X.}\ \bibnamefont {{Adhikari}}}, \bibinfo
  {author} {\bibfnamefont {V.~B.}\ \bibnamefont {{Adya}}}, \bibinfo {author}
  {\bibfnamefont {C.}~\bibnamefont {{Affeldt}}}, \bibinfo {author}
  {\bibfnamefont {D.}~\bibnamefont {{Agarwal}}},\ and\ \bibinfo {author}
  {\bibnamefont {et~al.}},\ }\href {https://doi.org/10.1103/PhysRevX.13.041039}
  {\bibfield  {journal} {\bibinfo  {journal} {Phys. Rev. X}\ }\textbf {\bibinfo
  {volume} {13}},\ \bibinfo {eid} {041039} (\bibinfo {year}
  {2023}{\natexlab{a}})},\ \Eprint {https://arxiv.org/abs/2111.03606}
  {arXiv:2111.03606 [gr-qc]} \BibitemShut {NoStop}%
\bibitem [{\citenamefont {{Mapelli}}(2021)}]{2021hgwa.bookE..16M}%
  \BibitemOpen
  \bibfield  {author} {\bibinfo {author} {\bibfnamefont {M.}~\bibnamefont
  {{Mapelli}}},\ }in\ \href {https://doi.org/10.1007/978-981-15-4702-7_16-1}
  {\emph {\bibinfo {booktitle} {Handbook of Gravitational Wave Astronomy}}},\
  \bibinfo {editor} {edited by\ \bibinfo {editor} {\bibfnamefont
  {C.}~\bibnamefont {{Bambi}}}, \bibinfo {editor} {\bibfnamefont
  {S.}~\bibnamefont {{Katsanevas}}},\ and\ \bibinfo {editor} {\bibfnamefont
  {K.~D.}\ \bibnamefont {{Kokkotas}}}}\ (\bibinfo {year} {2021})\ p.~\bibinfo
  {pages} {16}\BibitemShut {NoStop}%
\bibitem [{\citenamefont {{Gerosa}}\ and\ \citenamefont
  {{Fishbach}}(2021)}]{2021NatAs...5..749G}%
  \BibitemOpen
  \bibfield  {author} {\bibinfo {author} {\bibfnamefont {D.}~\bibnamefont
  {{Gerosa}}}\ and\ \bibinfo {author} {\bibfnamefont {M.}~\bibnamefont
  {{Fishbach}}},\ }\href {https://doi.org/10.1038/s41550-021-01398-w}
  {\bibfield  {journal} {\bibinfo  {journal} {Nat. Astron.}\ }\textbf {\bibinfo
  {volume} {5}},\ \bibinfo {pages} {749} (\bibinfo {year} {2021})},\ \Eprint
  {https://arxiv.org/abs/2105.03439} {arXiv:2105.03439 [astro-ph.HE]}
  \BibitemShut {NoStop}%
\bibitem [{\citenamefont {{Mandel}}\ and\ \citenamefont
  {{Farmer}}(2022)}]{2022PhR...955....1M}%
  \BibitemOpen
  \bibfield  {author} {\bibinfo {author} {\bibfnamefont {I.}~\bibnamefont
  {{Mandel}}}\ and\ \bibinfo {author} {\bibfnamefont {A.}~\bibnamefont
  {{Farmer}}},\ }\href {https://doi.org/10.1016/j.physrep.2022.01.003}
  {\bibfield  {journal} {\bibinfo  {journal} {Phys. Rep.}\ }\textbf {\bibinfo
  {volume} {955}},\ \bibinfo {pages} {1} (\bibinfo {year} {2022})},\ \Eprint
  {https://arxiv.org/abs/1806.05820} {arXiv:1806.05820 [astro-ph.HE]}
  \BibitemShut {NoStop}%
\bibitem [{\citenamefont {{Mandel}}\ and\ \citenamefont
  {{Broekgaarden}}(2022)}]{2022LRR....25....1M}%
  \BibitemOpen
  \bibfield  {author} {\bibinfo {author} {\bibfnamefont {I.}~\bibnamefont
  {{Mandel}}}\ and\ \bibinfo {author} {\bibfnamefont {F.~S.}\ \bibnamefont
  {{Broekgaarden}}},\ }\href {https://doi.org/10.1007/s41114-021-00034-3}
  {\bibfield  {journal} {\bibinfo  {journal} {Living Rev. Relativ.}\ }\textbf
  {\bibinfo {volume} {25}},\ \bibinfo {eid} {1} (\bibinfo {year} {2022})},\
  \Eprint {https://arxiv.org/abs/2107.14239} {arXiv:2107.14239 [astro-ph.HE]}
  \BibitemShut {NoStop}%
\bibitem [{\citenamefont {{Zevin}}\ \emph {et~al.}(2017)\citenamefont
  {{Zevin}}, \citenamefont {{Pankow}}, \citenamefont {{Rodriguez}},
  \citenamefont {{Sampson}}, \citenamefont {{Chase}}, \citenamefont
  {{Kalogera}},\ and\ \citenamefont {{Rasio}}}]{2017ApJ...846...82Z}%
  \BibitemOpen
  \bibfield  {author} {\bibinfo {author} {\bibfnamefont {M.}~\bibnamefont
  {{Zevin}}}, \bibinfo {author} {\bibfnamefont {C.}~\bibnamefont {{Pankow}}},
  \bibinfo {author} {\bibfnamefont {C.~L.}\ \bibnamefont {{Rodriguez}}},
  \bibinfo {author} {\bibfnamefont {L.}~\bibnamefont {{Sampson}}}, \bibinfo
  {author} {\bibfnamefont {E.}~\bibnamefont {{Chase}}}, \bibinfo {author}
  {\bibfnamefont {V.}~\bibnamefont {{Kalogera}}},\ and\ \bibinfo {author}
  {\bibfnamefont {F.~A.}\ \bibnamefont {{Rasio}}},\ }\href
  {https://doi.org/10.3847/1538-4357/aa8408} {\bibfield  {journal} {\bibinfo
  {journal} {Astrophys. J.}\ }\textbf {\bibinfo {volume} {846}},\ \bibinfo
  {eid} {82} (\bibinfo {year} {2017})},\ \Eprint
  {https://arxiv.org/abs/1704.07379} {arXiv:1704.07379 [astro-ph.HE]}
  \BibitemShut {NoStop}%
\bibitem [{\citenamefont {{Cheng}}\ \emph {et~al.}(2023)\citenamefont
  {{Cheng}}, \citenamefont {{Zevin}},\ and\ \citenamefont
  {{Vitale}}}]{2023ApJ...955..127C}%
  \BibitemOpen
  \bibfield  {author} {\bibinfo {author} {\bibfnamefont {A.~Q.}\ \bibnamefont
  {{Cheng}}}, \bibinfo {author} {\bibfnamefont {M.}~\bibnamefont {{Zevin}}},\
  and\ \bibinfo {author} {\bibfnamefont {S.}~\bibnamefont {{Vitale}}},\ }\href
  {https://doi.org/10.3847/1538-4357/aced98} {\bibfield  {journal} {\bibinfo
  {journal} {Astrophys. J.}\ }\textbf {\bibinfo {volume} {955}},\ \bibinfo
  {eid} {127} (\bibinfo {year} {2023})},\ \Eprint
  {https://arxiv.org/abs/2307.03129} {arXiv:2307.03129 [astro-ph.HE]}
  \BibitemShut {NoStop}%
\bibitem [{\citenamefont {{Gerosa}}\ \emph {et~al.}(2013)\citenamefont
  {{Gerosa}}, \citenamefont {{Kesden}}, \citenamefont {{Berti}}, \citenamefont
  {{O'Shaughnessy}},\ and\ \citenamefont {{Sperhake}}}]{2013PhRvD..87j4028G}%
  \BibitemOpen
  \bibfield  {author} {\bibinfo {author} {\bibfnamefont {D.}~\bibnamefont
  {{Gerosa}}}, \bibinfo {author} {\bibfnamefont {M.}~\bibnamefont {{Kesden}}},
  \bibinfo {author} {\bibfnamefont {E.}~\bibnamefont {{Berti}}}, \bibinfo
  {author} {\bibfnamefont {R.}~\bibnamefont {{O'Shaughnessy}}},\ and\ \bibinfo
  {author} {\bibfnamefont {U.}~\bibnamefont {{Sperhake}}},\ }\href
  {https://doi.org/10.1103/PhysRevD.87.104028} {\bibfield  {journal} {\bibinfo
  {journal} {Phys. Rev. D}\ }\textbf {\bibinfo {volume} {87}},\ \bibinfo {eid}
  {104028} (\bibinfo {year} {2013})},\ \Eprint
  {https://arxiv.org/abs/1302.4442} {arXiv:1302.4442 [gr-qc]} \BibitemShut
  {NoStop}%
\bibitem [{\citenamefont {{Rodriguez}}\ \emph {et~al.}(2016)\citenamefont
  {{Rodriguez}}, \citenamefont {{Zevin}}, \citenamefont {{Pankow}},
  \citenamefont {{Kalogera}},\ and\ \citenamefont
  {{Rasio}}}]{2016ApJ...832L...2R}%
  \BibitemOpen
  \bibfield  {author} {\bibinfo {author} {\bibfnamefont {C.~L.}\ \bibnamefont
  {{Rodriguez}}}, \bibinfo {author} {\bibfnamefont {M.}~\bibnamefont
  {{Zevin}}}, \bibinfo {author} {\bibfnamefont {C.}~\bibnamefont {{Pankow}}},
  \bibinfo {author} {\bibfnamefont {V.}~\bibnamefont {{Kalogera}}},\ and\
  \bibinfo {author} {\bibfnamefont {F.~A.}\ \bibnamefont {{Rasio}}},\ }\href
  {https://doi.org/10.3847/2041-8205/832/1/L2} {\bibfield  {journal} {\bibinfo
  {journal} {Astrophys. J. Lett.}\ }\textbf {\bibinfo {volume} {832}},\
  \bibinfo {eid} {L2} (\bibinfo {year} {2016})},\ \Eprint
  {https://arxiv.org/abs/1609.05916} {arXiv:1609.05916 [astro-ph.HE]}
  \BibitemShut {NoStop}%
\bibitem [{\citenamefont {{Vitale}}\ \emph {et~al.}(2017)\citenamefont
  {{Vitale}}, \citenamefont {{Lynch}}, \citenamefont {{Sturani}},\ and\
  \citenamefont {{Graff}}}]{2017CQGra..34cLT01V}%
  \BibitemOpen
  \bibfield  {author} {\bibinfo {author} {\bibfnamefont {S.}~\bibnamefont
  {{Vitale}}}, \bibinfo {author} {\bibfnamefont {R.}~\bibnamefont {{Lynch}}},
  \bibinfo {author} {\bibfnamefont {R.}~\bibnamefont {{Sturani}}},\ and\
  \bibinfo {author} {\bibfnamefont {P.}~\bibnamefont {{Graff}}},\ }\href
  {https://doi.org/10.1088/1361-6382/aa552e} {\bibfield  {journal} {\bibinfo
  {journal} {Class. Quantum Grav.}\ }\textbf {\bibinfo {volume} {34}},\
  \bibinfo {eid} {03LT01} (\bibinfo {year} {2017})},\ \Eprint
  {https://arxiv.org/abs/1503.04307} {arXiv:1503.04307 [gr-qc]} \BibitemShut
  {NoStop}%
\bibitem [{\citenamefont {{Stevenson}}\ \emph {et~al.}(2017)\citenamefont
  {{Stevenson}}, \citenamefont {{Berry}},\ and\ \citenamefont
  {{Mandel}}}]{2017MNRAS.471.2801S}%
  \BibitemOpen
  \bibfield  {author} {\bibinfo {author} {\bibfnamefont {S.}~\bibnamefont
  {{Stevenson}}}, \bibinfo {author} {\bibfnamefont {C.~P.~L.}\ \bibnamefont
  {{Berry}}},\ and\ \bibinfo {author} {\bibfnamefont {I.}~\bibnamefont
  {{Mandel}}},\ }\href {https://doi.org/10.1093/mnras/stx1764} {\bibfield
  {journal} {\bibinfo  {journal} {Mon. Not. R. Astron. Soc.}\ }\textbf
  {\bibinfo {volume} {471}},\ \bibinfo {pages} {2801} (\bibinfo {year}
  {2017})},\ \Eprint {https://arxiv.org/abs/1703.06873} {arXiv:1703.06873
  [astro-ph.HE]} \BibitemShut {NoStop}%
\bibitem [{\citenamefont {{Gerosa}}\ \emph {et~al.}(2018)\citenamefont
  {{Gerosa}}, \citenamefont {{Berti}}, \citenamefont {{O'Shaughnessy}},
  \citenamefont {{Belczynski}}, \citenamefont {{Kesden}}, \citenamefont
  {{Wysocki}},\ and\ \citenamefont {{Gladysz}}}]{2018PhRvD..98h4036G}%
  \BibitemOpen
  \bibfield  {author} {\bibinfo {author} {\bibfnamefont {D.}~\bibnamefont
  {{Gerosa}}}, \bibinfo {author} {\bibfnamefont {E.}~\bibnamefont {{Berti}}},
  \bibinfo {author} {\bibfnamefont {R.}~\bibnamefont {{O'Shaughnessy}}},
  \bibinfo {author} {\bibfnamefont {K.}~\bibnamefont {{Belczynski}}}, \bibinfo
  {author} {\bibfnamefont {M.}~\bibnamefont {{Kesden}}}, \bibinfo {author}
  {\bibfnamefont {D.}~\bibnamefont {{Wysocki}}},\ and\ \bibinfo {author}
  {\bibfnamefont {W.}~\bibnamefont {{Gladysz}}},\ }\href
  {https://doi.org/10.1103/PhysRevD.98.084036} {\bibfield  {journal} {\bibinfo
  {journal} {Phys. Rev. D}\ }\textbf {\bibinfo {volume} {98}},\ \bibinfo {eid}
  {084036} (\bibinfo {year} {2018})},\ \Eprint
  {https://arxiv.org/abs/1808.02491} {arXiv:1808.02491 [astro-ph.HE]}
  \BibitemShut {NoStop}%
\bibitem [{\citenamefont {{Bavera}}\ \emph {et~al.}(2020)\citenamefont
  {{Bavera}}, \citenamefont {{Fragos}}, \citenamefont {{Qin}}, \citenamefont
  {{Zapartas}}, \citenamefont {{Neijssel}}, \citenamefont {{Mandel}},
  \citenamefont {{Batta}}, \citenamefont {{Gaebel}}, \citenamefont
  {{Kimball}},\ and\ \citenamefont {{Stevenson}}}]{2020A&A...635A..97B}%
  \BibitemOpen
  \bibfield  {author} {\bibinfo {author} {\bibfnamefont {S.~S.}\ \bibnamefont
  {{Bavera}}}, \bibinfo {author} {\bibfnamefont {T.}~\bibnamefont {{Fragos}}},
  \bibinfo {author} {\bibfnamefont {Y.}~\bibnamefont {{Qin}}}, \bibinfo
  {author} {\bibfnamefont {E.}~\bibnamefont {{Zapartas}}}, \bibinfo {author}
  {\bibfnamefont {C.~J.}\ \bibnamefont {{Neijssel}}}, \bibinfo {author}
  {\bibfnamefont {I.}~\bibnamefont {{Mandel}}}, \bibinfo {author}
  {\bibfnamefont {A.}~\bibnamefont {{Batta}}}, \bibinfo {author} {\bibfnamefont
  {S.~M.}\ \bibnamefont {{Gaebel}}}, \bibinfo {author} {\bibfnamefont
  {C.}~\bibnamefont {{Kimball}}},\ and\ \bibinfo {author} {\bibfnamefont
  {S.}~\bibnamefont {{Stevenson}}},\ }\href
  {https://doi.org/10.1051/0004-6361/201936204} {\bibfield  {journal} {\bibinfo
   {journal} {Astron. Astrophys.}\ }\textbf {\bibinfo {volume} {635}},\
  \bibinfo {eid} {A97} (\bibinfo {year} {2020})},\ \Eprint
  {https://arxiv.org/abs/1906.12257} {arXiv:1906.12257 [astro-ph.HE]}
  \BibitemShut {NoStop}%
\bibitem [{\citenamefont {{Olejak}}\ and\ \citenamefont
  {{Belczynski}}(2021)}]{2021ApJ...921L...2O}%
  \BibitemOpen
  \bibfield  {author} {\bibinfo {author} {\bibfnamefont {A.}~\bibnamefont
  {{Olejak}}}\ and\ \bibinfo {author} {\bibfnamefont {K.}~\bibnamefont
  {{Belczynski}}},\ }\href {https://doi.org/10.3847/2041-8213/ac2f48}
  {\bibfield  {journal} {\bibinfo  {journal} {Astrophys. J. Lett.}\ }\textbf
  {\bibinfo {volume} {921}},\ \bibinfo {eid} {L2} (\bibinfo {year} {2021})},\
  \Eprint {https://arxiv.org/abs/2109.06872} {arXiv:2109.06872 [astro-ph.HE]}
  \BibitemShut {NoStop}%
\bibitem [{\citenamefont {{Banerjee}}\ \emph {et~al.}(2023)\citenamefont
  {{Banerjee}}, \citenamefont {{Olejak}},\ and\ \citenamefont
  {{Belczynski}}}]{2023ApJ...953...80B}%
  \BibitemOpen
  \bibfield  {author} {\bibinfo {author} {\bibfnamefont {S.}~\bibnamefont
  {{Banerjee}}}, \bibinfo {author} {\bibfnamefont {A.}~\bibnamefont
  {{Olejak}}},\ and\ \bibinfo {author} {\bibfnamefont {K.}~\bibnamefont
  {{Belczynski}}},\ }\href {https://doi.org/10.3847/1538-4357/acdd59}
  {\bibfield  {journal} {\bibinfo  {journal} {Astrophys. J.}\ }\textbf
  {\bibinfo {volume} {953}},\ \bibinfo {eid} {80} (\bibinfo {year} {2023})},\
  \Eprint {https://arxiv.org/abs/2302.10851} {arXiv:2302.10851 [astro-ph.HE]}
  \BibitemShut {NoStop}%
\bibitem [{\citenamefont {{Samsing}}\ \emph {et~al.}(2014)\citenamefont
  {{Samsing}}, \citenamefont {{MacLeod}},\ and\ \citenamefont
  {{Ramirez-Ruiz}}}]{2014ApJ...784...71S}%
  \BibitemOpen
  \bibfield  {author} {\bibinfo {author} {\bibfnamefont {J.}~\bibnamefont
  {{Samsing}}}, \bibinfo {author} {\bibfnamefont {M.}~\bibnamefont
  {{MacLeod}}},\ and\ \bibinfo {author} {\bibfnamefont {E.}~\bibnamefont
  {{Ramirez-Ruiz}}},\ }\href {https://doi.org/10.1088/0004-637X/784/1/71}
  {\bibfield  {journal} {\bibinfo  {journal} {Astrophys. J.}\ }\textbf
  {\bibinfo {volume} {784}},\ \bibinfo {eid} {71} (\bibinfo {year} {2014})},\
  \Eprint {https://arxiv.org/abs/1308.2964} {arXiv:1308.2964 [astro-ph.HE]}
  \BibitemShut {NoStop}%
\bibitem [{\citenamefont {{Samsing}}(2018)}]{2018PhRvD..97j3014S}%
  \BibitemOpen
  \bibfield  {author} {\bibinfo {author} {\bibfnamefont {J.}~\bibnamefont
  {{Samsing}}},\ }\href {https://doi.org/10.1103/PhysRevD.97.103014} {\bibfield
   {journal} {\bibinfo  {journal} {Phys. Rev. D}\ }\textbf {\bibinfo {volume}
  {97}},\ \bibinfo {eid} {103014} (\bibinfo {year} {2018})},\ \Eprint
  {https://arxiv.org/abs/1711.07452} {arXiv:1711.07452 [astro-ph.HE]}
  \BibitemShut {NoStop}%
\bibitem [{\citenamefont {{Zevin}}\ \emph {et~al.}(2021)\citenamefont
  {{Zevin}}, \citenamefont {{Romero-Shaw}}, \citenamefont {{Kremer}},
  \citenamefont {{Thrane}},\ and\ \citenamefont
  {{Lasky}}}]{2021ApJ...921L..43Z}%
  \BibitemOpen
  \bibfield  {author} {\bibinfo {author} {\bibfnamefont {M.}~\bibnamefont
  {{Zevin}}}, \bibinfo {author} {\bibfnamefont {I.~M.}\ \bibnamefont
  {{Romero-Shaw}}}, \bibinfo {author} {\bibfnamefont {K.}~\bibnamefont
  {{Kremer}}}, \bibinfo {author} {\bibfnamefont {E.}~\bibnamefont {{Thrane}}},\
  and\ \bibinfo {author} {\bibfnamefont {P.~D.}\ \bibnamefont {{Lasky}}},\
  }\href {https://doi.org/10.3847/2041-8213/ac32dc} {\bibfield  {journal}
  {\bibinfo  {journal} {Astrophys. J. Lett.}\ }\textbf {\bibinfo {volume}
  {921}},\ \bibinfo {eid} {L43} (\bibinfo {year} {2021})},\ \Eprint
  {https://arxiv.org/abs/2106.09042} {arXiv:2106.09042 [astro-ph.HE]}
  \BibitemShut {NoStop}%
\bibitem [{\citenamefont {{Apostolatos}}\ \emph {et~al.}(1994)\citenamefont
  {{Apostolatos}}, \citenamefont {{Cutler}}, \citenamefont {{Sussman}},\ and\
  \citenamefont {{Thorne}}}]{1994PhRvD..49.6274A}%
  \BibitemOpen
  \bibfield  {author} {\bibinfo {author} {\bibfnamefont {T.~A.}\ \bibnamefont
  {{Apostolatos}}}, \bibinfo {author} {\bibfnamefont {C.}~\bibnamefont
  {{Cutler}}}, \bibinfo {author} {\bibfnamefont {G.~J.}\ \bibnamefont
  {{Sussman}}},\ and\ \bibinfo {author} {\bibfnamefont {K.~S.}\ \bibnamefont
  {{Thorne}}},\ }\href {https://doi.org/10.1103/PhysRevD.49.6274} {\bibfield
  {journal} {\bibinfo  {journal} {Phys. Rev. D}\ }\textbf {\bibinfo {volume}
  {49}},\ \bibinfo {pages} {6274} (\bibinfo {year} {1994})}\BibitemShut
  {NoStop}%
\bibitem [{\citenamefont {{Kidder}}(1995)}]{1995PhRvD..52..821K}%
  \BibitemOpen
  \bibfield  {author} {\bibinfo {author} {\bibfnamefont {L.~E.}\ \bibnamefont
  {{Kidder}}},\ }\href {https://doi.org/10.1103/PhysRevD.52.821} {\bibfield
  {journal} {\bibinfo  {journal} {Phys. Rev. D}\ }\textbf {\bibinfo {volume}
  {52}},\ \bibinfo {pages} {821} (\bibinfo {year} {1995})},\ \Eprint
  {https://arxiv.org/abs/gr-qc/9506022} {arXiv:gr-qc/9506022 [gr-qc]}
  \BibitemShut {NoStop}%
\bibitem [{\citenamefont {{Romero-Shaw}}\ \emph {et~al.}(2019)\citenamefont
  {{Romero-Shaw}}, \citenamefont {{Lasky}},\ and\ \citenamefont
  {{Thrane}}}]{2019MNRAS.490.5210R}%
  \BibitemOpen
  \bibfield  {author} {\bibinfo {author} {\bibfnamefont {I.~M.}\ \bibnamefont
  {{Romero-Shaw}}}, \bibinfo {author} {\bibfnamefont {P.~D.}\ \bibnamefont
  {{Lasky}}},\ and\ \bibinfo {author} {\bibfnamefont {E.}~\bibnamefont
  {{Thrane}}},\ }\href {https://doi.org/10.1093/mnras/stz2996} {\bibfield
  {journal} {\bibinfo  {journal} {Mon. Not. R. Astron. Soc.}\ }\textbf
  {\bibinfo {volume} {490}},\ \bibinfo {pages} {5210} (\bibinfo {year}
  {2019})},\ \Eprint {https://arxiv.org/abs/1909.05466} {arXiv:1909.05466
  [astro-ph.HE]} \BibitemShut {NoStop}%
\bibitem [{\citenamefont {{Romero-Shaw}}\ \emph {et~al.}(2023)\citenamefont
  {{Romero-Shaw}}, \citenamefont {{Gerosa}},\ and\ \citenamefont
  {{Loutrel}}}]{2023MNRAS.519.5352R}%
  \BibitemOpen
  \bibfield  {author} {\bibinfo {author} {\bibfnamefont {I.~M.}\ \bibnamefont
  {{Romero-Shaw}}}, \bibinfo {author} {\bibfnamefont {D.}~\bibnamefont
  {{Gerosa}}},\ and\ \bibinfo {author} {\bibfnamefont {N.}~\bibnamefont
  {{Loutrel}}},\ }\href {https://doi.org/10.1093/mnras/stad031} {\bibfield
  {journal} {\bibinfo  {journal} {Mon. Not. R. Astron. Soc.}\ }\textbf
  {\bibinfo {volume} {519}},\ \bibinfo {pages} {5352} (\bibinfo {year}
  {2023})},\ \Eprint {https://arxiv.org/abs/2211.07528} {arXiv:2211.07528
  [astro-ph.HE]} \BibitemShut {NoStop}%
\bibitem [{\citenamefont {{Fumagalli}}\ and\ \citenamefont
  {{Gerosa}}(2023)}]{2023PhRvD.108l4055F}%
  \BibitemOpen
  \bibfield  {author} {\bibinfo {author} {\bibfnamefont {G.}~\bibnamefont
  {{Fumagalli}}}\ and\ \bibinfo {author} {\bibfnamefont {D.}~\bibnamefont
  {{Gerosa}}},\ }\href {https://doi.org/10.1103/PhysRevD.108.124055} {\bibfield
   {journal} {\bibinfo  {journal} {Phys. Rev. D}\ }\textbf {\bibinfo {volume}
  {108}},\ \bibinfo {eid} {124055} (\bibinfo {year} {2023})},\ \Eprint
  {https://arxiv.org/abs/2310.16893} {arXiv:2310.16893 [gr-qc]} \BibitemShut
  {NoStop}%
\bibitem [{\citenamefont {{Gerosa}}\ \emph {et~al.}(2015)\citenamefont
  {{Gerosa}}, \citenamefont {{Kesden}}, \citenamefont {{Sperhake}},
  \citenamefont {{Berti}},\ and\ \citenamefont
  {{O'Shaughnessy}}}]{2015PhRvD..92f4016G}%
  \BibitemOpen
  \bibfield  {author} {\bibinfo {author} {\bibfnamefont {D.}~\bibnamefont
  {{Gerosa}}}, \bibinfo {author} {\bibfnamefont {M.}~\bibnamefont {{Kesden}}},
  \bibinfo {author} {\bibfnamefont {U.}~\bibnamefont {{Sperhake}}}, \bibinfo
  {author} {\bibfnamefont {E.}~\bibnamefont {{Berti}}},\ and\ \bibinfo {author}
  {\bibfnamefont {R.}~\bibnamefont {{O'Shaughnessy}}},\ }\href
  {https://doi.org/10.1103/PhysRevD.92.064016} {\bibfield  {journal} {\bibinfo
  {journal} {Phys. Rev. D}\ }\textbf {\bibinfo {volume} {92}},\ \bibinfo {eid}
  {064016} (\bibinfo {year} {2015})},\ \Eprint
  {https://arxiv.org/abs/1506.03492} {arXiv:1506.03492 [gr-qc]} \BibitemShut
  {NoStop}%
\bibitem [{\citenamefont {{Gerosa}}\ \emph {et~al.}(2023)\citenamefont
  {{Gerosa}}, \citenamefont {{Fumagalli}}, \citenamefont {{Mould}},
  \citenamefont {{Cavallotto}}, \citenamefont {{Monroy}}, \citenamefont
  {{Gangardt}},\ and\ \citenamefont {{De Renzis}}}]{2023PhRvD.108b4042G}%
  \BibitemOpen
  \bibfield  {author} {\bibinfo {author} {\bibfnamefont {D.}~\bibnamefont
  {{Gerosa}}}, \bibinfo {author} {\bibfnamefont {G.}~\bibnamefont
  {{Fumagalli}}}, \bibinfo {author} {\bibfnamefont {M.}~\bibnamefont
  {{Mould}}}, \bibinfo {author} {\bibfnamefont {G.}~\bibnamefont
  {{Cavallotto}}}, \bibinfo {author} {\bibfnamefont {D.~P.}\ \bibnamefont
  {{Monroy}}}, \bibinfo {author} {\bibfnamefont {D.}~\bibnamefont
  {{Gangardt}}},\ and\ \bibinfo {author} {\bibfnamefont {V.}~\bibnamefont {{De
  Renzis}}},\ }\href {https://doi.org/10.1103/PhysRevD.108.024042} {\bibfield
  {journal} {\bibinfo  {journal} {Phys. Rev. D}\ }\textbf {\bibinfo {volume}
  {108}},\ \bibinfo {eid} {024042} (\bibinfo {year} {2023})},\ \Eprint
  {https://arxiv.org/abs/2304.04801} {arXiv:2304.04801 [gr-qc]} \BibitemShut
  {NoStop}%
\bibitem [{\citenamefont {{Mould}}\ and\ \citenamefont
  {{Gerosa}}(2022)}]{2022PhRvD.105b4076M}%
  \BibitemOpen
  \bibfield  {author} {\bibinfo {author} {\bibfnamefont {M.}~\bibnamefont
  {{Mould}}}\ and\ \bibinfo {author} {\bibfnamefont {D.}~\bibnamefont
  {{Gerosa}}},\ }\href {https://doi.org/10.1103/PhysRevD.105.024076} {\bibfield
   {journal} {\bibinfo  {journal} {Phys. Rev. D}\ }\textbf {\bibinfo {volume}
  {105}},\ \bibinfo {eid} {024076} (\bibinfo {year} {2022})},\ \Eprint
  {https://arxiv.org/abs/2110.05507} {arXiv:2110.05507 [astro-ph.HE]}
  \BibitemShut {NoStop}%
\bibitem [{\citenamefont {{Johnson-McDaniel}}\ \emph
  {et~al.}(2022)\citenamefont {{Johnson-McDaniel}}, \citenamefont
  {{Kulkarni}},\ and\ \citenamefont {{Gupta}}}]{2022PhRvD.106b3001J}%
  \BibitemOpen
  \bibfield  {author} {\bibinfo {author} {\bibfnamefont {N.~K.}\ \bibnamefont
  {{Johnson-McDaniel}}}, \bibinfo {author} {\bibfnamefont {S.}~\bibnamefont
  {{Kulkarni}}},\ and\ \bibinfo {author} {\bibfnamefont {A.}~\bibnamefont
  {{Gupta}}},\ }\href {https://doi.org/10.1103/PhysRevD.106.023001} {\bibfield
  {journal} {\bibinfo  {journal} {Phys. Rev. D}\ }\textbf {\bibinfo {volume}
  {106}},\ \bibinfo {eid} {023001} (\bibinfo {year} {2022})},\ \Eprint
  {https://arxiv.org/abs/2107.11902} {arXiv:2107.11902 [astro-ph.HE]}
  \BibitemShut {NoStop}%
\bibitem [{\citenamefont {{Kulkarni}}\ \emph {et~al.}(2024)\citenamefont
  {{Kulkarni}}, \citenamefont {{Johnson-McDaniel}}, \citenamefont {{Phukon}},
  \citenamefont {{Krishnendu}},\ and\ \citenamefont
  {{Gupta}}}]{2024PhRvD.109d3002K}%
  \BibitemOpen
  \bibfield  {author} {\bibinfo {author} {\bibfnamefont {S.}~\bibnamefont
  {{Kulkarni}}}, \bibinfo {author} {\bibfnamefont {N.~K.}\ \bibnamefont
  {{Johnson-McDaniel}}}, \bibinfo {author} {\bibfnamefont {K.~S.}\ \bibnamefont
  {{Phukon}}}, \bibinfo {author} {\bibfnamefont {N.~V.}\ \bibnamefont
  {{Krishnendu}}},\ and\ \bibinfo {author} {\bibfnamefont {A.}~\bibnamefont
  {{Gupta}}},\ }\href {https://doi.org/10.1103/PhysRevD.109.043002} {\bibfield
  {journal} {\bibinfo  {journal} {Phys. Rev. D}\ }\textbf {\bibinfo {volume}
  {109}},\ \bibinfo {eid} {043002} (\bibinfo {year} {2024})},\ \Eprint
  {https://arxiv.org/abs/2308.05098} {arXiv:2308.05098 [astro-ph.HE]}
  \BibitemShut {NoStop}%
\bibitem [{\citenamefont {{Abbott}}\ \emph
  {et~al.}(2023{\natexlab{b}})\citenamefont {{Abbott}}, \citenamefont
  {{Abbott}}, \citenamefont {{Acernese}}, \citenamefont {{Ackley}},
  \citenamefont {{Adams}}, \citenamefont {{Adhikari}}, \citenamefont
  {{Adhikari}}, \citenamefont {{Adya}}, \citenamefont {{Affeldt}},
  \citenamefont {{Agarwal}},\ and\ \citenamefont
  {et~al.}}]{2023PhRvX..13a1048A}%
  \BibitemOpen
  \bibfield  {author} {\bibinfo {author} {\bibfnamefont {R.}~\bibnamefont
  {{Abbott}}}, \bibinfo {author} {\bibfnamefont {T.~D.}\ \bibnamefont
  {{Abbott}}}, \bibinfo {author} {\bibfnamefont {F.}~\bibnamefont
  {{Acernese}}}, \bibinfo {author} {\bibfnamefont {K.}~\bibnamefont
  {{Ackley}}}, \bibinfo {author} {\bibfnamefont {C.}~\bibnamefont {{Adams}}},
  \bibinfo {author} {\bibfnamefont {N.}~\bibnamefont {{Adhikari}}}, \bibinfo
  {author} {\bibfnamefont {R.~X.}\ \bibnamefont {{Adhikari}}}, \bibinfo
  {author} {\bibfnamefont {V.~B.}\ \bibnamefont {{Adya}}}, \bibinfo {author}
  {\bibfnamefont {C.}~\bibnamefont {{Affeldt}}}, \bibinfo {author}
  {\bibfnamefont {D.}~\bibnamefont {{Agarwal}}},\ and\ \bibinfo {author}
  {\bibnamefont {et~al.}},\ }\href {https://doi.org/10.1103/PhysRevX.13.011048}
  {\bibfield  {journal} {\bibinfo  {journal} {Phys. Rev. X}\ }\textbf {\bibinfo
  {volume} {13}},\ \bibinfo {eid} {011048} (\bibinfo {year}
  {2023}{\natexlab{b}})},\ \Eprint {https://arxiv.org/abs/2111.03634}
  {arXiv:2111.03634 [astro-ph.HE]} \BibitemShut {NoStop}%
\bibitem [{\citenamefont {{Wong}}\ \emph {et~al.}(2018)\citenamefont {{Wong}},
  \citenamefont {{Kovetz}}, \citenamefont {{Cutler}},\ and\ \citenamefont
  {{Berti}}}]{2018PhRvL.121y1102W}%
  \BibitemOpen
  \bibfield  {author} {\bibinfo {author} {\bibfnamefont {K.~W.~K.}\
  \bibnamefont {{Wong}}}, \bibinfo {author} {\bibfnamefont {E.~D.}\
  \bibnamefont {{Kovetz}}}, \bibinfo {author} {\bibfnamefont {C.}~\bibnamefont
  {{Cutler}}},\ and\ \bibinfo {author} {\bibfnamefont {E.}~\bibnamefont
  {{Berti}}},\ }\href {https://doi.org/10.1103/PhysRevLett.121.251102}
  {\bibfield  {journal} {\bibinfo  {journal} {Phys. Rev. Lett.}\ }\textbf
  {\bibinfo {volume} {121}},\ \bibinfo {eid} {251102} (\bibinfo {year}
  {2018})},\ \Eprint {https://arxiv.org/abs/1808.08247} {arXiv:1808.08247
  [astro-ph.HE]} \BibitemShut {NoStop}%
\bibitem [{\citenamefont {{Gerosa}}\ \emph {et~al.}(2019)\citenamefont
  {{Gerosa}}, \citenamefont {{Ma}}, \citenamefont {{Wong}}, \citenamefont
  {{Berti}}, \citenamefont {{O'Shaughnessy}}, \citenamefont {{Chen}},\ and\
  \citenamefont {{Belczynski}}}]{2019PhRvD..99j3004G}%
  \BibitemOpen
  \bibfield  {author} {\bibinfo {author} {\bibfnamefont {D.}~\bibnamefont
  {{Gerosa}}}, \bibinfo {author} {\bibfnamefont {S.}~\bibnamefont {{Ma}}},
  \bibinfo {author} {\bibfnamefont {K.~W.~K.}\ \bibnamefont {{Wong}}}, \bibinfo
  {author} {\bibfnamefont {E.}~\bibnamefont {{Berti}}}, \bibinfo {author}
  {\bibfnamefont {R.}~\bibnamefont {{O'Shaughnessy}}}, \bibinfo {author}
  {\bibfnamefont {Y.}~\bibnamefont {{Chen}}},\ and\ \bibinfo {author}
  {\bibfnamefont {K.}~\bibnamefont {{Belczynski}}},\ }\href
  {https://doi.org/10.1103/PhysRevD.99.103004} {\bibfield  {journal} {\bibinfo
  {journal} {Phys. Rev. D}\ }\textbf {\bibinfo {volume} {99}},\ \bibinfo {eid}
  {103004} (\bibinfo {year} {2019})},\ \Eprint
  {https://arxiv.org/abs/1902.00021} {arXiv:1902.00021 [astro-ph.HE]}
  \BibitemShut {NoStop}%
\bibitem [{\citenamefont {{Ewing}}\ \emph {et~al.}(2021)\citenamefont
  {{Ewing}}, \citenamefont {{Sachdev}}, \citenamefont {{Borhanian}},\ and\
  \citenamefont {{Sathyaprakash}}}]{2021PhRvD.103b3025E}%
  \BibitemOpen
  \bibfield  {author} {\bibinfo {author} {\bibfnamefont {B.}~\bibnamefont
  {{Ewing}}}, \bibinfo {author} {\bibfnamefont {S.}~\bibnamefont {{Sachdev}}},
  \bibinfo {author} {\bibfnamefont {S.}~\bibnamefont {{Borhanian}}},\ and\
  \bibinfo {author} {\bibfnamefont {B.~S.}\ \bibnamefont {{Sathyaprakash}}},\
  }\href {https://doi.org/10.1103/PhysRevD.103.023025} {\bibfield  {journal}
  {\bibinfo  {journal} {Phys. Rev. D}\ }\textbf {\bibinfo {volume} {103}},\
  \bibinfo {eid} {023025} (\bibinfo {year} {2021})},\ \Eprint
  {https://arxiv.org/abs/2011.03036} {arXiv:2011.03036 [gr-qc]} \BibitemShut
  {NoStop}%
\bibitem [{\citenamefont {{Toubiana}}\ \emph {et~al.}(2022)\citenamefont
  {{Toubiana}}, \citenamefont {{Babak}}, \citenamefont {{Marsat}},\ and\
  \citenamefont {{Ossokine}}}]{2022PhRvD.106j4034T}%
  \BibitemOpen
  \bibfield  {author} {\bibinfo {author} {\bibfnamefont {A.}~\bibnamefont
  {{Toubiana}}}, \bibinfo {author} {\bibfnamefont {S.}~\bibnamefont {{Babak}}},
  \bibinfo {author} {\bibfnamefont {S.}~\bibnamefont {{Marsat}}},\ and\
  \bibinfo {author} {\bibfnamefont {S.}~\bibnamefont {{Ossokine}}},\ }\href
  {https://doi.org/10.1103/PhysRevD.106.104034} {\bibfield  {journal} {\bibinfo
   {journal} {Phys. Rev. D}\ }\textbf {\bibinfo {volume} {106}},\ \bibinfo
  {eid} {104034} (\bibinfo {year} {2022})},\ \Eprint
  {https://arxiv.org/abs/2206.12439} {arXiv:2206.12439 [gr-qc]} \BibitemShut
  {NoStop}%
\bibitem [{\citenamefont {{Abbott}}\ \emph
  {et~al.}(2019{\natexlab{b}})\citenamefont {{Abbott}}, \citenamefont
  {{Abbott}}, \citenamefont {{Abbott}}, \citenamefont {{Abraham}},
  \citenamefont {{Acernese}}, \citenamefont {{Ackley}}, \citenamefont
  {{Adams}}, \citenamefont {{Adhikari}}, \citenamefont {{Adya}}, \citenamefont
  {{Affeldt}},\ and\ \citenamefont {et~al.}}]{2019ApJ...883..149A}%
  \BibitemOpen
  \bibfield  {author} {\bibinfo {author} {\bibfnamefont {B.~P.}\ \bibnamefont
  {{Abbott}}}, \bibinfo {author} {\bibfnamefont {R.}~\bibnamefont {{Abbott}}},
  \bibinfo {author} {\bibfnamefont {T.~D.}\ \bibnamefont {{Abbott}}}, \bibinfo
  {author} {\bibfnamefont {S.}~\bibnamefont {{Abraham}}}, \bibinfo {author}
  {\bibfnamefont {F.}~\bibnamefont {{Acernese}}}, \bibinfo {author}
  {\bibfnamefont {K.}~\bibnamefont {{Ackley}}}, \bibinfo {author}
  {\bibfnamefont {C.}~\bibnamefont {{Adams}}}, \bibinfo {author} {\bibfnamefont
  {R.~X.}\ \bibnamefont {{Adhikari}}}, \bibinfo {author} {\bibfnamefont
  {V.~B.}\ \bibnamefont {{Adya}}}, \bibinfo {author} {\bibfnamefont
  {C.}~\bibnamefont {{Affeldt}}},\ and\ \bibinfo {author} {\bibnamefont
  {et~al.}},\ }\href {https://doi.org/10.3847/1538-4357/ab3c2d} {\bibfield
  {journal} {\bibinfo  {journal} {Astrophys. J.}\ }\textbf {\bibinfo {volume}
  {883}},\ \bibinfo {eid} {149} (\bibinfo {year}
  {2019}{\natexlab{b}})}\BibitemShut {NoStop}%
\bibitem [{\citenamefont {{The LIGO Scientific Collaboration}}\ \emph
  {et~al.}(2023)\citenamefont {{The LIGO Scientific Collaboration}},
  \citenamefont {{the Virgo Collaboration}}, \citenamefont {{the KAGRA
  Collaboration}}, \citenamefont {{Abac}}, \citenamefont {{Abbott}},
  \citenamefont {{Abe}}, \citenamefont {{Acernese}}, \citenamefont {{Ackley}},
  \citenamefont {{Adamcewicz}}, \citenamefont {{Adhicary}},\ and\ \citenamefont
  {et~al.}}]{2023arXiv230803822T}%
  \BibitemOpen
  \bibfield  {author} {\bibinfo {author} {\bibnamefont {{The LIGO Scientific
  Collaboration}}}, \bibinfo {author} {\bibnamefont {{the Virgo
  Collaboration}}}, \bibinfo {author} {\bibnamefont {{the KAGRA
  Collaboration}}}, \bibinfo {author} {\bibfnamefont {A.~G.}\ \bibnamefont
  {{Abac}}}, \bibinfo {author} {\bibfnamefont {R.}~\bibnamefont {{Abbott}}},
  \bibinfo {author} {\bibfnamefont {H.}~\bibnamefont {{Abe}}}, \bibinfo
  {author} {\bibfnamefont {F.}~\bibnamefont {{Acernese}}}, \bibinfo {author}
  {\bibfnamefont {K.}~\bibnamefont {{Ackley}}}, \bibinfo {author}
  {\bibfnamefont {C.}~\bibnamefont {{Adamcewicz}}}, \bibinfo {author}
  {\bibfnamefont {S.}~\bibnamefont {{Adhicary}}},\ and\ \bibinfo {author}
  {\bibnamefont {et~al.}},\ }\href@noop {} {\bibfield  {journal} {\bibinfo
  {journal} {{}}\ } (\bibinfo {year} {2023})},\ \Eprint
  {https://arxiv.org/abs/2308.03822} {arXiv:2308.03822 [astro-ph.HE]}
  \BibitemShut {NoStop}%
\bibitem [{\citenamefont {{Romero-Shaw}}\ \emph {et~al.}(2022)\citenamefont
  {{Romero-Shaw}}, \citenamefont {{Lasky}},\ and\ \citenamefont
  {{Thrane}}}]{2022ApJ...940..171R}%
  \BibitemOpen
  \bibfield  {author} {\bibinfo {author} {\bibfnamefont {I.}~\bibnamefont
  {{Romero-Shaw}}}, \bibinfo {author} {\bibfnamefont {P.~D.}\ \bibnamefont
  {{Lasky}}},\ and\ \bibinfo {author} {\bibfnamefont {E.}~\bibnamefont
  {{Thrane}}},\ }\href {https://doi.org/10.3847/1538-4357/ac9798} {\bibfield
  {journal} {\bibinfo  {journal} {Astrophys. J.}\ }\textbf {\bibinfo {volume}
  {940}},\ \bibinfo {eid} {171} (\bibinfo {year} {2022})},\ \Eprint
  {https://arxiv.org/abs/2206.14695} {arXiv:2206.14695 [astro-ph.HE]}
  \BibitemShut {NoStop}%
\bibitem [{\citenamefont {{Gupte}}\ \emph {et~al.}(2024)\citenamefont
  {{Gupte}}, \citenamefont {{Ramos-Buades}}, \citenamefont {{Buonanno}},
  \citenamefont {{Gair}}, \citenamefont {{Miller}}, \citenamefont {{Dax}},
  \citenamefont {{Green}}, \citenamefont {{P{\"u}rrer}}, \citenamefont
  {{Wildberger}}, \citenamefont {{Macke}},\ and\ \citenamefont
  {{Sch{\"o}lkopf}}}]{2024arXiv240414286G}%
  \BibitemOpen
  \bibfield  {author} {\bibinfo {author} {\bibfnamefont {N.}~\bibnamefont
  {{Gupte}}}, \bibinfo {author} {\bibfnamefont {A.}~\bibnamefont
  {{Ramos-Buades}}}, \bibinfo {author} {\bibfnamefont {A.}~\bibnamefont
  {{Buonanno}}}, \bibinfo {author} {\bibfnamefont {J.}~\bibnamefont {{Gair}}},
  \bibinfo {author} {\bibfnamefont {M.~C.}\ \bibnamefont {{Miller}}}, \bibinfo
  {author} {\bibfnamefont {M.}~\bibnamefont {{Dax}}}, \bibinfo {author}
  {\bibfnamefont {S.~R.}\ \bibnamefont {{Green}}}, \bibinfo {author}
  {\bibfnamefont {M.}~\bibnamefont {{P{\"u}rrer}}}, \bibinfo {author}
  {\bibfnamefont {J.}~\bibnamefont {{Wildberger}}}, \bibinfo {author}
  {\bibfnamefont {J.}~\bibnamefont {{Macke}}},\ and\ \bibinfo {author}
  {\bibfnamefont {B.}~\bibnamefont {{Sch{\"o}lkopf}}},\ }\href@noop {}
  {\bibfield  {journal} {\bibinfo  {journal} {{}}\ } (\bibinfo {year}
  {2024})},\ \Eprint {https://arxiv.org/abs/2404.14286} {arXiv:2404.14286
  [gr-qc]} \BibitemShut {NoStop}%
\bibitem [{\citenamefont {{Peters}}(1964)}]{1964PhRv..136.1224P}%
  \BibitemOpen
  \bibfield  {author} {\bibinfo {author} {\bibfnamefont {P.~C.}\ \bibnamefont
  {{Peters}}},\ }\href {https://doi.org/10.1103/PhysRev.136.B1224} {\bibfield
  {journal} {\bibinfo  {journal} {Phys. Rev.}\ }\textbf {\bibinfo {volume}
  {136}},\ \bibinfo {pages} {1224} (\bibinfo {year} {1964})}\BibitemShut
  {NoStop}%
\bibitem [{\citenamefont {{Lower}}\ \emph {et~al.}(2018)\citenamefont
  {{Lower}}, \citenamefont {{Thrane}}, \citenamefont {{Lasky}},\ and\
  \citenamefont {{Smith}}}]{2018PhRvD..98h3028L}%
  \BibitemOpen
  \bibfield  {author} {\bibinfo {author} {\bibfnamefont {M.~E.}\ \bibnamefont
  {{Lower}}}, \bibinfo {author} {\bibfnamefont {E.}~\bibnamefont {{Thrane}}},
  \bibinfo {author} {\bibfnamefont {P.~D.}\ \bibnamefont {{Lasky}}},\ and\
  \bibinfo {author} {\bibfnamefont {R.}~\bibnamefont {{Smith}}},\ }\href
  {https://doi.org/10.1103/PhysRevD.98.083028} {\bibfield  {journal} {\bibinfo
  {journal} {Phys. Rev. D}\ }\textbf {\bibinfo {volume} {98}},\ \bibinfo {eid}
  {083028} (\bibinfo {year} {2018})},\ \Eprint
  {https://arxiv.org/abs/1806.05350} {arXiv:1806.05350 [astro-ph.HE]}
  \BibitemShut {NoStop}%
\bibitem [{\citenamefont {{De Renzis}}\ \emph {et~al.}(2022)\citenamefont {{De
  Renzis}}, \citenamefont {{Gerosa}}, \citenamefont {{Pratten}}, \citenamefont
  {{Schmidt}},\ and\ \citenamefont {{Mould}}}]{2022PhRvD.106h4040D}%
  \BibitemOpen
  \bibfield  {author} {\bibinfo {author} {\bibfnamefont {V.}~\bibnamefont {{De
  Renzis}}}, \bibinfo {author} {\bibfnamefont {D.}~\bibnamefont {{Gerosa}}},
  \bibinfo {author} {\bibfnamefont {G.}~\bibnamefont {{Pratten}}}, \bibinfo
  {author} {\bibfnamefont {P.}~\bibnamefont {{Schmidt}}},\ and\ \bibinfo
  {author} {\bibfnamefont {M.}~\bibnamefont {{Mould}}},\ }\href
  {https://doi.org/10.1103/PhysRevD.106.084040} {\bibfield  {journal} {\bibinfo
   {journal} {Phys. Rev. D}\ }\textbf {\bibinfo {volume} {106}},\ \bibinfo
  {eid} {084040} (\bibinfo {year} {2022})},\ \Eprint
  {https://arxiv.org/abs/2207.00030} {arXiv:2207.00030 [gr-qc]} \BibitemShut
  {NoStop}%
\bibitem [{\citenamefont {{Olejak}}\ \emph {et~al.}(2022)\citenamefont
  {{Olejak}}, \citenamefont {{Fryer}}, \citenamefont {{Belczynski}},\ and\
  \citenamefont {{Baibhav}}}]{2022MNRAS.516.2252O}%
  \BibitemOpen
  \bibfield  {author} {\bibinfo {author} {\bibfnamefont {A.}~\bibnamefont
  {{Olejak}}}, \bibinfo {author} {\bibfnamefont {C.~L.}\ \bibnamefont
  {{Fryer}}}, \bibinfo {author} {\bibfnamefont {K.}~\bibnamefont
  {{Belczynski}}},\ and\ \bibinfo {author} {\bibfnamefont {V.}~\bibnamefont
  {{Baibhav}}},\ }\href {https://doi.org/10.1093/mnras/stac2359} {\bibfield
  {journal} {\bibinfo  {journal} {Mon. Not. R. Astron. Soc.}\ }\textbf
  {\bibinfo {volume} {516}},\ \bibinfo {pages} {2252} (\bibinfo {year}
  {2022})},\ \Eprint {https://arxiv.org/abs/2204.09061} {arXiv:2204.09061
  [astro-ph.HE]} \BibitemShut {NoStop}%
\bibitem [{\citenamefont {{Kremer}}\ \emph {et~al.}(2020)\citenamefont
  {{Kremer}}, \citenamefont {{Ye}}, \citenamefont {{Rui}}, \citenamefont
  {{Weatherford}}, \citenamefont {{Chatterjee}}, \citenamefont {{Fragione}},
  \citenamefont {{Rodriguez}}, \citenamefont {{Spera}},\ and\ \citenamefont
  {{Rasio}}}]{2020ApJS..247...48K}%
  \BibitemOpen
  \bibfield  {author} {\bibinfo {author} {\bibfnamefont {K.}~\bibnamefont
  {{Kremer}}}, \bibinfo {author} {\bibfnamefont {C.~S.}\ \bibnamefont {{Ye}}},
  \bibinfo {author} {\bibfnamefont {N.~Z.}\ \bibnamefont {{Rui}}}, \bibinfo
  {author} {\bibfnamefont {N.~C.}\ \bibnamefont {{Weatherford}}}, \bibinfo
  {author} {\bibfnamefont {S.}~\bibnamefont {{Chatterjee}}}, \bibinfo {author}
  {\bibfnamefont {G.}~\bibnamefont {{Fragione}}}, \bibinfo {author}
  {\bibfnamefont {C.~L.}\ \bibnamefont {{Rodriguez}}}, \bibinfo {author}
  {\bibfnamefont {M.}~\bibnamefont {{Spera}}},\ and\ \bibinfo {author}
  {\bibfnamefont {F.~A.}\ \bibnamefont {{Rasio}}},\ }\href
  {https://doi.org/10.3847/1538-4365/ab7919} {\bibfield  {journal} {\bibinfo
  {journal} {Astrophys. J. Supp. S.}\ }\textbf {\bibinfo {volume} {247}},\
  \bibinfo {eid} {48} (\bibinfo {year} {2020})},\ \Eprint
  {https://arxiv.org/abs/1911.00018} {arXiv:1911.00018 [astro-ph.HE]}
  \BibitemShut {NoStop}%
\bibitem [{\citenamefont {{Kritos}}\ \emph {et~al.}(2022)\citenamefont
  {{Kritos}}, \citenamefont {{Strokov}}, \citenamefont {{Baibhav}},\ and\
  \citenamefont {{Berti}}}]{2022arXiv221010055K}%
  \BibitemOpen
  \bibfield  {author} {\bibinfo {author} {\bibfnamefont {K.}~\bibnamefont
  {{Kritos}}}, \bibinfo {author} {\bibfnamefont {V.}~\bibnamefont {{Strokov}}},
  \bibinfo {author} {\bibfnamefont {V.}~\bibnamefont {{Baibhav}}},\ and\
  \bibinfo {author} {\bibfnamefont {E.}~\bibnamefont {{Berti}}},\ }\href@noop
  {} {\bibfield  {journal} {\bibinfo  {journal} {{}}\ } (\bibinfo {year}
  {2022})},\ \Eprint {https://arxiv.org/abs/2210.10055} {arXiv:2210.10055
  [astro-ph.HE]} \BibitemShut {NoStop}%
\bibitem [{\citenamefont {{Racine}}(2008)}]{2008PhRvD..78d4021R}%
  \BibitemOpen
  \bibfield  {author} {\bibinfo {author} {\bibfnamefont {{\'E}.}~\bibnamefont
  {{Racine}}},\ }\href {https://doi.org/10.1103/PhysRevD.78.044021} {\bibfield
  {journal} {\bibinfo  {journal} {Phys. Rev. D}\ }\textbf {\bibinfo {volume}
  {78}},\ \bibinfo {eid} {044021} (\bibinfo {year} {2008})},\ \Eprint
  {https://arxiv.org/abs/0803.1820} {arXiv:0803.1820 [gr-qc]} \BibitemShut
  {NoStop}%
\bibitem [{\citenamefont {{Schmidt}}\ \emph {et~al.}(2015)\citenamefont
  {{Schmidt}}, \citenamefont {{Ohme}},\ and\ \citenamefont
  {{Hannam}}}]{2015PhRvD..91b4043S}%
  \BibitemOpen
  \bibfield  {author} {\bibinfo {author} {\bibfnamefont {P.}~\bibnamefont
  {{Schmidt}}}, \bibinfo {author} {\bibfnamefont {F.}~\bibnamefont {{Ohme}}},\
  and\ \bibinfo {author} {\bibfnamefont {M.}~\bibnamefont {{Hannam}}},\ }\href
  {https://doi.org/10.1103/PhysRevD.91.024043} {\bibfield  {journal} {\bibinfo
  {journal} {Phys. Rev. D}\ }\textbf {\bibinfo {volume} {91}},\ \bibinfo {eid}
  {024043} (\bibinfo {year} {2015})},\ \Eprint
  {https://arxiv.org/abs/1408.1810} {arXiv:1408.1810 [gr-qc]} \BibitemShut
  {NoStop}%
\bibitem [{\citenamefont {{Gerosa}}\ \emph {et~al.}(2021)\citenamefont
  {{Gerosa}}, \citenamefont {{Mould}}, \citenamefont {{Gangardt}},
  \citenamefont {{Schmidt}}, \citenamefont {{Pratten}},\ and\ \citenamefont
  {{Thomas}}}]{2021PhRvD.103f4067G}%
  \BibitemOpen
  \bibfield  {author} {\bibinfo {author} {\bibfnamefont {D.}~\bibnamefont
  {{Gerosa}}}, \bibinfo {author} {\bibfnamefont {M.}~\bibnamefont {{Mould}}},
  \bibinfo {author} {\bibfnamefont {D.}~\bibnamefont {{Gangardt}}}, \bibinfo
  {author} {\bibfnamefont {P.}~\bibnamefont {{Schmidt}}}, \bibinfo {author}
  {\bibfnamefont {G.}~\bibnamefont {{Pratten}}},\ and\ \bibinfo {author}
  {\bibfnamefont {L.~M.}\ \bibnamefont {{Thomas}}},\ }\href
  {https://doi.org/10.1103/PhysRevD.103.064067} {\bibfield  {journal} {\bibinfo
   {journal} {Phys. Rev. D}\ }\textbf {\bibinfo {volume} {103}},\ \bibinfo
  {eid} {064067} (\bibinfo {year} {2021})},\ \Eprint
  {https://arxiv.org/abs/2011.11948} {arXiv:2011.11948 [gr-qc]} \BibitemShut
  {NoStop}%
\bibitem [{\citenamefont {{Kesden}}\ \emph {et~al.}(2015)\citenamefont
  {{Kesden}}, \citenamefont {{Gerosa}}, \citenamefont {{O'Shaughnessy}},
  \citenamefont {{Berti}},\ and\ \citenamefont
  {{Sperhake}}}]{2015PhRvL.114h1103K}%
  \BibitemOpen
  \bibfield  {author} {\bibinfo {author} {\bibfnamefont {M.}~\bibnamefont
  {{Kesden}}}, \bibinfo {author} {\bibfnamefont {D.}~\bibnamefont {{Gerosa}}},
  \bibinfo {author} {\bibfnamefont {R.}~\bibnamefont {{O'Shaughnessy}}},
  \bibinfo {author} {\bibfnamefont {E.}~\bibnamefont {{Berti}}},\ and\ \bibinfo
  {author} {\bibfnamefont {U.}~\bibnamefont {{Sperhake}}},\ }\href
  {https://doi.org/10.1103/PhysRevLett.114.081103} {\bibfield  {journal}
  {\bibinfo  {journal} {Phys. Rev. Lett.}\ }\textbf {\bibinfo {volume} {114}},\
  \bibinfo {eid} {081103} (\bibinfo {year} {2015})},\ \Eprint
  {https://arxiv.org/abs/1411.0674} {arXiv:1411.0674 [gr-qc]} \BibitemShut
  {NoStop}%
\bibitem [{\citenamefont {{Romero-Shaw}}\ \emph {et~al.}(2021)\citenamefont
  {{Romero-Shaw}}, \citenamefont {{Lasky}},\ and\ \citenamefont
  {{Thrane}}}]{2021ApJ...921L..31R}%
  \BibitemOpen
  \bibfield  {author} {\bibinfo {author} {\bibfnamefont {I.}~\bibnamefont
  {{Romero-Shaw}}}, \bibinfo {author} {\bibfnamefont {P.~D.}\ \bibnamefont
  {{Lasky}}},\ and\ \bibinfo {author} {\bibfnamefont {E.}~\bibnamefont
  {{Thrane}}},\ }\href {https://doi.org/10.3847/2041-8213/ac3138} {\bibfield
  {journal} {\bibinfo  {journal} {Astrophys. J. Lett.}\ }\textbf {\bibinfo
  {volume} {921}},\ \bibinfo {eid} {L31} (\bibinfo {year} {2021})},\ \Eprint
  {https://arxiv.org/abs/2108.01284} {arXiv:2108.01284 [astro-ph.HE]}
  \BibitemShut {NoStop}%
\bibitem [{\citenamefont {{Belczynski}}\ \emph {et~al.}(2020)\citenamefont
  {{Belczynski}}, \citenamefont {{Klencki}}, \citenamefont {{Fields}},
  \citenamefont {{Olejak}}, \citenamefont {{Berti}}, \citenamefont {{Meynet}},
  \citenamefont {{Fryer}}, \citenamefont {{Holz}}, \citenamefont
  {{O'Shaughnessy}}, \citenamefont {{Brown}}, \citenamefont {{Bulik}},
  \citenamefont {{Leung}}, \citenamefont {{Nomoto}}, \citenamefont {{Madau}},
  \citenamefont {{Hirschi}}, \citenamefont {{Kaiser}}, \citenamefont {{Jones}},
  \citenamefont {{Mondal}}, \citenamefont {{Chruslinska}}, \citenamefont
  {{Drozda}}, \citenamefont {{Gerosa}}, \citenamefont {{Doctor}}, \citenamefont
  {{Giersz}}, \citenamefont {{Ekstrom}}, \citenamefont {{Georgy}},
  \citenamefont {{Askar}}, \citenamefont {{Baibhav}}, \citenamefont
  {{Wysocki}}, \citenamefont {{Natan}}, \citenamefont {{Farr}}, \citenamefont
  {{Wiktorowicz}}, \citenamefont {{Coleman Miller}}, \citenamefont {{Farr}},\
  and\ \citenamefont {{Lasota}}}]{2020A&A...636A.104B}%
  \BibitemOpen
  \bibfield  {author} {\bibinfo {author} {\bibfnamefont {K.}~\bibnamefont
  {{Belczynski}}}, \bibinfo {author} {\bibfnamefont {J.}~\bibnamefont
  {{Klencki}}}, \bibinfo {author} {\bibfnamefont {C.~E.}\ \bibnamefont
  {{Fields}}}, \bibinfo {author} {\bibfnamefont {A.}~\bibnamefont {{Olejak}}},
  \bibinfo {author} {\bibfnamefont {E.}~\bibnamefont {{Berti}}}, \bibinfo
  {author} {\bibfnamefont {G.}~\bibnamefont {{Meynet}}}, \bibinfo {author}
  {\bibfnamefont {C.~L.}\ \bibnamefont {{Fryer}}}, \bibinfo {author}
  {\bibfnamefont {D.~E.}\ \bibnamefont {{Holz}}}, \bibinfo {author}
  {\bibfnamefont {R.}~\bibnamefont {{O'Shaughnessy}}}, \bibinfo {author}
  {\bibfnamefont {D.~A.}\ \bibnamefont {{Brown}}}, \bibinfo {author}
  {\bibfnamefont {T.}~\bibnamefont {{Bulik}}}, \bibinfo {author} {\bibfnamefont
  {S.~C.}\ \bibnamefont {{Leung}}}, \bibinfo {author} {\bibfnamefont
  {K.}~\bibnamefont {{Nomoto}}}, \bibinfo {author} {\bibfnamefont
  {P.}~\bibnamefont {{Madau}}}, \bibinfo {author} {\bibfnamefont
  {R.}~\bibnamefont {{Hirschi}}}, \bibinfo {author} {\bibfnamefont
  {E.}~\bibnamefont {{Kaiser}}}, \bibinfo {author} {\bibfnamefont
  {S.}~\bibnamefont {{Jones}}}, \bibinfo {author} {\bibfnamefont
  {S.}~\bibnamefont {{Mondal}}}, \bibinfo {author} {\bibfnamefont
  {M.}~\bibnamefont {{Chruslinska}}}, \bibinfo {author} {\bibfnamefont
  {P.}~\bibnamefont {{Drozda}}}, \bibinfo {author} {\bibfnamefont
  {D.}~\bibnamefont {{Gerosa}}}, \bibinfo {author} {\bibfnamefont
  {Z.}~\bibnamefont {{Doctor}}}, \bibinfo {author} {\bibfnamefont
  {M.}~\bibnamefont {{Giersz}}}, \bibinfo {author} {\bibfnamefont
  {S.}~\bibnamefont {{Ekstrom}}}, \bibinfo {author} {\bibfnamefont
  {C.}~\bibnamefont {{Georgy}}}, \bibinfo {author} {\bibfnamefont
  {A.}~\bibnamefont {{Askar}}}, \bibinfo {author} {\bibfnamefont
  {V.}~\bibnamefont {{Baibhav}}}, \bibinfo {author} {\bibfnamefont
  {D.}~\bibnamefont {{Wysocki}}}, \bibinfo {author} {\bibfnamefont
  {T.}~\bibnamefont {{Natan}}}, \bibinfo {author} {\bibfnamefont {W.~M.}\
  \bibnamefont {{Farr}}}, \bibinfo {author} {\bibfnamefont {G.}~\bibnamefont
  {{Wiktorowicz}}}, \bibinfo {author} {\bibfnamefont {M.}~\bibnamefont
  {{Coleman Miller}}}, \bibinfo {author} {\bibfnamefont {B.}~\bibnamefont
  {{Farr}}},\ and\ \bibinfo {author} {\bibfnamefont {J.~P.}\ \bibnamefont
  {{Lasota}}},\ }\href {https://doi.org/10.1051/0004-6361/201936528} {\bibfield
   {journal} {\bibinfo  {journal} {Astron. Astrophys.}\ }\textbf {\bibinfo
  {volume} {636}},\ \bibinfo {eid} {A104} (\bibinfo {year} {2020})},\ \Eprint
  {https://arxiv.org/abs/1706.07053} {arXiv:1706.07053 [astro-ph.HE]}
  \BibitemShut {NoStop}%
\bibitem [{\citenamefont {{Mapelli}}(2020)}]{2020FrASS...7...38M}%
  \BibitemOpen
  \bibfield  {author} {\bibinfo {author} {\bibfnamefont {M.}~\bibnamefont
  {{Mapelli}}},\ }\href {https://doi.org/10.3389/fspas.2020.00038} {\bibfield
  {journal} {\bibinfo  {journal} {Frontiers in Astronomy and Space Sciences}\
  }\textbf {\bibinfo {volume} {7}},\ \bibinfo {eid} {38} (\bibinfo {year}
  {2020})},\ \Eprint {https://arxiv.org/abs/2105.12455} {arXiv:2105.12455
  [astro-ph.HE]} \BibitemShut {NoStop}%
\bibitem [{\citenamefont {{Antonini}}\ and\ \citenamefont
  {{Gieles}}(2020)}]{2020MNRAS.492.2936A}%
  \BibitemOpen
  \bibfield  {author} {\bibinfo {author} {\bibfnamefont {F.}~\bibnamefont
  {{Antonini}}}\ and\ \bibinfo {author} {\bibfnamefont {M.}~\bibnamefont
  {{Gieles}}},\ }\href {https://doi.org/10.1093/mnras/stz3584} {\bibfield
  {journal} {\bibinfo  {journal} {Mon. Not. R. Astron. Soc.}\ }\textbf
  {\bibinfo {volume} {492}},\ \bibinfo {pages} {2936} (\bibinfo {year}
  {2020})},\ \Eprint {https://arxiv.org/abs/1906.11855} {arXiv:1906.11855
  [astro-ph.HE]} \BibitemShut {NoStop}%
\bibitem [{\citenamefont {{Fumagalli}}\ \emph {et~al.}()\citenamefont
  {{Fumagalli}}, \citenamefont {{Loutrel}}, \citenamefont {{Gerosa}},\ and\
  \citenamefont {{Boschini}}}]{giulianickdavidematteo}%
  \BibitemOpen
  \bibfield  {author} {\bibinfo {author} {\bibfnamefont {G.}~\bibnamefont
  {{Fumagalli}}}, \bibinfo {author} {\bibfnamefont {N.}~\bibnamefont
  {{Loutrel}}}, \bibinfo {author} {\bibfnamefont {D.}~\bibnamefont
  {{Gerosa}}},\ and\ \bibinfo {author} {\bibfnamefont {M.}~\bibnamefont
  {{Boschini}}},\ }\href@noop {} {\bibinfo  {journal} {to be published}\
  }\BibitemShut {NoStop}%
\bibitem [{\citenamefont {{Pratten}}\ \emph {et~al.}(2021)\citenamefont
  {{Pratten}}, \citenamefont {{Garc{\'\i}a-Quir{\'o}s}}, \citenamefont
  {{Colleoni}}, \citenamefont {{Ramos-Buades}}, \citenamefont {{Estell{\'e}s}},
  \citenamefont {{Mateu-Lucena}}, \citenamefont {{Jaume}}, \citenamefont
  {{Haney}}, \citenamefont {{Keitel}}, \citenamefont {{Thompson}},\ and\
  \citenamefont {{Husa}}}]{2021PhRvD.103j4056P}%
  \BibitemOpen
\bibfield  {journal} {  }\bibfield  {author} {\bibinfo {author} {\bibfnamefont
  {G.}~\bibnamefont {{Pratten}}}, \bibinfo {author} {\bibfnamefont
  {C.}~\bibnamefont {{Garc{\'\i}a-Quir{\'o}s}}}, \bibinfo {author}
  {\bibfnamefont {M.}~\bibnamefont {{Colleoni}}}, \bibinfo {author}
  {\bibfnamefont {A.}~\bibnamefont {{Ramos-Buades}}}, \bibinfo {author}
  {\bibfnamefont {H.}~\bibnamefont {{Estell{\'e}s}}}, \bibinfo {author}
  {\bibfnamefont {M.}~\bibnamefont {{Mateu-Lucena}}}, \bibinfo {author}
  {\bibfnamefont {R.}~\bibnamefont {{Jaume}}}, \bibinfo {author} {\bibfnamefont
  {M.}~\bibnamefont {{Haney}}}, \bibinfo {author} {\bibfnamefont
  {D.}~\bibnamefont {{Keitel}}}, \bibinfo {author} {\bibfnamefont {J.~E.}\
  \bibnamefont {{Thompson}}},\ and\ \bibinfo {author} {\bibfnamefont
  {S.}~\bibnamefont {{Husa}}},\ }\href
  {https://doi.org/10.1103/PhysRevD.103.104056} {\bibfield  {journal} {\bibinfo
   {journal} {Phys. Rev. D}\ }\textbf {\bibinfo {volume} {103}},\ \bibinfo
  {eid} {104056} (\bibinfo {year} {2021})},\ \Eprint
  {https://arxiv.org/abs/2004.06503} {arXiv:2004.06503 [gr-qc]} \BibitemShut
  {NoStop}%
\bibitem [{\citenamefont {{Ossokine}}\ \emph {et~al.}(2020)\citenamefont
  {{Ossokine}}, \citenamefont {{Buonanno}}, \citenamefont {{Marsat}},
  \citenamefont {{Cotesta}}, \citenamefont {{Babak}}, \citenamefont
  {{Dietrich}}, \citenamefont {{Haas}}, \citenamefont {{Hinder}}, \citenamefont
  {{Pfeiffer}}, \citenamefont {{P{\"u}rrer}}, \citenamefont {{Woodford}},
  \citenamefont {{Boyle}}, \citenamefont {{Kidder}}, \citenamefont {{Scheel}},\
  and\ \citenamefont {{Szil{\'a}gyi}}}]{2020PhRvD.102d4055O}%
  \BibitemOpen
  \bibfield  {author} {\bibinfo {author} {\bibfnamefont {S.}~\bibnamefont
  {{Ossokine}}}, \bibinfo {author} {\bibfnamefont {A.}~\bibnamefont
  {{Buonanno}}}, \bibinfo {author} {\bibfnamefont {S.}~\bibnamefont
  {{Marsat}}}, \bibinfo {author} {\bibfnamefont {R.}~\bibnamefont {{Cotesta}}},
  \bibinfo {author} {\bibfnamefont {S.}~\bibnamefont {{Babak}}}, \bibinfo
  {author} {\bibfnamefont {T.}~\bibnamefont {{Dietrich}}}, \bibinfo {author}
  {\bibfnamefont {R.}~\bibnamefont {{Haas}}}, \bibinfo {author} {\bibfnamefont
  {I.}~\bibnamefont {{Hinder}}}, \bibinfo {author} {\bibfnamefont {H.~P.}\
  \bibnamefont {{Pfeiffer}}}, \bibinfo {author} {\bibfnamefont
  {M.}~\bibnamefont {{P{\"u}rrer}}}, \bibinfo {author} {\bibfnamefont {C.~J.}\
  \bibnamefont {{Woodford}}}, \bibinfo {author} {\bibfnamefont
  {M.}~\bibnamefont {{Boyle}}}, \bibinfo {author} {\bibfnamefont {L.~E.}\
  \bibnamefont {{Kidder}}}, \bibinfo {author} {\bibfnamefont {M.~A.}\
  \bibnamefont {{Scheel}}},\ and\ \bibinfo {author} {\bibfnamefont
  {B.}~\bibnamefont {{Szil{\'a}gyi}}},\ }\href
  {https://doi.org/10.1103/PhysRevD.102.044055} {\bibfield  {journal} {\bibinfo
   {journal} {Phys. Rev. D}\ }\textbf {\bibinfo {volume} {102}},\ \bibinfo
  {eid} {044055} (\bibinfo {year} {2020})},\ \Eprint
  {https://arxiv.org/abs/2004.09442} {arXiv:2004.09442 [gr-qc]} \BibitemShut
  {NoStop}%
\bibitem [{\citenamefont {{Varma}}\ \emph {et~al.}(2019)\citenamefont
  {{Varma}}, \citenamefont {{Field}}, \citenamefont {{Scheel}}, \citenamefont
  {{Blackman}}, \citenamefont {{Gerosa}}, \citenamefont {{Stein}},
  \citenamefont {{Kidder}},\ and\ \citenamefont
  {{Pfeiffer}}}]{2019PhRvR...1c3015V}%
  \BibitemOpen
  \bibfield  {author} {\bibinfo {author} {\bibfnamefont {V.}~\bibnamefont
  {{Varma}}}, \bibinfo {author} {\bibfnamefont {S.~E.}\ \bibnamefont
  {{Field}}}, \bibinfo {author} {\bibfnamefont {M.~A.}\ \bibnamefont
  {{Scheel}}}, \bibinfo {author} {\bibfnamefont {J.}~\bibnamefont
  {{Blackman}}}, \bibinfo {author} {\bibfnamefont {D.}~\bibnamefont
  {{Gerosa}}}, \bibinfo {author} {\bibfnamefont {L.~C.}\ \bibnamefont
  {{Stein}}}, \bibinfo {author} {\bibfnamefont {L.~E.}\ \bibnamefont
  {{Kidder}}},\ and\ \bibinfo {author} {\bibfnamefont {H.~P.}\ \bibnamefont
  {{Pfeiffer}}},\ }\href {https://doi.org/10.1103/PhysRevResearch.1.033015}
  {\bibfield  {journal} {\bibinfo  {journal} {Phys. Rev. Res.}\ }\textbf
  {\bibinfo {volume} {1}},\ \bibinfo {eid} {033015} (\bibinfo {year} {2019})},\
  \Eprint {https://arxiv.org/abs/1905.09300} {arXiv:1905.09300 [gr-qc]}
  \BibitemShut {NoStop}%
\bibitem [{\citenamefont {{Hannam}}\ \emph {et~al.}(2022)\citenamefont
  {{Hannam}}, \citenamefont {{Hoy}}, \citenamefont {{Thompson}}, \citenamefont
  {{Fairhurst}}, \citenamefont {{Raymond}}, \citenamefont {{Colleoni}},
  \citenamefont {{Davis}}, \citenamefont {{Estell{\'e}s}}, \citenamefont
  {{Haster}}, \citenamefont {{Helmling-Cornell}}, \citenamefont {{Husa}},
  \citenamefont {{Keitel}}, \citenamefont {{Massinger}}, \citenamefont
  {{Men{\'e}ndez-V{\'a}zquez}}, \citenamefont {{Mogushi}}, \citenamefont
  {{Ossokine}}, \citenamefont {{Payne}}, \citenamefont {{Pratten}},
  \citenamefont {{Romero-Shaw}}, \citenamefont {{Sadiq}}, \citenamefont
  {{Schmidt}}, \citenamefont {{Tenorio}}, \citenamefont {{Udall}},
  \citenamefont {{Veitch}}, \citenamefont {{Williams}}, \citenamefont
  {{Yelikar}},\ and\ \citenamefont {{Zimmerman}}}]{2022Natur.610..652H}%
  \BibitemOpen
  \bibfield  {author} {\bibinfo {author} {\bibfnamefont {M.}~\bibnamefont
  {{Hannam}}}, \bibinfo {author} {\bibfnamefont {C.}~\bibnamefont {{Hoy}}},
  \bibinfo {author} {\bibfnamefont {J.~E.}\ \bibnamefont {{Thompson}}},
  \bibinfo {author} {\bibfnamefont {S.}~\bibnamefont {{Fairhurst}}}, \bibinfo
  {author} {\bibfnamefont {V.}~\bibnamefont {{Raymond}}}, \bibinfo {author}
  {\bibfnamefont {M.}~\bibnamefont {{Colleoni}}}, \bibinfo {author}
  {\bibfnamefont {D.}~\bibnamefont {{Davis}}}, \bibinfo {author} {\bibfnamefont
  {H.}~\bibnamefont {{Estell{\'e}s}}}, \bibinfo {author} {\bibfnamefont
  {C.-J.}\ \bibnamefont {{Haster}}}, \bibinfo {author} {\bibfnamefont
  {A.}~\bibnamefont {{Helmling-Cornell}}}, \bibinfo {author} {\bibfnamefont
  {S.}~\bibnamefont {{Husa}}}, \bibinfo {author} {\bibfnamefont
  {D.}~\bibnamefont {{Keitel}}}, \bibinfo {author} {\bibfnamefont {T.~J.}\
  \bibnamefont {{Massinger}}}, \bibinfo {author} {\bibfnamefont
  {A.}~\bibnamefont {{Men{\'e}ndez-V{\'a}zquez}}}, \bibinfo {author}
  {\bibfnamefont {K.}~\bibnamefont {{Mogushi}}}, \bibinfo {author}
  {\bibfnamefont {S.}~\bibnamefont {{Ossokine}}}, \bibinfo {author}
  {\bibfnamefont {E.}~\bibnamefont {{Payne}}}, \bibinfo {author} {\bibfnamefont
  {G.}~\bibnamefont {{Pratten}}}, \bibinfo {author} {\bibfnamefont
  {I.}~\bibnamefont {{Romero-Shaw}}}, \bibinfo {author} {\bibfnamefont
  {J.}~\bibnamefont {{Sadiq}}}, \bibinfo {author} {\bibfnamefont
  {P.}~\bibnamefont {{Schmidt}}}, \bibinfo {author} {\bibfnamefont
  {R.}~\bibnamefont {{Tenorio}}}, \bibinfo {author} {\bibfnamefont
  {R.}~\bibnamefont {{Udall}}}, \bibinfo {author} {\bibfnamefont
  {J.}~\bibnamefont {{Veitch}}}, \bibinfo {author} {\bibfnamefont
  {D.}~\bibnamefont {{Williams}}}, \bibinfo {author} {\bibfnamefont {A.~B.}\
  \bibnamefont {{Yelikar}}},\ and\ \bibinfo {author} {\bibfnamefont
  {A.}~\bibnamefont {{Zimmerman}}},\ }\href
  {https://doi.org/10.1038/s41586-022-05212-z} {\bibfield  {journal} {\bibinfo
  {journal} {Nature}\ }\textbf {\bibinfo {volume} {610}},\ \bibinfo {pages}
  {652} (\bibinfo {year} {2022})},\ \Eprint {https://arxiv.org/abs/2112.11300}
  {arXiv:2112.11300 [gr-qc]} \BibitemShut {NoStop}%
\bibitem [{\citenamefont {{Varma}}\ \emph {et~al.}(2022)\citenamefont
  {{Varma}}, \citenamefont {{Biscoveanu}}, \citenamefont {{Islam}},
  \citenamefont {{Shaik}}, \citenamefont {{Haster}}, \citenamefont {{Isi}},
  \citenamefont {{Farr}}, \citenamefont {{Field}},\ and\ \citenamefont
  {{Vitale}}}]{2022PhRvL.128s1102V}%
  \BibitemOpen
  \bibfield  {author} {\bibinfo {author} {\bibfnamefont {V.}~\bibnamefont
  {{Varma}}}, \bibinfo {author} {\bibfnamefont {S.}~\bibnamefont
  {{Biscoveanu}}}, \bibinfo {author} {\bibfnamefont {T.}~\bibnamefont
  {{Islam}}}, \bibinfo {author} {\bibfnamefont {F.~H.}\ \bibnamefont
  {{Shaik}}}, \bibinfo {author} {\bibfnamefont {C.-J.}\ \bibnamefont
  {{Haster}}}, \bibinfo {author} {\bibfnamefont {M.}~\bibnamefont {{Isi}}},
  \bibinfo {author} {\bibfnamefont {W.~M.}\ \bibnamefont {{Farr}}}, \bibinfo
  {author} {\bibfnamefont {S.~E.}\ \bibnamefont {{Field}}},\ and\ \bibinfo
  {author} {\bibfnamefont {S.}~\bibnamefont {{Vitale}}},\ }\href
  {https://doi.org/10.1103/PhysRevLett.128.191102} {\bibfield  {journal}
  {\bibinfo  {journal} {Phys. Rev. Lett.}\ }\textbf {\bibinfo {volume} {128}},\
  \bibinfo {eid} {191102} (\bibinfo {year} {2022})},\ \Eprint
  {https://arxiv.org/abs/2201.01302} {arXiv:2201.01302 [astro-ph.HE]}
  \BibitemShut {NoStop}%
\bibitem [{\citenamefont {{Payne}}\ \emph {et~al.}(2022)\citenamefont
  {{Payne}}, \citenamefont {{Hourihane}}, \citenamefont {{Golomb}},
  \citenamefont {{Udall}}, \citenamefont {{Davis}},\ and\ \citenamefont
  {{Chatziioannou}}}]{2022PhRvD.106j4017P}%
  \BibitemOpen
  \bibfield  {author} {\bibinfo {author} {\bibfnamefont {E.}~\bibnamefont
  {{Payne}}}, \bibinfo {author} {\bibfnamefont {S.}~\bibnamefont
  {{Hourihane}}}, \bibinfo {author} {\bibfnamefont {J.}~\bibnamefont
  {{Golomb}}}, \bibinfo {author} {\bibfnamefont {R.}~\bibnamefont {{Udall}}},
  \bibinfo {author} {\bibfnamefont {D.}~\bibnamefont {{Davis}}},\ and\ \bibinfo
  {author} {\bibfnamefont {K.}~\bibnamefont {{Chatziioannou}}},\ }\href
  {https://doi.org/10.1103/PhysRevD.106.104017} {\bibfield  {journal} {\bibinfo
   {journal} {Phys. Rev. D}\ }\textbf {\bibinfo {volume} {106}},\ \bibinfo
  {eid} {104017} (\bibinfo {year} {2022})},\ \Eprint
  {https://arxiv.org/abs/2206.11932} {arXiv:2206.11932 [gr-qc]} \BibitemShut
  {NoStop}%
\bibitem [{\citenamefont {Fumagalli}\ \emph {et~al.}(2024)\citenamefont
  {Fumagalli}, \citenamefont {Romero-Shaw}, \citenamefont {Gerosa},
  \citenamefont {De~Renzis}, \citenamefont {Kritos},\ and\ \citenamefont
  {Olejak}}]{violadata}%
  \BibitemOpen
  \bibfield  {author} {\bibinfo {author} {\bibfnamefont {G.}~\bibnamefont
  {Fumagalli}}, \bibinfo {author} {\bibfnamefont {I.}~\bibnamefont
  {Romero-Shaw}}, \bibinfo {author} {\bibfnamefont {D.}~\bibnamefont {Gerosa}},
  \bibinfo {author} {\bibfnamefont {V.}~\bibnamefont {De~Renzis}}, \bibinfo
  {author} {\bibfnamefont {K.}~\bibnamefont {Kritos}},\ and\ \bibinfo {author}
  {\bibfnamefont {A.}~\bibnamefont {Olejak}},\ }\href@noop {} {\  (\bibinfo
  {year} {2024})},\ \Eprint {https://arxiv.org/abs/2405.14945}
  {arXiv:2405.14945 [astro-ph.HE]} \BibitemShut {NoStop}%
\bibitem [{\citenamefont {Chung}\ \emph {et~al.}(1989)\citenamefont {Chung},
  \citenamefont {Kannappan}, \citenamefont {Ng},\ and\ \citenamefont
  {Sahoo}}]{chung1989measures}%
  \BibitemOpen
  \bibfield  {author} {\bibinfo {author} {\bibfnamefont {J.}~\bibnamefont
  {Chung}}, \bibinfo {author} {\bibfnamefont {P.}~\bibnamefont {Kannappan}},
  \bibinfo {author} {\bibfnamefont {C.~T.}\ \bibnamefont {Ng}},\ and\ \bibinfo
  {author} {\bibfnamefont {P.}~\bibnamefont {Sahoo}},\ }\href@noop {}
  {\bibfield  {journal} {\bibinfo  {journal} {J. Math. Anal. Appl}\ }\textbf
  {\bibinfo {volume} {138}},\ \bibinfo {pages} {280} (\bibinfo {year}
  {1989})}\BibitemShut {NoStop}%
\bibitem [{\citenamefont {{Moore}}\ and\ \citenamefont
  {{Gerosa}}(2021)}]{2021PhRvD.104h3008M}%
  \BibitemOpen
  \bibfield  {author} {\bibinfo {author} {\bibfnamefont {C.~J.}\ \bibnamefont
  {{Moore}}}\ and\ \bibinfo {author} {\bibfnamefont {D.}~\bibnamefont
  {{Gerosa}}},\ }\href {https://doi.org/10.1103/PhysRevD.104.083008} {\bibfield
   {journal} {\bibinfo  {journal} {Phys. Rev. D}\ }\textbf {\bibinfo {volume}
  {104}},\ \bibinfo {eid} {083008} (\bibinfo {year} {2021})},\ \Eprint
  {https://arxiv.org/abs/2108.02462} {arXiv:2108.02462 [gr-qc]} \BibitemShut
  {NoStop}%
\bibitem [{\citenamefont {{Klein}}(2021)}]{2021arXiv210610291K}%
  \BibitemOpen
  \bibfield  {author} {\bibinfo {author} {\bibfnamefont {A.}~\bibnamefont
  {{Klein}}},\ }\href@noop {} {\bibfield  {journal} {\bibinfo  {journal} {{}}\
  } (\bibinfo {year} {2021})},\ \Eprint {https://arxiv.org/abs/2106.10291}
  {arXiv:2106.10291 [gr-qc]} \BibitemShut {NoStop}%
\bibitem [{\citenamefont {{Yi}}\ \emph {et~al.}(2024)\citenamefont {{Yi}},
  \citenamefont {{Kuntz}}, \citenamefont {{Barausse}}, \citenamefont {{Berti}},
  \citenamefont {{Cheung}}, \citenamefont {{Kritos}},\ and\ \citenamefont
  {{Maselli}}}]{2024PhRvD.109l4029Y}%
  \BibitemOpen
  \bibfield  {author} {\bibinfo {author} {\bibfnamefont {S.}~\bibnamefont
  {{Yi}}}, \bibinfo {author} {\bibfnamefont {A.}~\bibnamefont {{Kuntz}}},
  \bibinfo {author} {\bibfnamefont {E.}~\bibnamefont {{Barausse}}}, \bibinfo
  {author} {\bibfnamefont {E.}~\bibnamefont {{Berti}}}, \bibinfo {author}
  {\bibfnamefont {M.~H.-Y.}\ \bibnamefont {{Cheung}}}, \bibinfo {author}
  {\bibfnamefont {K.}~\bibnamefont {{Kritos}}},\ and\ \bibinfo {author}
  {\bibfnamefont {A.}~\bibnamefont {{Maselli}}},\ }\href
  {https://doi.org/10.1103/PhysRevD.109.124029} {\bibfield  {journal} {\bibinfo
   {journal} {Phys. Rev. D}\ }\textbf {\bibinfo {volume} {109}},\ \bibinfo
  {eid} {124029} (\bibinfo {year} {2024})},\ \Eprint
  {https://arxiv.org/abs/2403.09767} {arXiv:2403.09767 [gr-qc]} \BibitemShut
  {NoStop}%
\bibitem [{\citenamefont {{Olejak}}\ \emph {et~al.}(2024)\citenamefont
  {{Olejak}}, \citenamefont {{Klencki}}, \citenamefont {{Xu}}, \citenamefont
  {{Wang}}, \citenamefont {{Belczynski}},\ and\ \citenamefont
  {{Lasota}}}]{2024arXiv240412426O}%
  \BibitemOpen
  \bibfield  {author} {\bibinfo {author} {\bibfnamefont {A.}~\bibnamefont
  {{Olejak}}}, \bibinfo {author} {\bibfnamefont {J.}~\bibnamefont {{Klencki}}},
  \bibinfo {author} {\bibfnamefont {X.-T.}\ \bibnamefont {{Xu}}}, \bibinfo
  {author} {\bibfnamefont {C.}~\bibnamefont {{Wang}}}, \bibinfo {author}
  {\bibfnamefont {K.}~\bibnamefont {{Belczynski}}},\ and\ \bibinfo {author}
  {\bibfnamefont {J.-P.}\ \bibnamefont {{Lasota}}},\ }\href@noop {} {\bibfield
  {journal} {\bibinfo  {journal} {{}}\ } (\bibinfo {year} {2024})},\ \Eprint
  {https://arxiv.org/abs/2404.12426} {arXiv:2404.12426 [astro-ph.HE]}
  \BibitemShut {NoStop}%
\bibitem [{\citenamefont {{Finn}}\ and\ \citenamefont
  {{Chernoff}}(1993)}]{1993PhRvD..47.2198F}%
  \BibitemOpen
  \bibfield  {author} {\bibinfo {author} {\bibfnamefont {L.~S.}\ \bibnamefont
  {{Finn}}}\ and\ \bibinfo {author} {\bibfnamefont {D.~F.}\ \bibnamefont
  {{Chernoff}}},\ }\href {https://doi.org/10.1103/PhysRevD.47.2198} {\bibfield
  {journal} {\bibinfo  {journal} {Phys. Rev. D}\ }\textbf {\bibinfo {volume}
  {47}},\ \bibinfo {pages} {2198} (\bibinfo {year} {1993})},\ \Eprint
  {https://arxiv.org/abs/gr-qc/9301003} {arXiv:gr-qc/9301003 [gr-qc]}
  \BibitemShut {NoStop}%
\bibitem [{\citenamefont {{Gerosa}}\ and\ \citenamefont
  {{Bellotti}}(2024)}]{2024CQGra..41l5002G}%
  \BibitemOpen
  \bibfield  {author} {\bibinfo {author} {\bibfnamefont {D.}~\bibnamefont
  {{Gerosa}}}\ and\ \bibinfo {author} {\bibfnamefont {M.}~\bibnamefont
  {{Bellotti}}},\ }\href {https://doi.org/10.1088/1361-6382/ad4509} {\bibfield
  {journal} {\bibinfo  {journal} {Class. Quantum Grav.}\ }\textbf {\bibinfo
  {volume} {41}},\ \bibinfo {eid} {125002} (\bibinfo {year} {2024})},\ \Eprint
  {https://arxiv.org/abs/2404.16930} {arXiv:2404.16930 [astro-ph.HE]}
  \BibitemShut {NoStop}%
\bibitem [{\citenamefont {{Gerosa}}(2017)}]{gwdet}%
  \BibitemOpen
  \bibfield  {author} {\bibinfo {author} {\bibfnamefont {D.}~\bibnamefont
  {{Gerosa}}},\ }\href@noop {} {\bibfield  {journal} {\bibinfo  {journal}
  {\href{https://github.com/dgerosa/gwdet/tree/master}{github.com/dgerosa/gwdet},\\
  \href{https://doi.org/10.5281/zenodo.889966}{doi.org/10.5281/zenodo.889966}}\
  } (\bibinfo {year} {2017})}\BibitemShut {NoStop}%
\bibitem [{\citenamefont {{Saini}}(2024)}]{2024MNRAS.528..833S}%
  \BibitemOpen
  \bibfield  {author} {\bibinfo {author} {\bibfnamefont {P.}~\bibnamefont
  {{Saini}}},\ }\href {https://doi.org/10.1093/mnras/stae037} {\bibfield
  {journal} {\bibinfo  {journal} {Mon. Not. R. Astron. Soc.}\ }\textbf
  {\bibinfo {volume} {528}},\ \bibinfo {pages} {833} (\bibinfo {year}
  {2024})},\ \Eprint {https://arxiv.org/abs/2308.07565} {arXiv:2308.07565
  [astro-ph.HE]} \BibitemShut {NoStop}%
\bibitem [{\citenamefont {{Samsing}}\ \emph {et~al.}(2022)\citenamefont
  {{Samsing}}, \citenamefont {{Bartos}}, \citenamefont {{D'Orazio}},
  \citenamefont {{Haiman}}, \citenamefont {{Kocsis}}, \citenamefont {{Leigh}},
  \citenamefont {{Liu}}, \citenamefont {{Pessah}},\ and\ \citenamefont
  {{Tagawa}}}]{2022Natur.603..237S}%
  \BibitemOpen
  \bibfield  {author} {\bibinfo {author} {\bibfnamefont {J.}~\bibnamefont
  {{Samsing}}}, \bibinfo {author} {\bibfnamefont {I.}~\bibnamefont {{Bartos}}},
  \bibinfo {author} {\bibfnamefont {D.~J.}\ \bibnamefont {{D'Orazio}}},
  \bibinfo {author} {\bibfnamefont {Z.}~\bibnamefont {{Haiman}}}, \bibinfo
  {author} {\bibfnamefont {B.}~\bibnamefont {{Kocsis}}}, \bibinfo {author}
  {\bibfnamefont {N.~W.~C.}\ \bibnamefont {{Leigh}}}, \bibinfo {author}
  {\bibfnamefont {B.}~\bibnamefont {{Liu}}}, \bibinfo {author} {\bibfnamefont
  {M.~E.}\ \bibnamefont {{Pessah}}},\ and\ \bibinfo {author} {\bibfnamefont
  {H.}~\bibnamefont {{Tagawa}}},\ }\href
  {https://doi.org/10.1038/s41586-021-04333-1} {\bibfield  {journal} {\bibinfo
  {journal} {Nature}\ }\textbf {\bibinfo {volume} {603}},\ \bibinfo {pages}
  {237} (\bibinfo {year} {2022})},\ \Eprint {https://arxiv.org/abs/2010.09765}
  {arXiv:2010.09765 [astro-ph.HE]} \BibitemShut {NoStop}%
\bibitem [{\citenamefont {{Trani}}\ \emph {et~al.}(2022)\citenamefont
  {{Trani}}, \citenamefont {{Rastello}}, \citenamefont {{Di Carlo}},
  \citenamefont {{Santoliquido}}, \citenamefont {{Tanikawa}},\ and\
  \citenamefont {{Mapelli}}}]{2022MNRAS.511.1362T}%
  \BibitemOpen
  \bibfield  {author} {\bibinfo {author} {\bibfnamefont {A.~A.}\ \bibnamefont
  {{Trani}}}, \bibinfo {author} {\bibfnamefont {S.}~\bibnamefont {{Rastello}}},
  \bibinfo {author} {\bibfnamefont {U.~N.}\ \bibnamefont {{Di Carlo}}},
  \bibinfo {author} {\bibfnamefont {F.}~\bibnamefont {{Santoliquido}}},
  \bibinfo {author} {\bibfnamefont {A.}~\bibnamefont {{Tanikawa}}},\ and\
  \bibinfo {author} {\bibfnamefont {M.}~\bibnamefont {{Mapelli}}},\ }\href
  {https://doi.org/10.1093/mnras/stac122} {\bibfield  {journal} {\bibinfo
  {journal} {Mon. Not. R. Astron. Soc.}\ }\textbf {\bibinfo {volume} {511}},\
  \bibinfo {pages} {1362} (\bibinfo {year} {2022})},\ \Eprint
  {https://arxiv.org/abs/2111.06388} {arXiv:2111.06388 [astro-ph.HE]}
  \BibitemShut {NoStop}%
\bibitem [{\citenamefont {{Stegmann}}\ and\ \citenamefont
  {{Antonini}}(2021)}]{2021PhRvD.103f3007S}%
  \BibitemOpen
  \bibfield  {author} {\bibinfo {author} {\bibfnamefont {J.}~\bibnamefont
  {{Stegmann}}}\ and\ \bibinfo {author} {\bibfnamefont {F.}~\bibnamefont
  {{Antonini}}},\ }\href {https://doi.org/10.1103/PhysRevD.103.063007}
  {\bibfield  {journal} {\bibinfo  {journal} {Phys. Rev. D}\ }\textbf {\bibinfo
  {volume} {103}},\ \bibinfo {eid} {063007} (\bibinfo {year} {2021})},\ \Eprint
  {https://arxiv.org/abs/2012.06329} {arXiv:2012.06329 [astro-ph.HE]}
  \BibitemShut {NoStop}%
\bibitem [{\citenamefont {{Steinle}}\ and\ \citenamefont
  {{Kesden}}(2021)}]{2021PhRvD.103f3032S}%
  \BibitemOpen
  \bibfield  {author} {\bibinfo {author} {\bibfnamefont {N.}~\bibnamefont
  {{Steinle}}}\ and\ \bibinfo {author} {\bibfnamefont {M.}~\bibnamefont
  {{Kesden}}},\ }\href {https://doi.org/10.1103/PhysRevD.103.063032} {\bibfield
   {journal} {\bibinfo  {journal} {Phys. Rev. D}\ }\textbf {\bibinfo {volume}
  {103}},\ \bibinfo {eid} {063032} (\bibinfo {year} {2021})},\ \Eprint
  {https://arxiv.org/abs/2010.00078} {arXiv:2010.00078 [astro-ph.HE]}
  \BibitemShut {NoStop}%
\bibitem [{\citenamefont {{McKernan}}\ \emph {et~al.}(2022)\citenamefont
  {{McKernan}}, \citenamefont {{Ford}}, \citenamefont {{Callister}},
  \citenamefont {{Farr}}, \citenamefont {{O'Shaughnessy}}, \citenamefont
  {{Smith}}, \citenamefont {{Thrane}},\ and\ \citenamefont
  {{Vajpeyi}}}]{2022MNRAS.514.3886M}%
  \BibitemOpen
  \bibfield  {author} {\bibinfo {author} {\bibfnamefont {B.}~\bibnamefont
  {{McKernan}}}, \bibinfo {author} {\bibfnamefont {K.~E.~S.}\ \bibnamefont
  {{Ford}}}, \bibinfo {author} {\bibfnamefont {T.}~\bibnamefont {{Callister}}},
  \bibinfo {author} {\bibfnamefont {W.~M.}\ \bibnamefont {{Farr}}}, \bibinfo
  {author} {\bibfnamefont {R.}~\bibnamefont {{O'Shaughnessy}}}, \bibinfo
  {author} {\bibfnamefont {R.}~\bibnamefont {{Smith}}}, \bibinfo {author}
  {\bibfnamefont {E.}~\bibnamefont {{Thrane}}},\ and\ \bibinfo {author}
  {\bibfnamefont {A.}~\bibnamefont {{Vajpeyi}}},\ }\href
  {https://doi.org/10.1093/mnras/stac1570} {\bibfield  {journal} {\bibinfo
  {journal} {Mon. Not. R. Astron. Soc.}\ }\textbf {\bibinfo {volume} {514}},\
  \bibinfo {pages} {3886} (\bibinfo {year} {2022})},\ \Eprint
  {https://arxiv.org/abs/2107.07551} {arXiv:2107.07551 [astro-ph.HE]}
  \BibitemShut {NoStop}%
\bibitem [{\citenamefont {{Tauris}}(2022)}]{2022ApJ...938...66T}%
  \BibitemOpen
  \bibfield  {author} {\bibinfo {author} {\bibfnamefont {T.~M.}\ \bibnamefont
  {{Tauris}}},\ }\href {https://doi.org/10.3847/1538-4357/ac86c8} {\bibfield
  {journal} {\bibinfo  {journal} {Astrophys. J.}\ }\textbf {\bibinfo {volume}
  {938}},\ \bibinfo {eid} {66} (\bibinfo {year} {2022})},\ \Eprint
  {https://arxiv.org/abs/2205.02541} {arXiv:2205.02541 [astro-ph.HE]}
  \BibitemShut {NoStop}%
\bibitem [{\citenamefont {{Mould}}\ \emph {et~al.}(2022)\citenamefont
  {{Mould}}, \citenamefont {{Gerosa}}, \citenamefont {{Broekgaarden}},\ and\
  \citenamefont {{Steinle}}}]{2022MNRAS.517.2738M}%
  \BibitemOpen
  \bibfield  {author} {\bibinfo {author} {\bibfnamefont {M.}~\bibnamefont
  {{Mould}}}, \bibinfo {author} {\bibfnamefont {D.}~\bibnamefont {{Gerosa}}},
  \bibinfo {author} {\bibfnamefont {F.~S.}\ \bibnamefont {{Broekgaarden}}},\
  and\ \bibinfo {author} {\bibfnamefont {N.}~\bibnamefont {{Steinle}}},\ }\href
  {https://doi.org/10.1093/mnras/stac2859} {\bibfield  {journal} {\bibinfo
  {journal} {Mon. Not. R. Astron. Soc.}\ }\textbf {\bibinfo {volume} {517}},\
  \bibinfo {pages} {2738} (\bibinfo {year} {2022})},\ \Eprint
  {https://arxiv.org/abs/2205.12329} {arXiv:2205.12329 [astro-ph.HE]}
  \BibitemShut {NoStop}%
\bibitem [{\citenamefont {{Santini}}\ \emph {et~al.}(2023)\citenamefont
  {{Santini}}, \citenamefont {{Gerosa}}, \citenamefont {{Cotesta}},\ and\
  \citenamefont {{Berti}}}]{2023PhRvD.108h3033S}%
  \BibitemOpen
  \bibfield  {author} {\bibinfo {author} {\bibfnamefont {A.}~\bibnamefont
  {{Santini}}}, \bibinfo {author} {\bibfnamefont {D.}~\bibnamefont {{Gerosa}}},
  \bibinfo {author} {\bibfnamefont {R.}~\bibnamefont {{Cotesta}}},\ and\
  \bibinfo {author} {\bibfnamefont {E.}~\bibnamefont {{Berti}}},\ }\href
  {https://doi.org/10.1103/PhysRevD.108.083033} {\bibfield  {journal} {\bibinfo
   {journal} {Phys. Rev. D}\ }\textbf {\bibinfo {volume} {108}},\ \bibinfo
  {eid} {083033} (\bibinfo {year} {2023})},\ \Eprint
  {https://arxiv.org/abs/2308.12998} {arXiv:2308.12998 [astro-ph.HE]}
  \BibitemShut {NoStop}%
\bibitem [{\citenamefont {{Adamcewicz}}\ \emph {et~al.}(2024)\citenamefont
  {{Adamcewicz}}, \citenamefont {{Galaudage}}, \citenamefont {{Lasky}},\ and\
  \citenamefont {{Thrane}}}]{2024ApJ...964L...6A}%
  \BibitemOpen
  \bibfield  {author} {\bibinfo {author} {\bibfnamefont {C.}~\bibnamefont
  {{Adamcewicz}}}, \bibinfo {author} {\bibfnamefont {S.}~\bibnamefont
  {{Galaudage}}}, \bibinfo {author} {\bibfnamefont {P.~D.}\ \bibnamefont
  {{Lasky}}},\ and\ \bibinfo {author} {\bibfnamefont {E.}~\bibnamefont
  {{Thrane}}},\ }\href {https://doi.org/10.3847/2041-8213/ad2df2} {\bibfield
  {journal} {\bibinfo  {journal} {Astrophys. J. Lett.}\ }\textbf {\bibinfo
  {volume} {964}},\ \bibinfo {eid} {L6} (\bibinfo {year} {2024})},\ \Eprint
  {https://arxiv.org/abs/2311.05182} {arXiv:2311.05182 [astro-ph.HE]}
  \BibitemShut {NoStop}%
\bibitem [{\citenamefont {{Buscicchio}}\ \emph {et~al.}(2021)\citenamefont
  {{Buscicchio}}, \citenamefont {{Klein}}, \citenamefont {{Roebber}},
  \citenamefont {{Moore}}, \citenamefont {{Gerosa}}, \citenamefont {{Finch}},\
  and\ \citenamefont {{Vecchio}}}]{2021PhRvD.104d4065B}%
  \BibitemOpen
  \bibfield  {author} {\bibinfo {author} {\bibfnamefont {R.}~\bibnamefont
  {{Buscicchio}}}, \bibinfo {author} {\bibfnamefont {A.}~\bibnamefont
  {{Klein}}}, \bibinfo {author} {\bibfnamefont {E.}~\bibnamefont {{Roebber}}},
  \bibinfo {author} {\bibfnamefont {C.~J.}\ \bibnamefont {{Moore}}}, \bibinfo
  {author} {\bibfnamefont {D.}~\bibnamefont {{Gerosa}}}, \bibinfo {author}
  {\bibfnamefont {E.}~\bibnamefont {{Finch}}},\ and\ \bibinfo {author}
  {\bibfnamefont {A.}~\bibnamefont {{Vecchio}}},\ }\href
  {https://doi.org/10.1103/PhysRevD.104.044065} {\bibfield  {journal} {\bibinfo
   {journal} {Phys. Rev. D}\ }\textbf {\bibinfo {volume} {104}},\ \bibinfo
  {eid} {044065} (\bibinfo {year} {2021})},\ \Eprint
  {https://arxiv.org/abs/2106.05259} {arXiv:2106.05259 [astro-ph.HE]}
  \BibitemShut {NoStop}%
\bibitem [{\citenamefont {{Colpi}}\ \emph {et~al.}(2024)\citenamefont
  {{Colpi}}, \citenamefont {{Danzmann}}, \citenamefont {{Hewitson}},
  \citenamefont {{Holley-Bockelmann}}, \citenamefont {{Jetzer}}, \citenamefont
  {{Nelemans}}, \citenamefont {{Petiteau}}, \citenamefont {{Shoemaker}},
  \citenamefont {{Sopuerta}}, \citenamefont {{Stebbins}}, \citenamefont
  {{Tanvir}}, \citenamefont {{Ward}}, \citenamefont {{Weber}}, \citenamefont
  {{Thorpe}}, \citenamefont {{Daurskikh}}, \citenamefont {{Deep}},
  \citenamefont {{Fern{\'a}ndez N{\'u}{\~n}ez}}, \citenamefont {{Garc{\'\i}a
  Marirrodriga}}, \citenamefont {{Gehler}}, \citenamefont {{Halain}},
  \citenamefont {{Jennrich}}, \citenamefont {{Lammers}}, \citenamefont
  {{Larra{\~n}aga}}, \citenamefont {{Lieser}}, \citenamefont
  {{L{\"u}tzgendorf}}, \citenamefont {{Martens}}, \citenamefont {{Mondin}},
  \citenamefont {{Piris Ni{\~n}o}}, \citenamefont {{Amaro-Seoane}},
  \citenamefont {{Arca Sedda}}, \citenamefont {{Auclair}}, \citenamefont
  {{Babak}}, \citenamefont {{Baghi}}, \citenamefont {{Baibhav}}, \citenamefont
  {{Baker}}, \citenamefont {{Bayle}}, \citenamefont {{Berry}}, \citenamefont
  {{Berti}}, \citenamefont {{Boileau}}, \citenamefont {{Bonetti}},
  \citenamefont {{Brito}}, \citenamefont {{Buscicchio}}, \citenamefont
  {{Calcagni}}, \citenamefont {{Capelo}}, \citenamefont {{Caprini}},
  \citenamefont {{Caputo}}, \citenamefont {{Castelli}}, \citenamefont {{Chen}},
  \citenamefont {{Chen}}, \citenamefont {{Chua}}, \citenamefont {{Davies}},
  \citenamefont {{Derdzinski}}, \citenamefont {{Domcke}}, \citenamefont
  {{Doneva}}, \citenamefont {{Dvorkin}}, \citenamefont {{Mar{\'\i}a Ezquiaga}},
  \citenamefont {{Gair}}, \citenamefont {{Haiman}}, \citenamefont {{Harry}},
  \citenamefont {{Hartwig}}, \citenamefont {{Hees}}, \citenamefont
  {{Heffernan}}, \citenamefont {{Husa}}, \citenamefont {{Izquierdo-Villalba}},
  \citenamefont {{Karnesis}}, \citenamefont {{Klein}}, \citenamefont {{Korol}},
  \citenamefont {{Korsakova}}, \citenamefont {{Kupfer}}, \citenamefont
  {{Laghi}}, \citenamefont {{Lamberts}}, \citenamefont {{Larson}},
  \citenamefont {{Le Jeune}}, \citenamefont {{Lewicki}}, \citenamefont
  {{Littenberg}}, \citenamefont {{Madge}}, \citenamefont {{Mangiagli}},
  \citenamefont {{Marsat}}, \citenamefont {{Vilchez}}, \citenamefont
  {{Maselli}}, \citenamefont {{Mathews}}, \citenamefont {{van de Meent}},
  \citenamefont {{Muratore}}, \citenamefont {{Nardini}}, \citenamefont
  {{Pani}}, \citenamefont {{Peloso}}, \citenamefont {{Pieroni}}, \citenamefont
  {{Pound}}, \citenamefont {{Quelquejay-Leclere}}, \citenamefont
  {{Ricciardone}}, \citenamefont {{Rossi}}, \citenamefont {{Sartirana}},
  \citenamefont {{Savalle}}, \citenamefont {{Sberna}}, \citenamefont
  {{Sesana}}, \citenamefont {{Shoemaker}}, \citenamefont {{Slutsky}},
  \citenamefont {{Sotiriou}}, \citenamefont {{Speri}}, \citenamefont {{Staab}},
  \citenamefont {{Steer}}, \citenamefont {{Tamanini}}, \citenamefont
  {{Tasinato}}, \citenamefont {{Torrado}}, \citenamefont {{Torres-Orjuela}},
  \citenamefont {{Toubiana}}, \citenamefont {{Vallisneri}}, \citenamefont
  {{Vecchio}}, \citenamefont {{Volonteri}}, \citenamefont {{Yagi}},\ and\
  \citenamefont {{Zwick}}}]{2024arXiv240207571C}%
  \BibitemOpen
  \bibfield  {author} {\bibinfo {author} {\bibfnamefont {M.}~\bibnamefont
  {{Colpi}}}, \bibinfo {author} {\bibfnamefont {K.}~\bibnamefont {{Danzmann}}},
  \bibinfo {author} {\bibfnamefont {M.}~\bibnamefont {{Hewitson}}}, \bibinfo
  {author} {\bibfnamefont {K.}~\bibnamefont {{Holley-Bockelmann}}}, \bibinfo
  {author} {\bibfnamefont {P.}~\bibnamefont {{Jetzer}}}, \bibinfo {author}
  {\bibfnamefont {G.}~\bibnamefont {{Nelemans}}}, \bibinfo {author}
  {\bibfnamefont {A.}~\bibnamefont {{Petiteau}}}, \bibinfo {author}
  {\bibfnamefont {D.}~\bibnamefont {{Shoemaker}}}, \bibinfo {author}
  {\bibfnamefont {C.}~\bibnamefont {{Sopuerta}}}, \bibinfo {author}
  {\bibfnamefont {R.}~\bibnamefont {{Stebbins}}}, \bibinfo {author}
  {\bibfnamefont {N.}~\bibnamefont {{Tanvir}}}, \bibinfo {author}
  {\bibfnamefont {H.}~\bibnamefont {{Ward}}}, \bibinfo {author} {\bibfnamefont
  {W.~J.}\ \bibnamefont {{Weber}}}, \bibinfo {author} {\bibfnamefont
  {I.}~\bibnamefont {{Thorpe}}}, \bibinfo {author} {\bibfnamefont
  {A.}~\bibnamefont {{Daurskikh}}}, \bibinfo {author} {\bibfnamefont
  {A.}~\bibnamefont {{Deep}}}, \bibinfo {author} {\bibfnamefont
  {I.}~\bibnamefont {{Fern{\'a}ndez N{\'u}{\~n}ez}}}, \bibinfo {author}
  {\bibfnamefont {C.}~\bibnamefont {{Garc{\'\i}a Marirrodriga}}}, \bibinfo
  {author} {\bibfnamefont {M.}~\bibnamefont {{Gehler}}}, \bibinfo {author}
  {\bibfnamefont {J.-P.}\ \bibnamefont {{Halain}}}, \bibinfo {author}
  {\bibfnamefont {O.}~\bibnamefont {{Jennrich}}}, \bibinfo {author}
  {\bibfnamefont {U.}~\bibnamefont {{Lammers}}}, \bibinfo {author}
  {\bibfnamefont {J.}~\bibnamefont {{Larra{\~n}aga}}}, \bibinfo {author}
  {\bibfnamefont {M.}~\bibnamefont {{Lieser}}}, \bibinfo {author}
  {\bibfnamefont {N.}~\bibnamefont {{L{\"u}tzgendorf}}}, \bibinfo {author}
  {\bibfnamefont {W.}~\bibnamefont {{Martens}}}, \bibinfo {author}
  {\bibfnamefont {L.}~\bibnamefont {{Mondin}}}, \bibinfo {author}
  {\bibfnamefont {A.}~\bibnamefont {{Piris Ni{\~n}o}}}, \bibinfo {author}
  {\bibfnamefont {P.}~\bibnamefont {{Amaro-Seoane}}}, \bibinfo {author}
  {\bibfnamefont {M.}~\bibnamefont {{Arca Sedda}}}, \bibinfo {author}
  {\bibfnamefont {P.}~\bibnamefont {{Auclair}}}, \bibinfo {author}
  {\bibfnamefont {S.}~\bibnamefont {{Babak}}}, \bibinfo {author} {\bibfnamefont
  {Q.}~\bibnamefont {{Baghi}}}, \bibinfo {author} {\bibfnamefont
  {V.}~\bibnamefont {{Baibhav}}}, \bibinfo {author} {\bibfnamefont
  {T.}~\bibnamefont {{Baker}}}, \bibinfo {author} {\bibfnamefont {J.-B.}\
  \bibnamefont {{Bayle}}}, \bibinfo {author} {\bibfnamefont {C.}~\bibnamefont
  {{Berry}}}, \bibinfo {author} {\bibfnamefont {E.}~\bibnamefont {{Berti}}},
  \bibinfo {author} {\bibfnamefont {G.}~\bibnamefont {{Boileau}}}, \bibinfo
  {author} {\bibfnamefont {M.}~\bibnamefont {{Bonetti}}}, \bibinfo {author}
  {\bibfnamefont {R.}~\bibnamefont {{Brito}}}, \bibinfo {author} {\bibfnamefont
  {R.}~\bibnamefont {{Buscicchio}}}, \bibinfo {author} {\bibfnamefont
  {G.}~\bibnamefont {{Calcagni}}}, \bibinfo {author} {\bibfnamefont {P.~R.}\
  \bibnamefont {{Capelo}}}, \bibinfo {author} {\bibfnamefont {C.}~\bibnamefont
  {{Caprini}}}, \bibinfo {author} {\bibfnamefont {A.}~\bibnamefont {{Caputo}}},
  \bibinfo {author} {\bibfnamefont {E.}~\bibnamefont {{Castelli}}}, \bibinfo
  {author} {\bibfnamefont {H.-Y.}\ \bibnamefont {{Chen}}}, \bibinfo {author}
  {\bibfnamefont {X.}~\bibnamefont {{Chen}}}, \bibinfo {author} {\bibfnamefont
  {A.}~\bibnamefont {{Chua}}}, \bibinfo {author} {\bibfnamefont
  {G.}~\bibnamefont {{Davies}}}, \bibinfo {author} {\bibfnamefont
  {A.}~\bibnamefont {{Derdzinski}}}, \bibinfo {author} {\bibfnamefont {V.~F.}\
  \bibnamefont {{Domcke}}}, \bibinfo {author} {\bibfnamefont {D.}~\bibnamefont
  {{Doneva}}}, \bibinfo {author} {\bibfnamefont {I.}~\bibnamefont {{Dvorkin}}},
  \bibinfo {author} {\bibfnamefont {J.}~\bibnamefont {{Mar{\'\i}a Ezquiaga}}},
  \bibinfo {author} {\bibfnamefont {J.}~\bibnamefont {{Gair}}}, \bibinfo
  {author} {\bibfnamefont {Z.}~\bibnamefont {{Haiman}}}, \bibinfo {author}
  {\bibfnamefont {I.}~\bibnamefont {{Harry}}}, \bibinfo {author} {\bibfnamefont
  {O.}~\bibnamefont {{Hartwig}}}, \bibinfo {author} {\bibfnamefont
  {A.}~\bibnamefont {{Hees}}}, \bibinfo {author} {\bibfnamefont
  {A.}~\bibnamefont {{Heffernan}}}, \bibinfo {author} {\bibfnamefont
  {S.}~\bibnamefont {{Husa}}}, \bibinfo {author} {\bibfnamefont
  {D.}~\bibnamefont {{Izquierdo-Villalba}}}, \bibinfo {author} {\bibfnamefont
  {N.}~\bibnamefont {{Karnesis}}}, \bibinfo {author} {\bibfnamefont
  {A.}~\bibnamefont {{Klein}}}, \bibinfo {author} {\bibfnamefont
  {V.}~\bibnamefont {{Korol}}}, \bibinfo {author} {\bibfnamefont
  {N.}~\bibnamefont {{Korsakova}}}, \bibinfo {author} {\bibfnamefont
  {T.}~\bibnamefont {{Kupfer}}}, \bibinfo {author} {\bibfnamefont
  {D.}~\bibnamefont {{Laghi}}}, \bibinfo {author} {\bibfnamefont
  {A.}~\bibnamefont {{Lamberts}}}, \bibinfo {author} {\bibfnamefont
  {S.}~\bibnamefont {{Larson}}}, \bibinfo {author} {\bibfnamefont
  {M.}~\bibnamefont {{Le Jeune}}}, \bibinfo {author} {\bibfnamefont
  {M.}~\bibnamefont {{Lewicki}}}, \bibinfo {author} {\bibfnamefont
  {T.}~\bibnamefont {{Littenberg}}}, \bibinfo {author} {\bibfnamefont
  {E.}~\bibnamefont {{Madge}}}, \bibinfo {author} {\bibfnamefont
  {A.}~\bibnamefont {{Mangiagli}}}, \bibinfo {author} {\bibfnamefont
  {S.}~\bibnamefont {{Marsat}}}, \bibinfo {author} {\bibfnamefont {I.~M.}\
  \bibnamefont {{Vilchez}}}, \bibinfo {author} {\bibfnamefont {A.}~\bibnamefont
  {{Maselli}}}, \bibinfo {author} {\bibfnamefont {J.}~\bibnamefont
  {{Mathews}}}, \bibinfo {author} {\bibfnamefont {M.}~\bibnamefont {{van de
  Meent}}}, \bibinfo {author} {\bibfnamefont {M.}~\bibnamefont {{Muratore}}},
  \bibinfo {author} {\bibfnamefont {G.}~\bibnamefont {{Nardini}}}, \bibinfo
  {author} {\bibfnamefont {P.}~\bibnamefont {{Pani}}}, \bibinfo {author}
  {\bibfnamefont {M.}~\bibnamefont {{Peloso}}}, \bibinfo {author}
  {\bibfnamefont {M.}~\bibnamefont {{Pieroni}}}, \bibinfo {author}
  {\bibfnamefont {A.}~\bibnamefont {{Pound}}}, \bibinfo {author} {\bibfnamefont
  {H.}~\bibnamefont {{Quelquejay-Leclere}}}, \bibinfo {author} {\bibfnamefont
  {A.}~\bibnamefont {{Ricciardone}}}, \bibinfo {author} {\bibfnamefont {E.~M.}\
  \bibnamefont {{Rossi}}}, \bibinfo {author} {\bibfnamefont {A.}~\bibnamefont
  {{Sartirana}}}, \bibinfo {author} {\bibfnamefont {E.}~\bibnamefont
  {{Savalle}}}, \bibinfo {author} {\bibfnamefont {L.}~\bibnamefont {{Sberna}}},
  \bibinfo {author} {\bibfnamefont {A.}~\bibnamefont {{Sesana}}}, \bibinfo
  {author} {\bibfnamefont {D.}~\bibnamefont {{Shoemaker}}}, \bibinfo {author}
  {\bibfnamefont {J.}~\bibnamefont {{Slutsky}}}, \bibinfo {author}
  {\bibfnamefont {T.}~\bibnamefont {{Sotiriou}}}, \bibinfo {author}
  {\bibfnamefont {L.}~\bibnamefont {{Speri}}}, \bibinfo {author} {\bibfnamefont
  {M.}~\bibnamefont {{Staab}}}, \bibinfo {author} {\bibfnamefont
  {D.}~\bibnamefont {{Steer}}}, \bibinfo {author} {\bibfnamefont
  {N.}~\bibnamefont {{Tamanini}}}, \bibinfo {author} {\bibfnamefont
  {G.}~\bibnamefont {{Tasinato}}}, \bibinfo {author} {\bibfnamefont
  {J.}~\bibnamefont {{Torrado}}}, \bibinfo {author} {\bibfnamefont
  {A.}~\bibnamefont {{Torres-Orjuela}}}, \bibinfo {author} {\bibfnamefont
  {A.}~\bibnamefont {{Toubiana}}}, \bibinfo {author} {\bibfnamefont
  {M.}~\bibnamefont {{Vallisneri}}}, \bibinfo {author} {\bibfnamefont
  {A.}~\bibnamefont {{Vecchio}}}, \bibinfo {author} {\bibfnamefont
  {M.}~\bibnamefont {{Volonteri}}}, \bibinfo {author} {\bibfnamefont
  {K.}~\bibnamefont {{Yagi}}},\ and\ \bibinfo {author} {\bibfnamefont
  {L.}~\bibnamefont {{Zwick}}},\ }\href@noop {} {\bibfield  {journal} {\bibinfo
   {journal} {{}}\ } (\bibinfo {year} {2024})},\ \Eprint
  {https://arxiv.org/abs/2402.07571} {arXiv:2402.07571 [astro-ph.CO]}
  \BibitemShut {NoStop}%
\bibitem [{\citenamefont {{Reitze}}\ \emph {et~al.}(2019)\citenamefont
  {{Reitze}}, \citenamefont {{Adhikari}}, \citenamefont {{Ballmer}},
  \citenamefont {{Barish}}, \citenamefont {{Barsotti}}, \citenamefont
  {{Billingsley}}, \citenamefont {{Brown}}, \citenamefont {{Chen}},
  \citenamefont {{Coyne}}, \citenamefont {{Eisenstein}}, \citenamefont
  {{Evans}}, \citenamefont {{Fritschel}}, \citenamefont {{Hall}}, \citenamefont
  {{Lazzarini}}, \citenamefont {{Lovelace}}, \citenamefont {{Read}},
  \citenamefont {{Sathyaprakash}}, \citenamefont {{Shoemaker}}, \citenamefont
  {{Smith}}, \citenamefont {{Torrie}}, \citenamefont {{Vitale}}, \citenamefont
  {{Weiss}}, \citenamefont {{Wipf}},\ and\ \citenamefont
  {{Zucker}}}]{2019BAAS...51g..35R}%
  \BibitemOpen
  \bibfield  {author} {\bibinfo {author} {\bibfnamefont {D.}~\bibnamefont
  {{Reitze}}}, \bibinfo {author} {\bibfnamefont {R.~X.}\ \bibnamefont
  {{Adhikari}}}, \bibinfo {author} {\bibfnamefont {S.}~\bibnamefont
  {{Ballmer}}}, \bibinfo {author} {\bibfnamefont {B.}~\bibnamefont {{Barish}}},
  \bibinfo {author} {\bibfnamefont {L.}~\bibnamefont {{Barsotti}}}, \bibinfo
  {author} {\bibfnamefont {G.}~\bibnamefont {{Billingsley}}}, \bibinfo {author}
  {\bibfnamefont {D.~A.}\ \bibnamefont {{Brown}}}, \bibinfo {author}
  {\bibfnamefont {Y.}~\bibnamefont {{Chen}}}, \bibinfo {author} {\bibfnamefont
  {D.}~\bibnamefont {{Coyne}}}, \bibinfo {author} {\bibfnamefont
  {R.}~\bibnamefont {{Eisenstein}}}, \bibinfo {author} {\bibfnamefont
  {M.}~\bibnamefont {{Evans}}}, \bibinfo {author} {\bibfnamefont
  {P.}~\bibnamefont {{Fritschel}}}, \bibinfo {author} {\bibfnamefont {E.~D.}\
  \bibnamefont {{Hall}}}, \bibinfo {author} {\bibfnamefont {A.}~\bibnamefont
  {{Lazzarini}}}, \bibinfo {author} {\bibfnamefont {G.}~\bibnamefont
  {{Lovelace}}}, \bibinfo {author} {\bibfnamefont {J.}~\bibnamefont {{Read}}},
  \bibinfo {author} {\bibfnamefont {B.~S.}\ \bibnamefont {{Sathyaprakash}}},
  \bibinfo {author} {\bibfnamefont {D.}~\bibnamefont {{Shoemaker}}}, \bibinfo
  {author} {\bibfnamefont {J.}~\bibnamefont {{Smith}}}, \bibinfo {author}
  {\bibfnamefont {C.}~\bibnamefont {{Torrie}}}, \bibinfo {author}
  {\bibfnamefont {S.}~\bibnamefont {{Vitale}}}, \bibinfo {author}
  {\bibfnamefont {R.}~\bibnamefont {{Weiss}}}, \bibinfo {author} {\bibfnamefont
  {C.}~\bibnamefont {{Wipf}}},\ and\ \bibinfo {author} {\bibfnamefont
  {M.}~\bibnamefont {{Zucker}}},\ }in\ \href@noop {} {\emph {\bibinfo
  {booktitle} {Bulletin of the American Astronomical Society}}},\ Vol.~\bibinfo
  {volume} {51}\ (\bibinfo {year} {2019})\ p.~\bibinfo {pages} {35},\ \Eprint
  {https://arxiv.org/abs/1907.04833} {arXiv:1907.04833 [astro-ph.IM]}
  \BibitemShut {NoStop}%
\end{thebibliography}%

\end{document}